\documentclass{aa}

\usepackage{hyperref}
\usepackage{graphicx}
\usepackage{natbib}
\usepackage{xcolor}
\usepackage{url}
\usepackage{multirow}
\usepackage{orcidlink}
\usepackage{placeins}
\usepackage{txfonts}
\graphicspath{{./}{figures/}}
\begin{document} 

\title{Surface dynamics and geophysical environment of asteroid (3200) Phaethon}

\author{Hangbin Jo \inst{1,2} \orcidlink{0009-0004-9591-8646} \and
          Masateru Ishiguro \inst{1,2} \orcidlink{0000-0002-7332-2479} \and
          Derek C. Richardson \inst{3} \orcidlink{0000-0002-0054-6850} \and
          Sean E. Marshall \inst{4} \orcidlink{0000-0002-8144-7570} \and
          Tomoko Arai \inst{5} \orcidlink{0000-0002-8835-2013} \and
          Ko Ishibashi \inst{5} \orcidlink{0000-0003-1805-1503}
          }

\institute{Department of Physics and Astronomy, Seoul National University, 1 Gwanak-ro, Gwanak-gu, Seoul 08826, Republic of Korea
   \and SNU Astronomy Research Center, Department of Physics and Astronomy, Seoul National University, 1 Gwanak-ro, Gwanak-gu, Seoul 08826, Republic of Korea
   \and Department of Astronomy, University of Maryland, College Park, MD 20742, USA
   \and Florida Space Institute, University of Central Florida, Orlando, FL 32826, USA
   \and Planetary Exploration Research Center, Chiba Institute of Technology, Narashino, Japan
   \\
              \email{hangbin9@snu.ac.kr, ishiguro@snu.ac.kr}
             }
\date{Received <date> /
Accepted <date>}

\abstract
   {(3200) Phaethon is a $\sim 5$-km-diameter near-Earth asteroid with a small perihelion distance of 0.14 au and is the parent body of the Geminids. JAXA's DESTINY\textsuperscript{+} mission will fly by Phaethon in the near future.}
   {We aim to support the pre-flight planning for the DESTINY\textsuperscript{+} mission by performing a geophysical analysis on Phaethon's surface and near-surface environment utilizing the latest shape model from numerous observations.}
   {We employed the soft-sphere discrete element method code PKDGRAV to construct a "mascon" model of Phaethon and determine its gravity. We then computed the geopotential on Phaethon and derived various physical quantities related to its surface and near-surface dynamics.}
   {We calculated geophysical quantities for the surface, including surface acceleration and slope. To assess whether surface objects could be launched off the surface, we computed the escape speed, return speed, Jacobi speed, and the location and stability of equilibrium points around Phaethon, and conducted a simple dynamical simulation of launched particles.}
   {Our results suggest that a large depression feature in the northern hemisphere could harbor exposed subsurface material and the freshest material on Phaethon. We propose that this depression be considered a key area  for observation by the DESTINY\textsuperscript{+} mission.}
   

\maketitle
%

\section{Introduction} \label{sec:intro}

The physical behavior of asteroids near the Sun is mysterious. \citet{2016Natur.530..303G} compared a detected asteroid population with the debiased near-Earth object population model and suggested a lack of near-Sun asteroid detections, implying that objects with low perihelion distances may be prone to catastrophic disruptions. A near-Sun B-type asteroid (3200) Phaethon may hold a hint to answer this question because its observed mass loss has been reported.  It has an orbit with large eccentricity ($e=0.89$) and a short perihelion distance ($q=0.14$ au). Although the exact mechanism driving the near-Sun "supercatastrophic" disruption is unknown, thermal processes are suspected \citep{2016Natur.530..303G}. In this context, studying Phaethon may provide insight into understanding the effects of the extreme variation of diurnal and seasonal temperature on asteroids \citep{2021A&A...654A.113B,2022PSJ.....3..187H}. Phaethon is also notable for being the parent body of the Geminid meteoroid stream, which is responsible for one of the most prominent annual meteor showers. The proposed mechanisms for the formation of Geminids include cometary activity \citep{1989A&A...225..533G}, thermal fracture \citep{2010AJ....140.1519J,2012AJ....143...66J}, and rotational instability \citep{2020ApJ...892L..22N,2024A&A...683A..68J}. On the other hand, Phaethon has exhibited near-perihelion activity in multiple perihelion passages.  These activities were initially interpreted as dust ejection \citep{2010AJ....140.1519J,2013AJ....145..154L,2013ApJ...771L..36J,2017AJ....153...23H}. However, recent studies favor the idea that the observed activities may instead be due to the outgassing of sodium or iron \citep{2022Icar..38114995L,2023AJ....165...94H,2023PSJ.....4...70Z}. In any case, the connection between the present-day activity of Phaethon and the Geminid formation event remains uncertain. Phaethon is chosen as the fly-by target for the JAXA DESTINY\textsuperscript{+} mission \citep{2018LPI....49.2570A}. Accordingly, this space mission may provide clues to the various mysteries of asteroids near the Sun.

Phaethon has a diameter of $\sim 5$ km and a rotation period of 3.603957 hours \citep{2018AA...619A.123K,2018AA...620L...8H,2022Icar..38815226M}. Although this rotation period is above the critical spin barrier for asteroids \citep{2012AJ....143...66J}, the centrifugal force is comparable to its gravity, creating a complex surface dynamical environment. For this reason, fast-rotating asteroids selected as targets for space missions have been subjected to pre-flight geophysical analysis. For example, \citet{2016Icar..276..116S} provided the pre-encounter surface dynamics analysis of OSIRIS-REx target (101955) Bennu, which was later followed up by the post-encounter analysis of \citet{2019NatAs...3..352S}. Similarly, \citet{2019MNRAS.482.4243Y} studied the surface stability of (65803) Didymos prior to the DART mission. The surface dynamical state of (16) Psyche was investigated by \citet{2020MNRAS.491.3120M} as part of the preparation for the recently launched Psyche mission. Thus, it is crucial to perform a geophysical study on Phaethon to anticipate observational features that DESTINY\textsuperscript{+} will detect during the fly-by phase and to consider a link between Phaethon's current and past activities with this geological information.

\begin{table*}
 \caption{Physical parameters of Phaethon}              
 \label{tab:phaethon}      
 \centering                                      
 \begin{tabular}{ccc}          
 \hline\hline                        
    Parameter [units] & Value & References \\    
 \hline                                   
    Rotation period [hr] & 3.603957 & \citet{2018AA...619A.123K} \\
    Equivalent diameter [km] &  $5.4 \pm 0.1$ & \citet{2022Icar..38815226M} \\
    Mass density [g/cm\textsuperscript{3}] & $1.58 \pm 0.45$ &  \citet{2022Icar..38815226M} \\
    Perihelion distance [au] & 0.14 & JPL Horizons (JD 2460600.5)\tablefootmark{a} \\
    Eccentricity & 0.89 & JPL Horizons (JD 2460600.5)\tablefootmark{a}  \\
    Gravitational escape speed [m/s] & 2.54 & Used nominal value, assuming spherical, non-rotating body \\
 \hline
 \hline                                             
 \end{tabular}
 \tablefoottext{a}{\url{https://ssd.jpl.nasa.gov/horizons.cgi}}
\end{table*}

This work aims to investigate the geophysical environment of Phaethon using the latest shape model as part of the pre-flight preparation efforts for the DESTINY\textsuperscript{+} mission. Since DESTINY\textsuperscript{+} is a flyby mission that can only observe about half of Phaethon’s surface, pre-flight planning and the selection of the closest encounter timing are crucial, making this study a timely contribution to the mission. Section \ref{sec:method} outlines our methods and parameters used to calculate the key physical quantity: the geopotential. In Section \ref{sec:result}, we present the results of these calculations and various derivations related to the dynamical environment on and around Phaethon. The uncertainties and implications of our findings are discussed in Section \ref{sec:discussion}. In Section \ref{sec:conclusion}, we summarized this work. 

\section{Method} \label{sec:method}
In order to calculate and interpret the physical quantities related to the surface dynamics of Phaethon, we must first establish the physical characteristics of the asteroid. In this section, we describe our method for deriving the geopotential of Phaethon.

The geopotential is the sum of the gravitational potential and rotational potential, forming the effective potential in the co-rotating frame of the body. The geopotential equation comprises two parts: the centrifugal and gravitational terms. For the centrifugal term, following observational reports, we assumed that Phaethon is a principal axis rotator with a period of 3.603957 hours \citep{2018AA...619A.123K,2018AA...620L...8H}. In this assumption, the reference frame has its origin at the mass center of Phaethon, with the z-axis aligned with the rotation axis. With this setup, the geopotential at a point $\boldsymbol{r}$ is given as follows:
\begin{equation} \label{eq:geopot}
 V(\boldsymbol{r}) = -\frac{1}{2}(\boldsymbol{\omega} \times \boldsymbol{r}) \cdot (\boldsymbol{\omega} \times \boldsymbol{r}) + U(\boldsymbol{r}),
\end{equation}
where $U(\boldsymbol{r})$ is the gravitational potential at $\boldsymbol{r}$ and $\boldsymbol{\omega} = \omega \hat{z}$ is the spin vector of Phaethon with angular velocity $\omega$.
Although the centrifugal term can be uniquely computed once the rotational period is given, the calculation of the gravitational potential is more complicated. This is because, like most small bodies, Phaethon has an irregular shape. The analytical formula for the gravitational potential of an irregular shaped polyhedron was established with the key assumption that the density of the polyhedron is homogeneous \citep{1996CeMDA..65..313W}. Another method is the mass concentration (mascon) method \citep{1996Icar..120..140G}, where the body is filled with point masses of equal mass along a cubic grid space, whose combined gravity represents the gravity of the body. Although this configuration was criticized for its inaccuracy \citep{1996CeMDA..65..313W}, the spatial distribution of the point masses can be adjusted to reduce this limitation and emulate potential heterogeneity in asteroids \citep{2016odgf.book.....Y,2022CeMDA.134...58P}. 

In this work, we used the soft-sphere discrete element method (SSDEM) in the parallel N-body code PKDGRAV \citep{2000Icar..143...45R,2001PhDT........21S,schwartz2012implementation} to fill Phaethon's shape model with spherical particles. We divided the asteroid into two regions: the core and the envelope. The core consists of a 1.5-km radius spherical area where the particles, each 80 m in radius, are arranged in a hexagonal close packing (HCP) configuration. After generating the core, we surrounded it with a cloud of randomly distributed particles. By allowing these particles gravitate naturally toward the core, we covered the core with a random close packing (RCP) envelope. The radii of the envelope particles are within the range of 10 - 80 m and follow the boulder size-frequency distribution of the asteroid (162173) Ryugu, as investigated by Hayabusa2 \citep{2019Icar..331..179M}. This two-layer structure represents a rubble-pile asteroid with a dense core, chosen to account for the possibility that the Geminids formed from Yarkovsky–O'Keefe–Radzievskii–Paddack (YORP)-induced rotational instability \citep{2024A&A...683A..68J}. Previous numerical studies have shown that surface mass shedding due to rotational instability is more likely on bodies with a dense core or particles of a wide size range  \citep{2015ApJ...808...63H, 2017Icar..294...98Z,2021AsDyn...5..293Z}. 

\begin{table*}
 \caption{PKDGRAV simulation parameters}              
 \label{tab:param}      
 \centering                                      
 \begin{tabular}{ccc}          
 \hline\hline                        
    Parameter [units] & Value & Reference \\    
 \hline                                   
    Particle mass density [g/cm\textsuperscript{3}] &  2.38 & \citet{2022Icar..38815226M} \\
    Number of particles (core) & 4583 & - \\
    Number of particles (envelope) & 48287 & - \\
    Timestep [s] &  0.05  &  - \\
    Shape parameter $\beta$ & 0.8 & \citet{2020AA...640A.102Z, 2024PSJ.....5...54A}  \\
    Normal spring constant $k_n$ [kg/s\textsuperscript{2}] & $9.22 \times 10^6$ & - \\
    Tangential spring constant $k_t$ & $\frac{2}{7} k_n$ & \citet{schwartz2012implementation} \\
    Static friction coefficient $\mu_s$ & 1.0 & \citet{JIANG2015147,2020AA...640A.102Z,2024PSJ.....5...54A} \\
    Rolling friction coefficient $\mu_r$ & 1.05 & \citet{JIANG2015147} \\
    Twisting friction coefficient $\mu_t$ & 1.3 & \citet{JIANG2015147} \\
    Normal coefficient of restitution $\eta_n$ & 0.55 & \citet{CHAU200269} \\
    Tangential coefficient of restitution $\eta_t$ & 0.55 & \citet{CHAU200269} \\
    Cohesion coefficient $c$ [Pa] & 0 & \citet{2020Sci...368...67A,2022SciA....8M6229W} \\
 \hline
 \hline                                             
 \end{tabular}
\end{table*}

The material parameters used in this study follow conventional values used in previous studies (e.g. \citealt{2020AA...640A.102Z}). We summarize these values, along with their references, in Table \ref{tab:param}. For the static friction coefficient $\mu_n$ and the shape parameter $\beta$, we selected values that align with the approximate range of friction angles observed in terrestrial materials \citep{2015ApJ...808...63H, 2017Icar..294...98Z}. The normal spring constant $k_n$ and the simulation timestep were chosen to ensure that the maximum velocity, equivalent to the freefall speed, would limit interparticle overlaps to no more than 1\% of the minimum particle radius during the simulation. We note that this $k_n$ value is roughly equivalent to the elastic Young's modulus of $\sim 1$ MPa, which is approximately analogous to porous terrestrial gravels \citep{2019Icar..328...93D}. Studies of Ryugu by the Hayabusa2 mission and (101955) Bennu by the OSIRIS-REx mission suggest that the regoliths of both asteroids have nearly zero cohesion \citep{2020Sci...368...67A, 2022SciA....8M6229W}. \citet{2021P&SS..20405268R} and \citet{2022NatCo..13.4589Z} proposed that the interiors of these asteroids also have low cohesion. In accord with these findings, we assumed that the particles in our model are cohesionless. 

\begin{figure}
 \resizebox{\hsize}{!}{\includegraphics{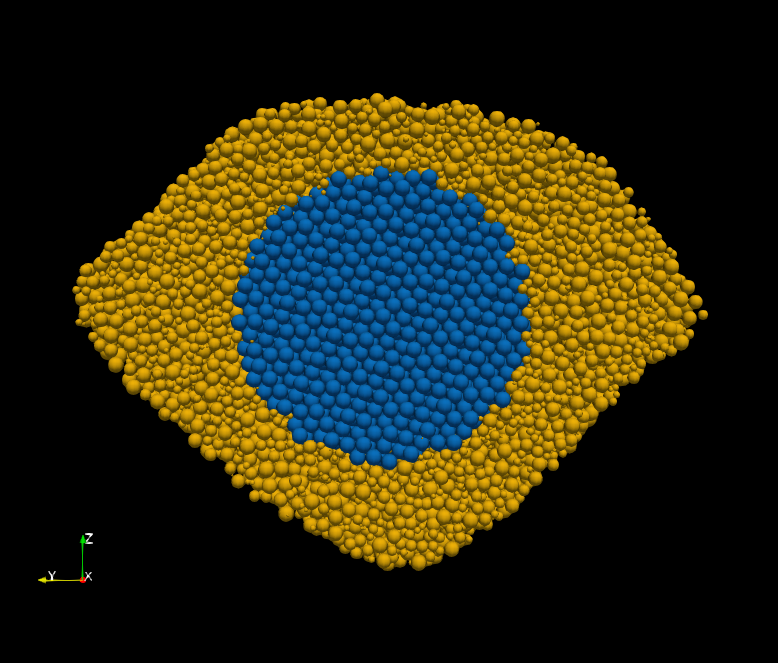}}
 \caption{Cross-section of the mascon model built using PKDGRAV. The white grid overlay shows the original 3D shape model of Phaethon. The core (blue) and envelope (yellow) particles are separately colored.
 \label{fig:pkdgrav_model}}
\end{figure}

Once particle aggregation was complete, the modeled asteroid was rotated at Phaethon's rotation period until the particles stabilized in the rotating environment. Next, we carved the aggregate into the shape model of Phaethon. The Phaethon shape model used in this work is based on twelve days of radar observations (from two apparitions), 195 lightcurves (from 18 oppositions), and eight occultations (from three apparitions). Its volumetric mean diameter is 5.0 km, with the $x$--, $y$-- and $z$--diameters of the dynamically equivalent, equal volume ellipsoid (DEEVE) measuring 5.88, 5.43 and 4.01 km respectively (Marshall et al., in prep.).

Finally, the carved model was rotated again at Phaethon's rotation period to complete the model preparation process. A cross-sectional view of the modeled asteroid is shown in Fig. \ref{fig:pkdgrav_model}. The envelope has a packing efficiency of 60.2\%. For the core, the theoretical HCP packing efficiency is $\frac{\pi}{3 \sqrt{2}} \approx 74 \%$, leading to a total packing efficiency of about 63.1\%. This corresponds to a bulk porosity of 36.9\% for the model. We adjusted the mass density of each particle to match Phaethon's bulk density of 1.58 g / cm\textsuperscript{3} \citep{2022Icar..38815226M}. In summary, we created a Phaethon discrete element model with a two-layer structure consisting of a robust HCP core and a polydisperse RCP envelope. Moving forward, we will apply this model in the mascon method, approximating the total gravitational potential of the body by summing the gravitational potential of each particle.

\section{Results} \label{sec:result}
In this section, we present our results on the geopotential in Sect. \ref{subsec:geopot}, the dynamical state on the surface in Sect. \ref{subsec:surf_dyn} and above the surface in Sect. \ref{subsec:off_surf_dyn}.

\subsection{Geopotential} \label{subsec:geopot}
\begin{figure*}
  \centering 
  \begin{tabular}{ccc}
   \includegraphics[width=0.3\textwidth ]{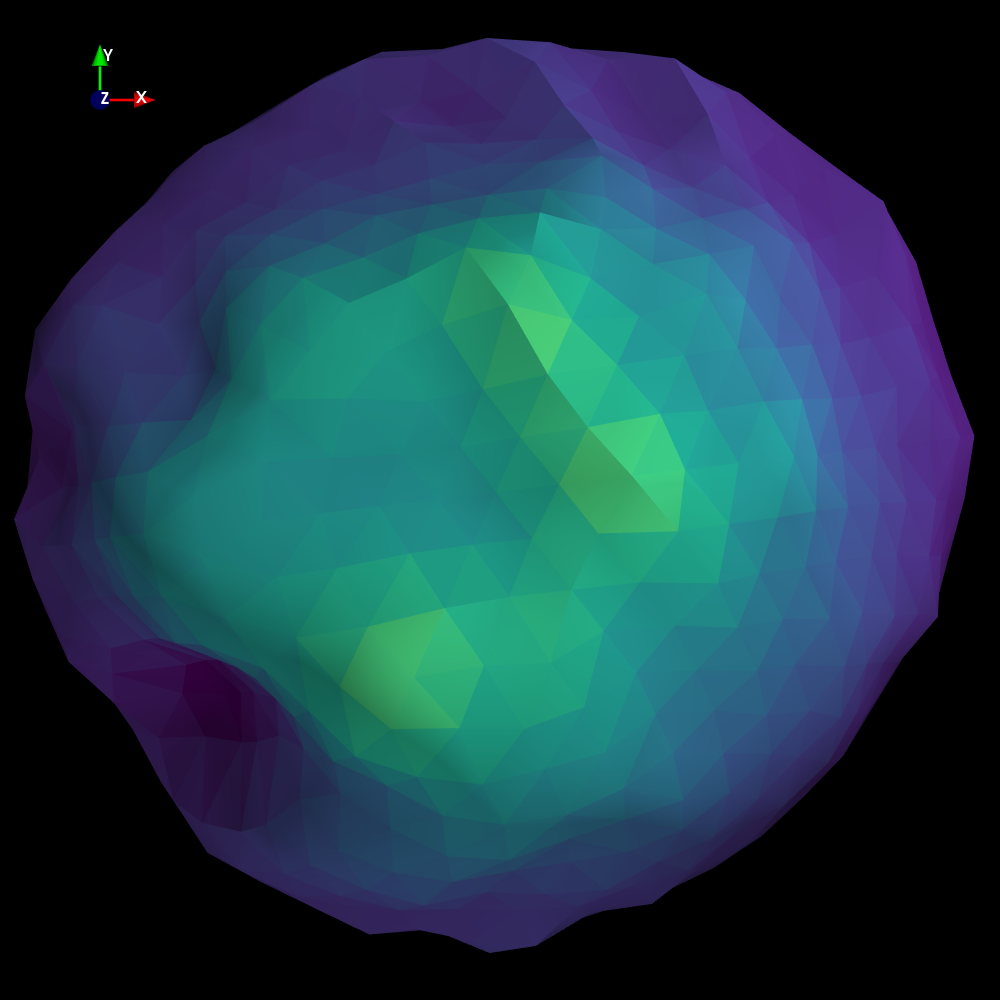} & 
   \includegraphics[width=0.3\textwidth ]{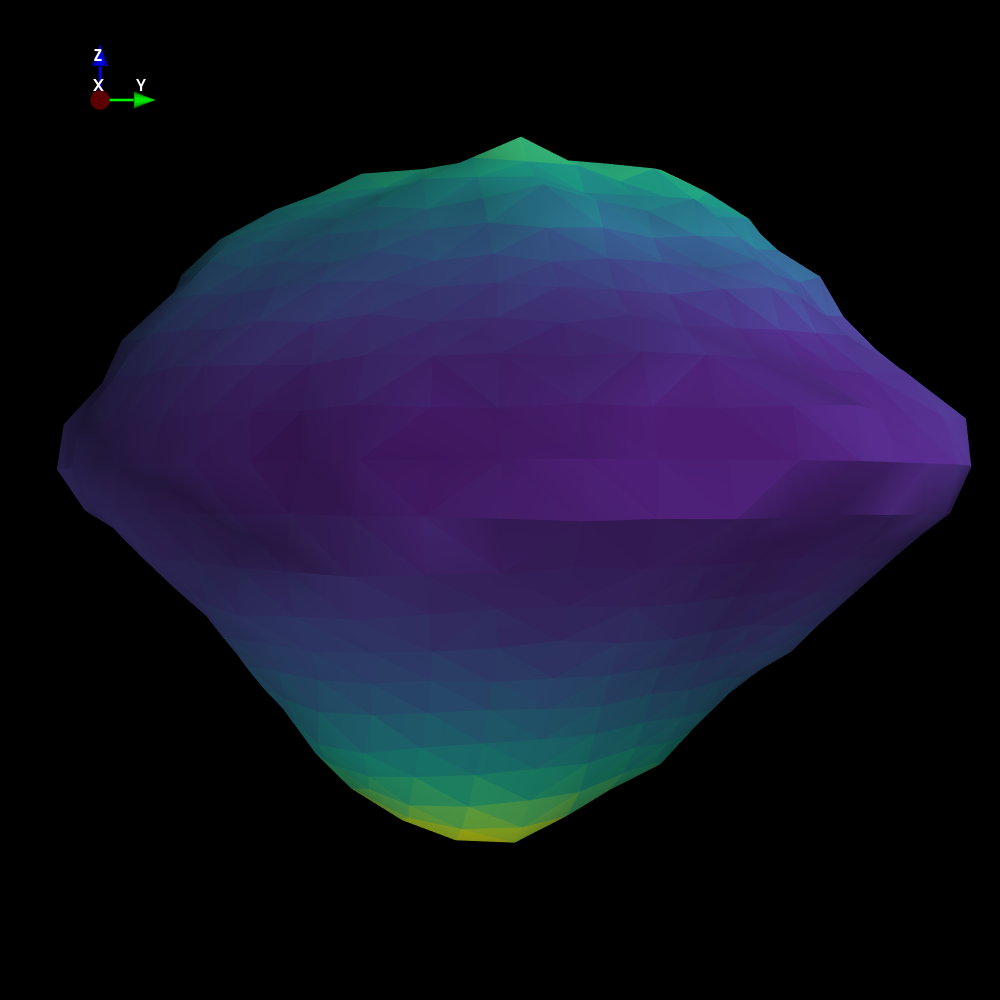} &
   \includegraphics[width=0.3\textwidth ]{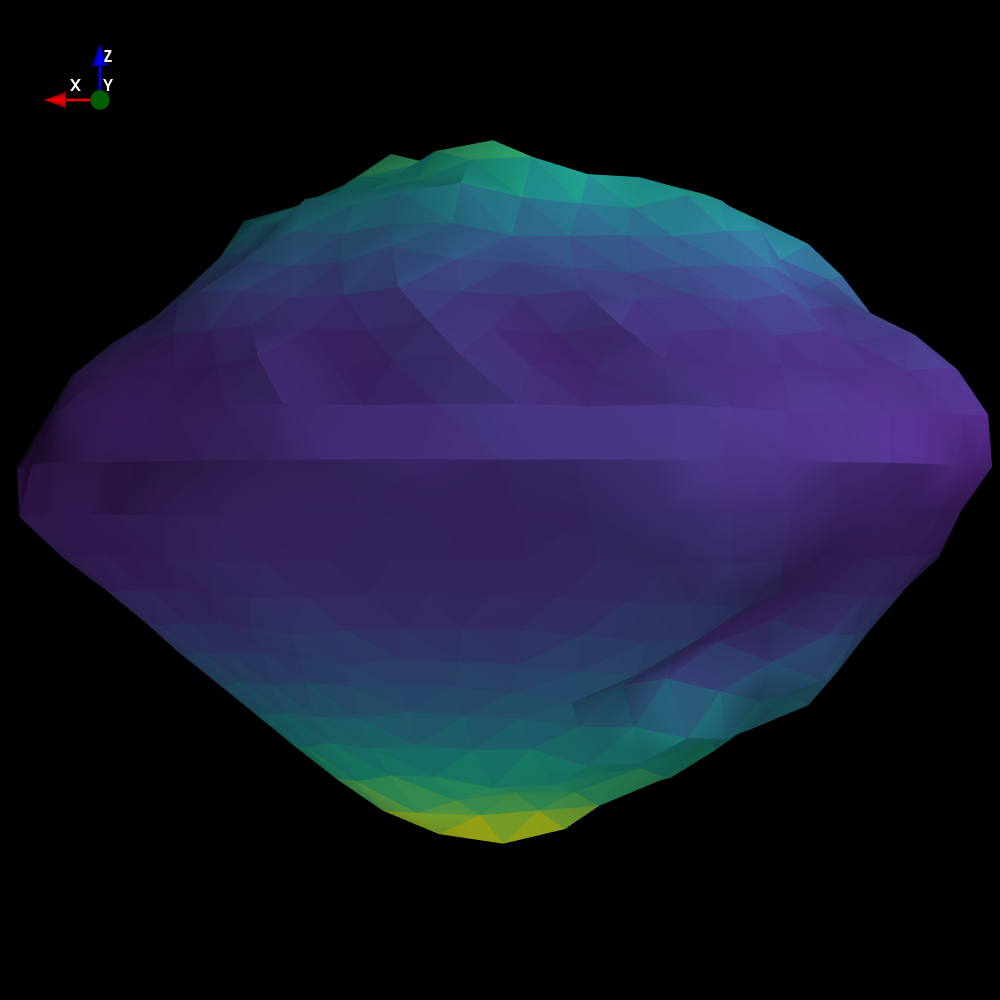}\\
   \includegraphics[width=0.3\textwidth ]{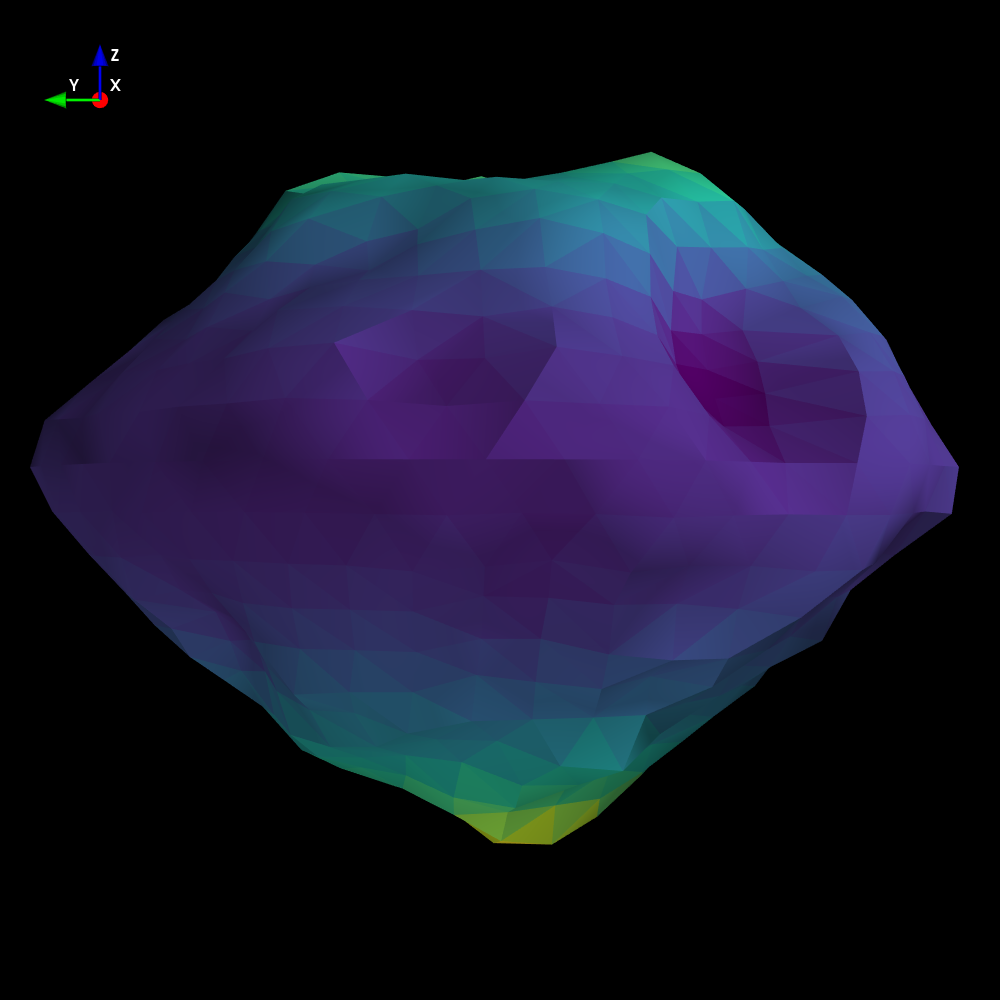} & 
   \includegraphics[width=0.3\textwidth ]{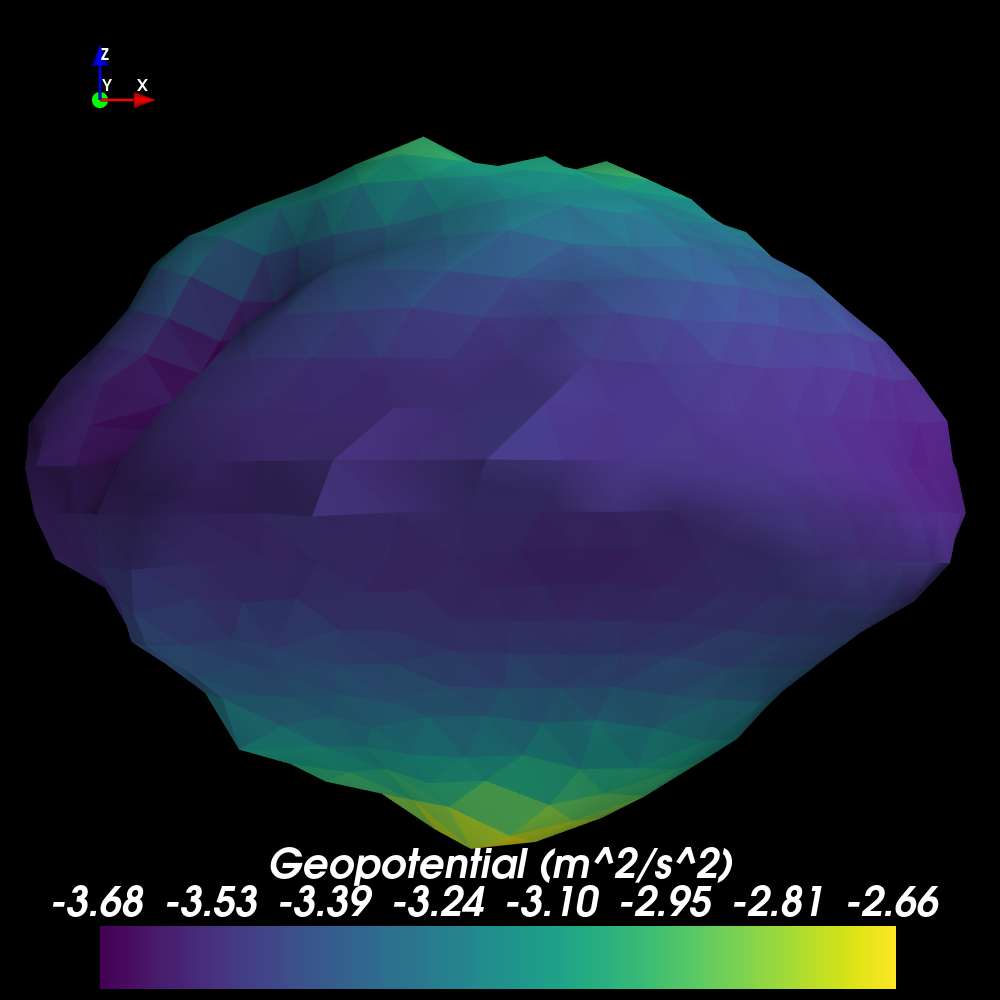} &
   \includegraphics[width=0.3\textwidth ]{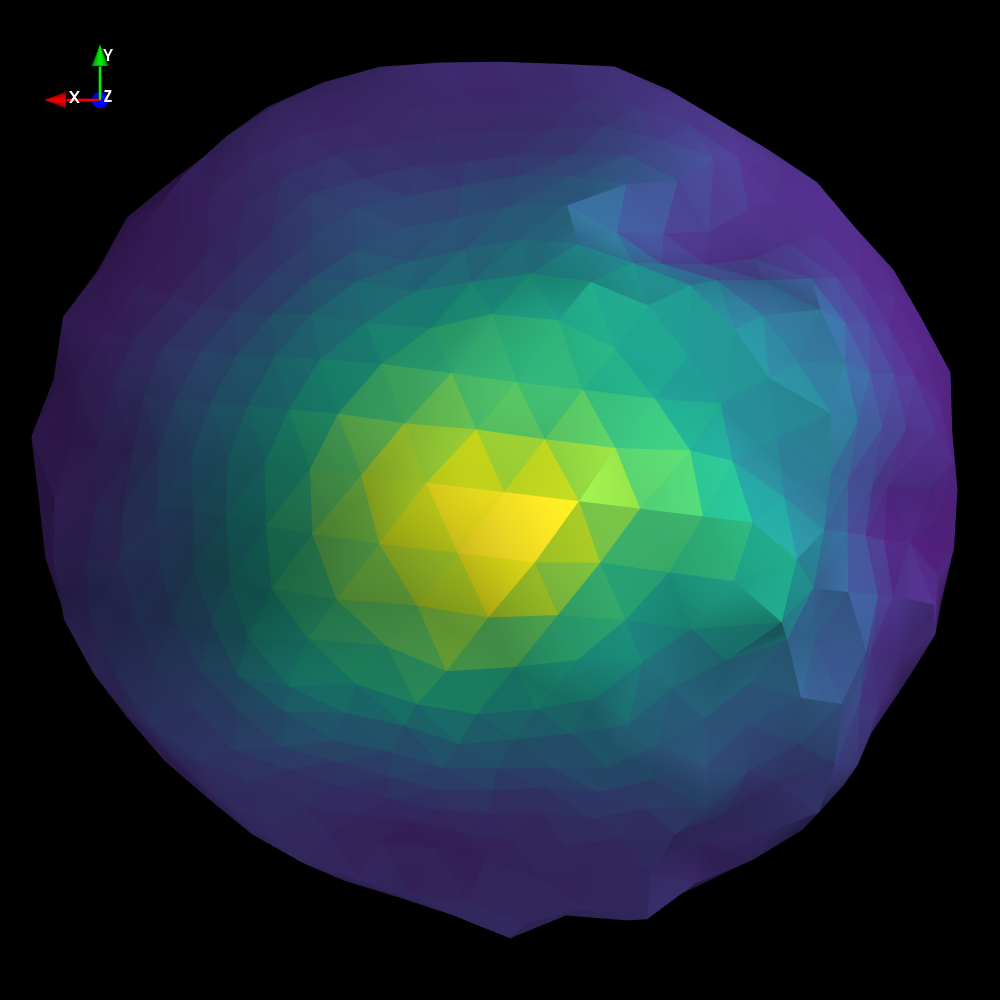}\\
  \end{tabular}
  \caption{Geopotential maps calculated using the mascon method. From left to right and from top to bottom, each panel shows the geopotential map oriented in the $-z, -x, -y, +x, +y,$ and $+z$ directions.
  \label{fig:geopotential}}
\end{figure*}

Using the mascon model, we calculated the geopotential at the centroid of each triangular face of the shape model, representing the geopotential of the face itself. In some faces, the absolute geopotential values were unusually high due to certain particle mass centers being too close to the surface. In these cases, we adjusted the distances from the problematic particles to the surface to be equal to their radii. Fig. \ref{fig:geopotential} shows the resulting geopotential map of Phaethon.

To confirm the accuracy of the mascon method, we calculated the geopotential using the polyhedron method for comparison and present the results in the Appendix. Fig. \ref{fig:pot_diff_simple} shows the relative difference between the two methods, assuming a uniform density that is equal to the bulk density of the mascon model for the polyhedron method. To account for the different structure assumptions between the two methods, we also calculated the polyhedron method by setting the bulk density to that of the mascon envelope (1.40 g/cm\textsuperscript{3}), then replacing the density of the central 1.5-km-radius spherical volume with the core bulk density (1.77 g/cm\textsuperscript{3}). The result is shown in Fig. \ref{fig:pot_diff_core}, where one can see that the relative difference between the two methods is reduced.

In both cases, while some of the relative error can be attributed to the mascon resolution, the mascon geopotential tends to be higher in the northern mid-to-high-latitude region compared to the polyhedron method. It is important to note that the mascon model was carved from a larger spherical aggregate, where the particles that are closer to the core would have been more compressed, causing them to be more tightly packed than those near the surface. This feature may have remained after carving it into Phaethon’s shape model, as the higher-latitude regions in the northern hemisphere are radially closer to the core, they correspond to the deeper and denser part of the original sphere model. Nonetheless, our "realistic" mascon model produces geopotential values consistent with the polyhedron method’s analytical solution, with only minor differences likely due to local heterogeneity and model resolution.

From the geopotential, we derived several quantities to understand the possible dynamical state of Phaethon's surface and near-surface materials, and eventually to consider science cases before the DESTINY\textsuperscript{+} flyby. The rest of this section mainly employs the methods in \citet{2012omsp.book.....S} and \citet{2016Icar..276..116S}. Unless otherwise specified, the following results are based on the mascon method calculation.

\subsection{Surface dynamical state} \label{subsec:surf_dyn}
Based on our model, we calculated the acceleration on the surface, the tilt, and the slope to characterize the dynamical environment on the Phaethon surface. Each result is reported in the following subsections.

\subsubsection{Surface acceleration} \label{subsec:surf_acc}

The gradient of the geopotential $\nabla V$ at the position $\boldsymbol{r}$ corresponds to the acceleration due to both gravity and the rotation of the body. By extracting the component orthogonal to the normal surface vector, we calculated the surface acceleration $a_t$.

\begin{equation} \label{eq:surfacc}
 a_t= \nabla V - (\nabla V \cdot \hat{n})\hat{n},
\end{equation}
where $\hat{n}$ is the surface normal unit vector. Fig. \ref{fig:surf_acc} illustrates the direction and magnitude of Phaethon's surface acceleration, which is higher in the high-latitude regions and lower in the equatorial region, implying that, at the current state, surface materials tend to migrate from polar to equatorial areas.

In realistic conditions, the acceleration of the surface by the geopotential is countered by other forces, particularly friction. For static friction, it is given by
\begin{equation} \label{eq:friction}
 f =  (\mu_s \nabla V \cdot \hat{n})\frac{a_t}{|a_t|},
\end{equation}
where $\mu_s$ denotes the static friction coefficient, which we assumed to be $\mu_s$=1.0 (see Table \ref{tab:param}). Taking friction into account, the net surface acceleration is then given by
\begin{equation} \label{eq:net_acceleration}
 a_\mathrm{net}= \begin{cases}
			a_t - f , & \text{if $|a_t| \geq |f|$}\\
            0, & \text{otherwise.}
		 \end{cases}
\end{equation}

Fig. \ref{fig:net_acc} shows the revised acceleration maps after accounting for friction. On most faces, static friction offsets surface acceleration. In areas where surface acceleration is strong enough to overcome friction, the resulting net surface acceleration reach up to $\sim 1 \times 10^{-4}$ m/s\textsuperscript{2}.

In addition to friction, cohesion can help counter surface acceleration. The relation between cohesion $c$ and the cohesive force $F_c$ for a spherical particle of radius $r$ can be approximated as $c \sim \frac{F_c}{2 \pi r^2}$ \citep{2014M&PS...49..788S}. Therefore, for cohesion to fully counteract the net surface acceleration, 
\begin{equation} \label{eq:cohesion}
 c \gtrsim \frac{2}{3} \rho s a_\mathrm{net},
\end{equation}
where $\rho$ and $s$ are the mass density and radius of a spherical surface object, respectively. Polarimetric observations of Phaethon estimate a characteristic grain diameter of $\sim 0.36$ mm \citep{2018NatCo...9.2486I}, while thermal conductivity modeling suggests a range from tens of microns to centimeters \citep{2022Icar..38815226M}. This grain size range is comparable to that observed on Ryugu through polarimetry \citep{2021ApJ...911L..24K}, as well as the measurements from returned samples of Ryugu and Bennu \citep{2023EP&S...75..171M,2024M&PS...59.2453L}. 
As such, it is not unreasonable to assume that the grain size range on Phaethon is similar to those of Ryugu and Bennu. Based on the returned samples from Ryugu and Bennu, we consider a representative spherical grain particle of 1-mm radius and 1.8 g/cm\textsuperscript{3} \citep{2023EP&S...75..171M,2024M&PS...59.2453L}. In this case, the cohesion needed to counter the net surface acceleration in Fig. \ref{fig:net_acc} is $c \gtrsim 10^{-4}$ Pa. The estimated cohesion on Ryugu and Bennu is reported to be $\lesssim 1$ Pa \citep{2020Sci...368...67A,2022SciA....8M6229W}. Thus, if Phaethon's surface exhibits a similar cohesion level, even minimal cohesion of $\sim 10^{-4}$ Pa should be sufficient to counteract the net surface acceleration.

\subsubsection{Slope} \label{subsec:slope}

On a massive, slowly rotating body like Earth, the geopotential is primarily governed by the gravitational term. As such, the geometric steepness of a surface strongly correlates with the geopotential gradient. However, this is not the case for fast-rotating asteroids, where the centrifugal term of the geopotential is comparable to the gravitational component in the equation. For such bodies, the steepness of a surface does not necessarily indicate larger changes in geopotential. To address this, \citet{2016Icar..276..116S} proposed distinct definitions for the geometric and geopotential orientations of a surface, referring to the former as "tilt" and the latter as "slope".

Tilt is defined as the angle between the normal vector of the surface and the radial vector from the center of mass to the surface. Meanwhile, slope is the angle between the normal vector of the surface and the geopotential gradient (or geopotential acceleration) vector. Essentially, the concept of slope aligns with the everyday terrestrial intuition that objects tend to move downhill more easily on steeper surfaces. 

Fig. \ref{fig:slope} displays the slope of each face in Phaethon’s shape model. Slope stability can be evaluated with the "factor of safety" \citep[or FS;][]{1997RvGeo..35..245I,2022JGRE..12706927B}. For slope angle $\theta$, a surface can be considered unstable if:
\begin{equation} \label{eq:FS}
 \tan \theta > \tan \theta_r \left(1 - \frac{c}{\rho_e |\nabla{V}| T \sin \theta_r}\right)^{-1},
\end{equation}
where $\theta_r$ is the angle of friction (or angle of repose), $\rho_e$ is the bulk density of the envelope, and $T$ is the depth of the surface layer that is prone to failure. Using our PKDGRAV model parameters, we find $\theta_r \sim 40 \degr$ \citep{2024PSJ.....5...54A}. Assuming $c=0$, Eq. \ref{eq:FS} simply tests whether the slope exceeds the angle of friction; if $\theta > 40 \degr$, the surface material on that face is susceptible to movement. For non-zero cohesion, we assume $T$ to be 10 m, similar to Bennu and Ryugu \citep{2020JGRE..12506475J,2022NatGe..15..447P,2022JGRE..12706927B, 2020Sci...368...67A}. However, given Phaethon's larger diameter ($\sim 5$ km) compared to Bennu and Ryugu (< 1 km), it is unclear whether this assumption holds but we adopt this value as an illustrative example. At the upper limit of $c = 1$ Pa, slopes become unstable when $\theta \gtrsim 44 \degr$. Thus, there is an uncertainty of $\sim 4 \degr$ in the slope stability criterion due to cohesion. Additionally, in reality, the angle of friction is influenced by particle geometry, roughness, size, etc. \citep{BEAKAWIALHASHEMI2018397}. In fact, the particles in our PKDGRAV model remain still throughout several rotation periods, indicating stability even with zero cohesion. Therefore, we view this analysis not as conclusive evidence of current surface instability but rather as a theoretical framework to identify regions potentially susceptible to mass movement if triggered \citep{2023MNRAS.520.3405K}.

\subsection{Dynamics off the surface} \label{subsec:off_surf_dyn}
\subsubsection{Equilibrium points} \label{subsec:equilibrium}

\begin{figure*}
 \resizebox{\hsize}{!}{\includegraphics{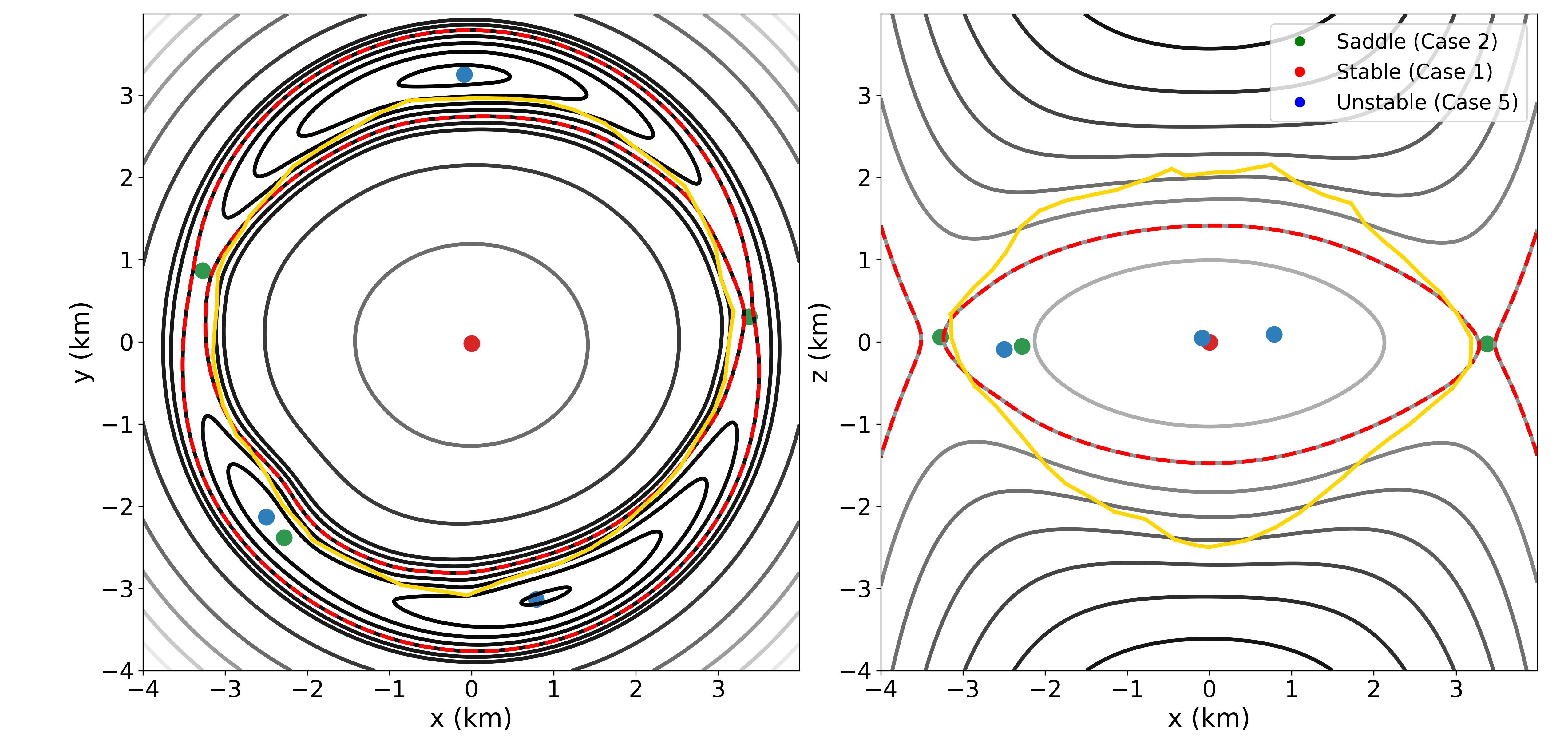}}
 \caption{Equilibrium and zero-velocity curves around Phaethon. The Phaethon's outline is drawn by the yellow lines. The zero-velocity curves (ZVCs) are represented by gray lines with larger $V$ having darker hues. The ZVC interval for the left panel was adjusted to enhance the geopotential around the equilibrium points The red dotted lines are the ZVCs of the minimum geopotential outside of Phaethon, also known as the "rotational Roche lobe" \citep{2012omsp.book.....S}.
 \label{fig:equilibrium}}
\end{figure*}

\begin{table*}
\caption{Equilibrium point (EP) coordinates, $V$ and stability}              
\label{tab:equil_coord}      
\centering                                      
\begin{tabular}{c c c c c c}          
\hline\hline                        
Point & x (km) & y (km) & z (km) & $V$ (m\textsuperscript{2}/s\textsuperscript{2}) & Stability type\\    
\hline                                   
    E1 &  0.78 & -3.13  & 0.09 & -3.48   &  Sink-Source-Center (Case 5)\\
    E2 & -2.50 & -2.13 & -0.09 & -3.52  &   Sink-Source-Center (Case 5)\\
    E3 &  -2.28 & -2.38 & -0.06 & -3.52 &  Saddle-Center-Center (Case 2)\\
    E4 &  0.00 & -0.02 & -0.01 & -4.15    &   Stable (Case 1)\\
    E5 &  3.38 &  0.31 & -0.03 & -3.58 &  Saddle-Center-Center (Case 2)\\
    E6 &  -3.27 &  0.87 &  0.06 & -3.58 &  Saddle-Center-Center (Case 2)\\
    E7 &  -0.09 &  3.25 &  0.05 & -3.48  &   Sink-Source-Center (Case 5)\\    
    \hline
\hline                                             
\end{tabular}
\end{table*}

We can infer from Fig. \ref{fig:geopotential} that in the tug-of-war between gravity and centrifugal force in the rotating frame, the former is dominant on the Phaethon's surface. This suggests the existence of a point above the surface where these forces balance each other. Such points, where $\frac{\partial V}{\partial \boldsymbol{r}} = 0$ are called equilibrium points \citep[EPs,][]{2012omsp.book.....S,2016Icar..276..116S}. Performing accurate calculations and characterization of EPs using our mascon model was computationally challenging. Therefore, for this section, we used the Minor-Equilibria-NR code \citep{2021JGRE..12606272A} and the homogeneous polyhedron method for this section's EP calculations. Fig. \ref{fig:equilibrium} summarizes the locations and types of EPs. Their detailed coordinates can be found in Table \ref{tab:equil_coord}, along with their $V$ values and stability types. The stability type is based on the classification of the topological structure by \citet{2014Ap&SS.349...83J} and \citet{2021JGRE..12606272A}. As shown in Sect. \ref{subsec:geopot}, we recall that the geopotential used in this code would have a < 2\% deviation from our mascon model on the surface. Nonetheless, we deem the results to be close approximations of the EPs. In summary, there are seven EPs found around Phaethon: one stable point inside Phaethon, three saddles, and three unstable center points outside Phaethon.

\subsubsection{Escape speed} \label{subsec:escape}

\begin{figure*}
  \centering 
  \begin{tabular}{ccc}
   \includegraphics[width=0.3\textwidth ]{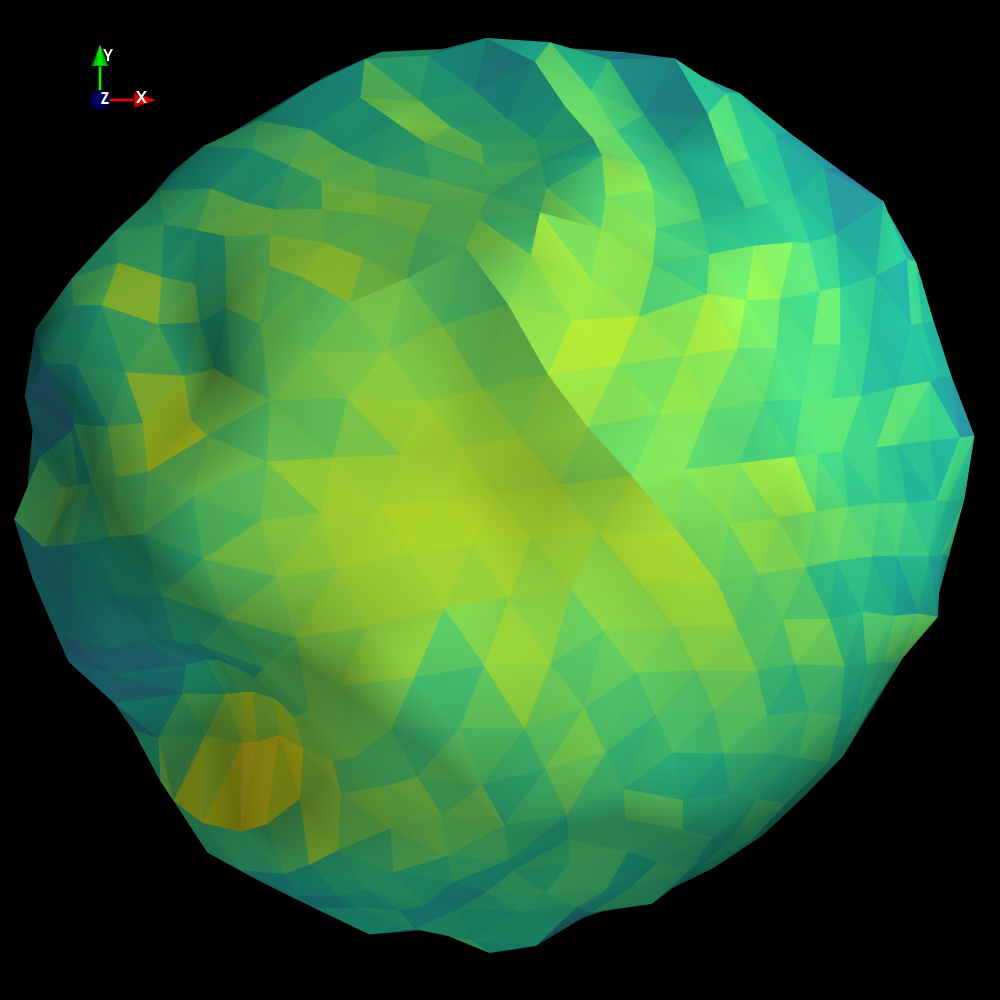} & 
   \includegraphics[width=0.3\textwidth ]{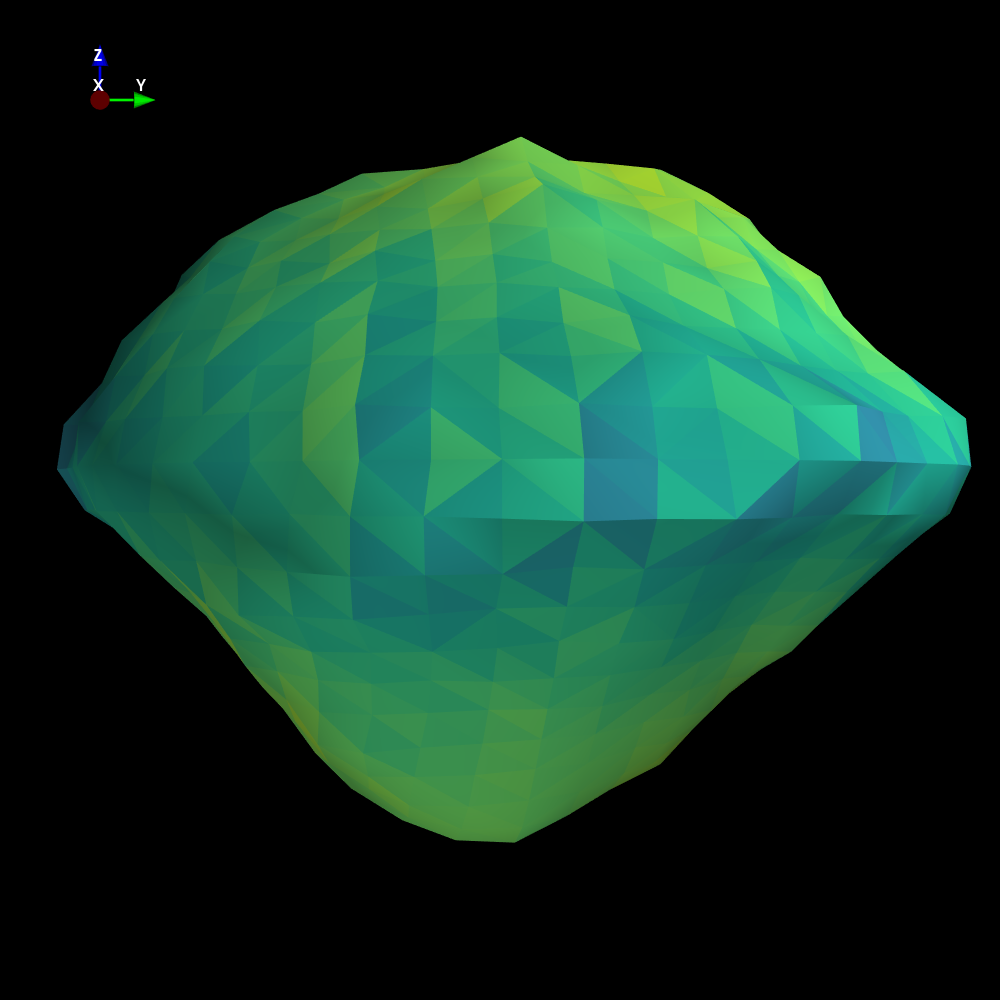} &
   \includegraphics[width=0.3\textwidth ]{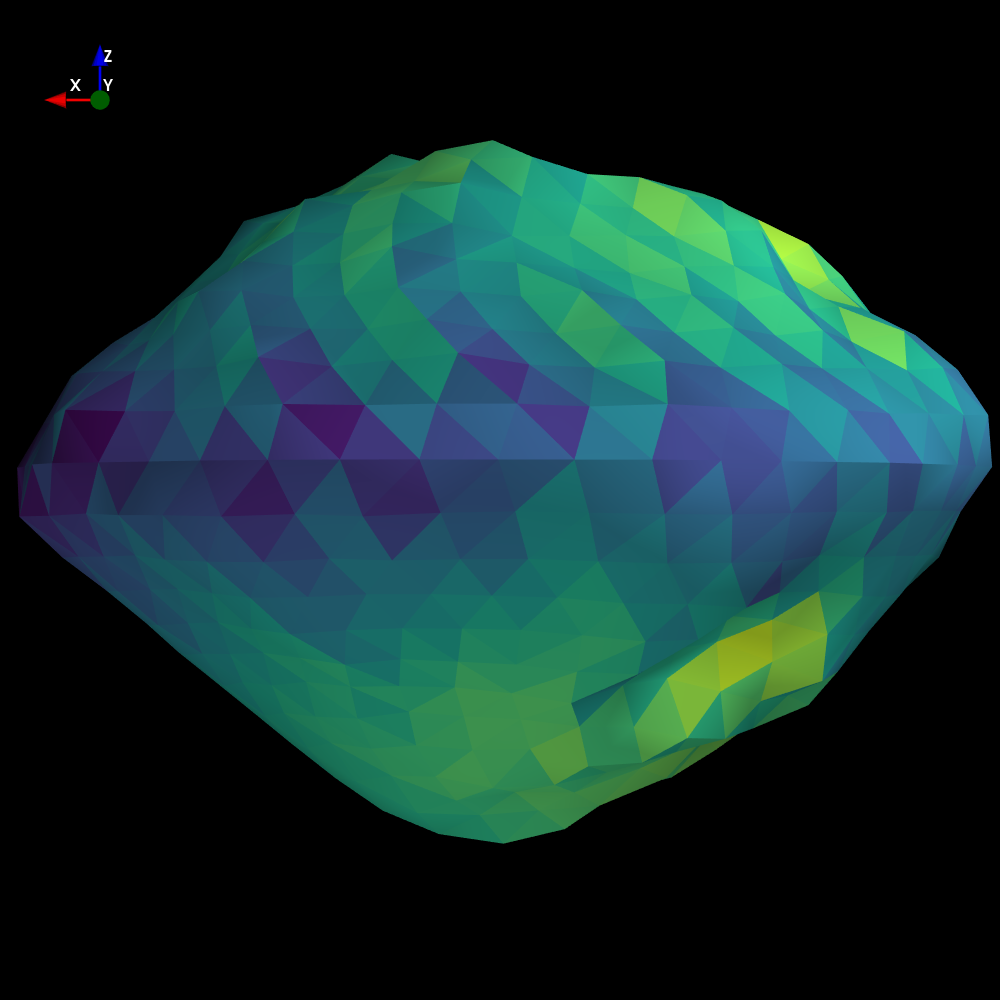}\\
   \includegraphics[width=0.3\textwidth ]{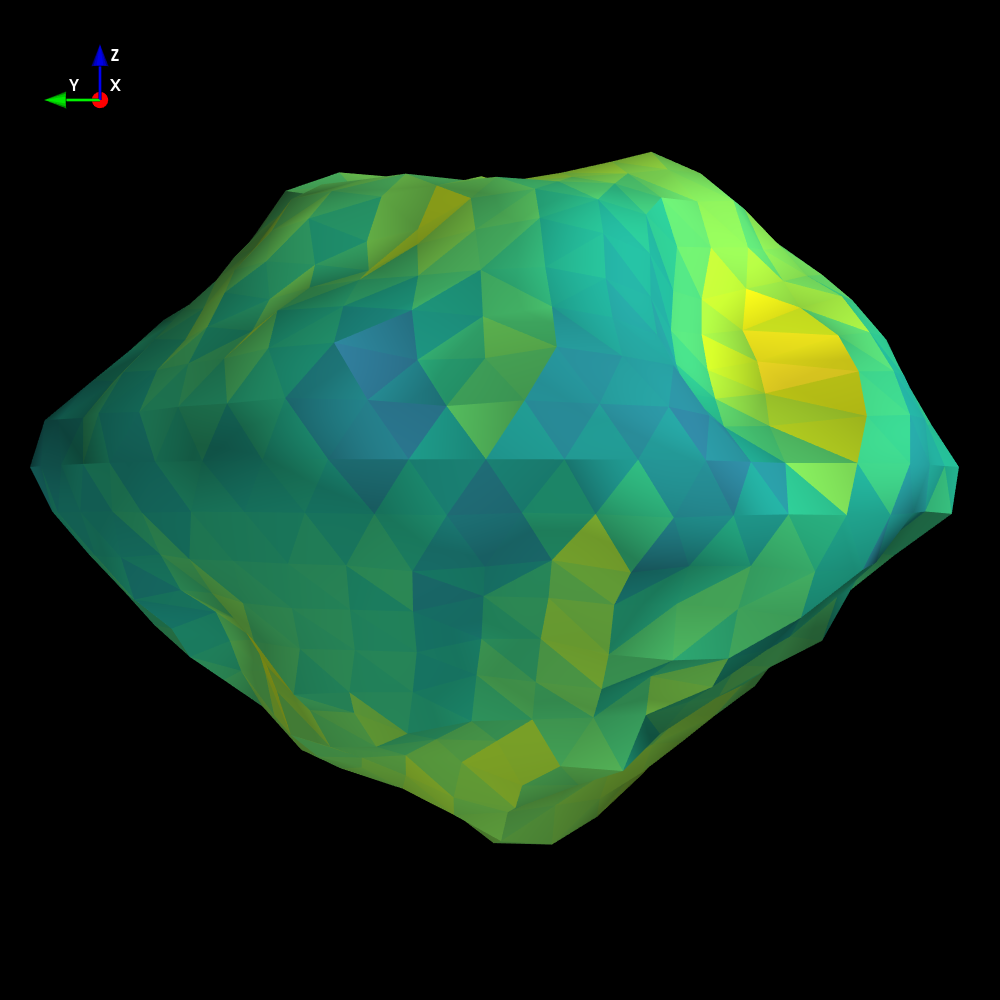} & 
   \includegraphics[width=0.3\textwidth ]{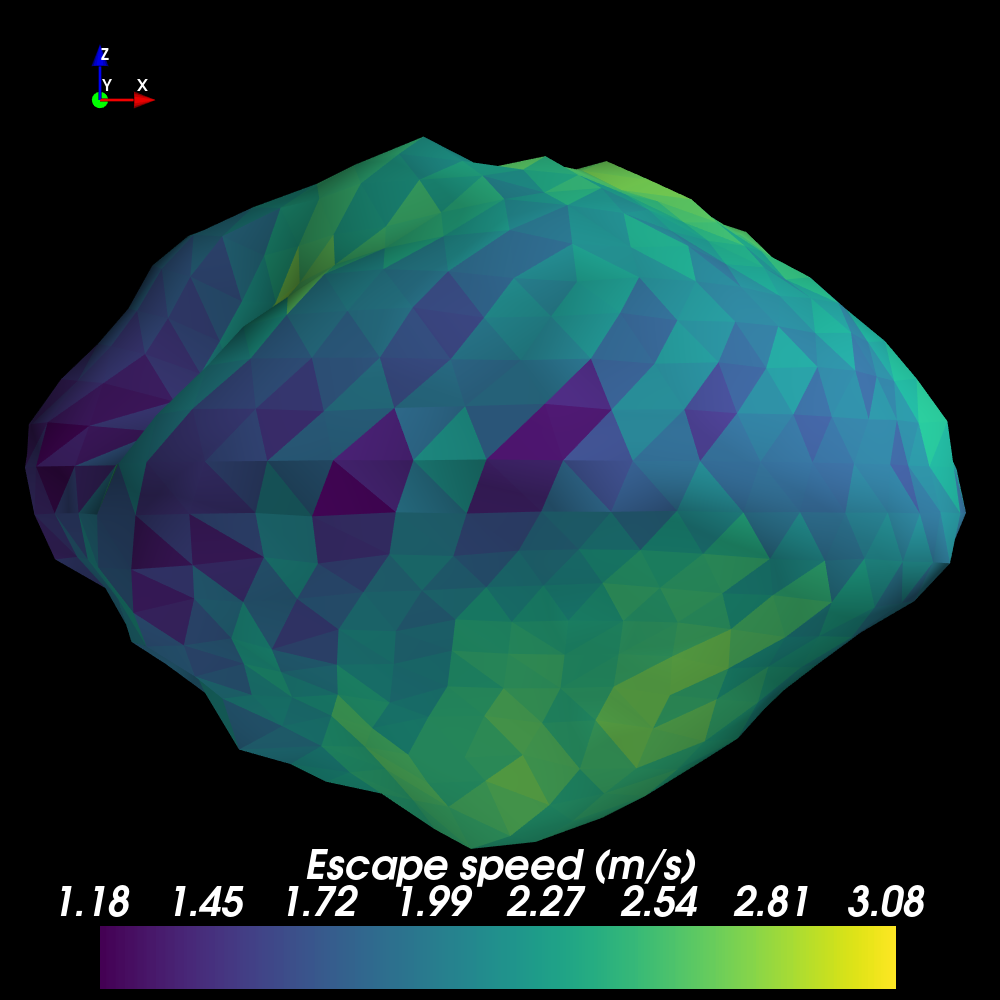} &
   \includegraphics[width=0.3\textwidth ]{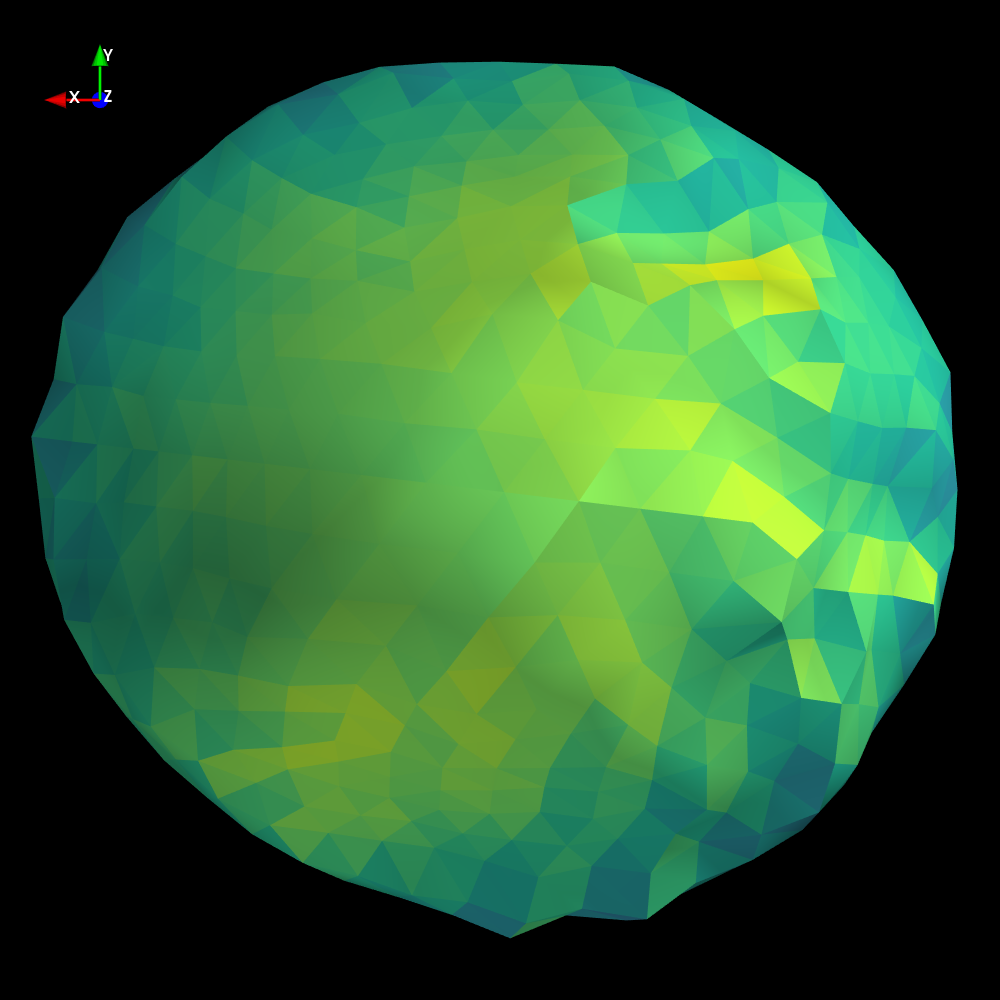}\\
  \end{tabular}
 \caption{Map of the escape speeds on Phaethon's surface.
 \label{fig:escape}}
\end{figure*}

If the velocity of a particle is sufficiently high, combined with the rotational velocity of the parent asteroid, the particle can overcome the asteroid's gravity. Following the definition by \citet{1998Icar..132...53S} and \citet{2012omsp.book.....S}, the escape speed is the minimum velocity in the surface-normal direction within the co-rotating frame required for an object to be immediately ejected on an escape trajectory from the parent body. After converting this definition to the inertial velocity, $v_I = v_\mathrm{esc}\hat{n} + (\boldsymbol{\omega} \times \boldsymbol{r})$, and combined with the escape condition $\frac{1}{2} {v_I}^2 \geq U$, the assumption that the ejection direction aligns with the surface normal simplifies the equations to a quadratic form. Solving the quadratic equation provides:

\begin{equation} \label{eq:v_esc}
 v_\mathrm{esc}(\boldsymbol{r}) = -\hat{n} \cdot (\boldsymbol{\omega} \times \boldsymbol{r}) + \sqrt{[\hat{n} \cdot (\boldsymbol{\omega} \times \boldsymbol{r})]^2 +2 U_{max}(\boldsymbol{r}) - (\boldsymbol{\omega} \times \boldsymbol{r})^2},
\end{equation}
where $U_\mathrm{max} = \mathrm{max}[U(\boldsymbol{r}), GM/r]$, with $G$ being the gravitational constant and $M$ the total mass of the asteroid. The map of the escape speed is displayed in Fig. \ref{fig:escape}.

\subsubsection{Rotational Roche lobe and return speed} \label{subsec:return}

\begin{figure*}
  \centering 
  \begin{tabular}{ccc}
   \includegraphics[width=0.3\textwidth ]{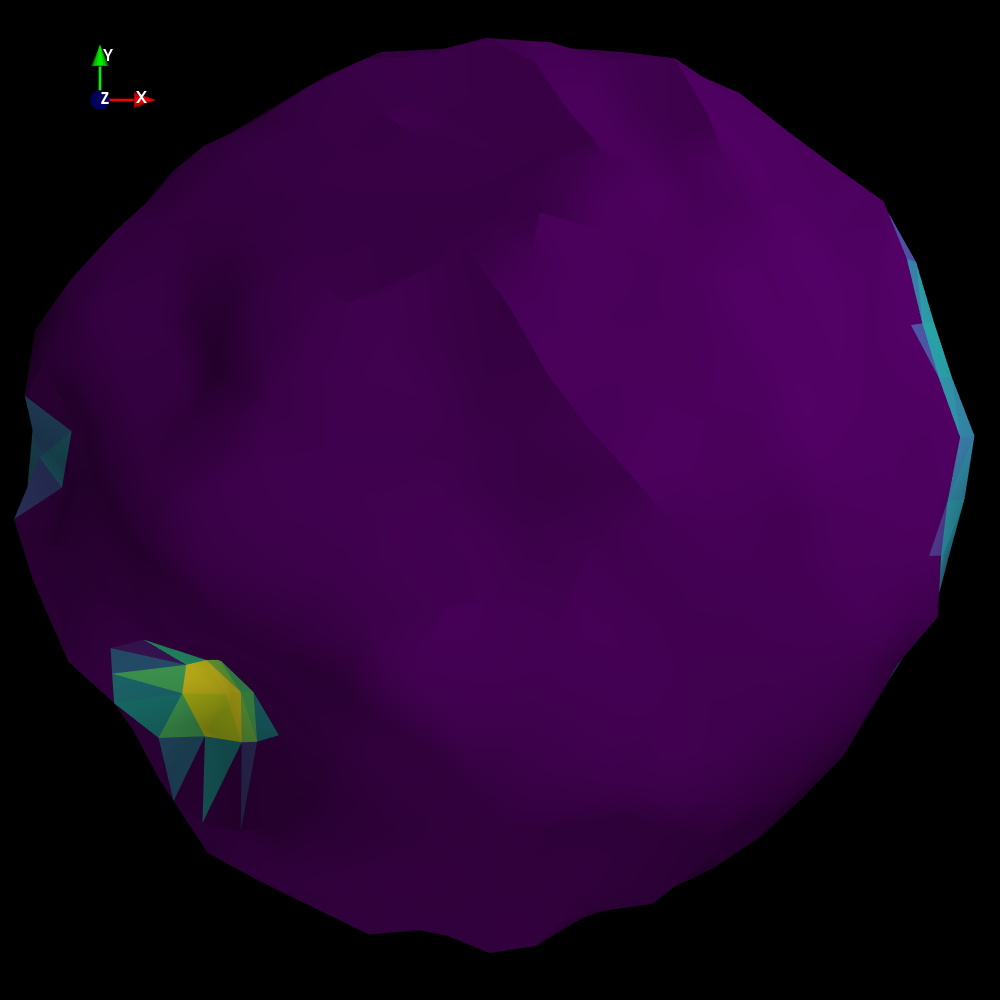} & 
   \includegraphics[width=0.3\textwidth ]{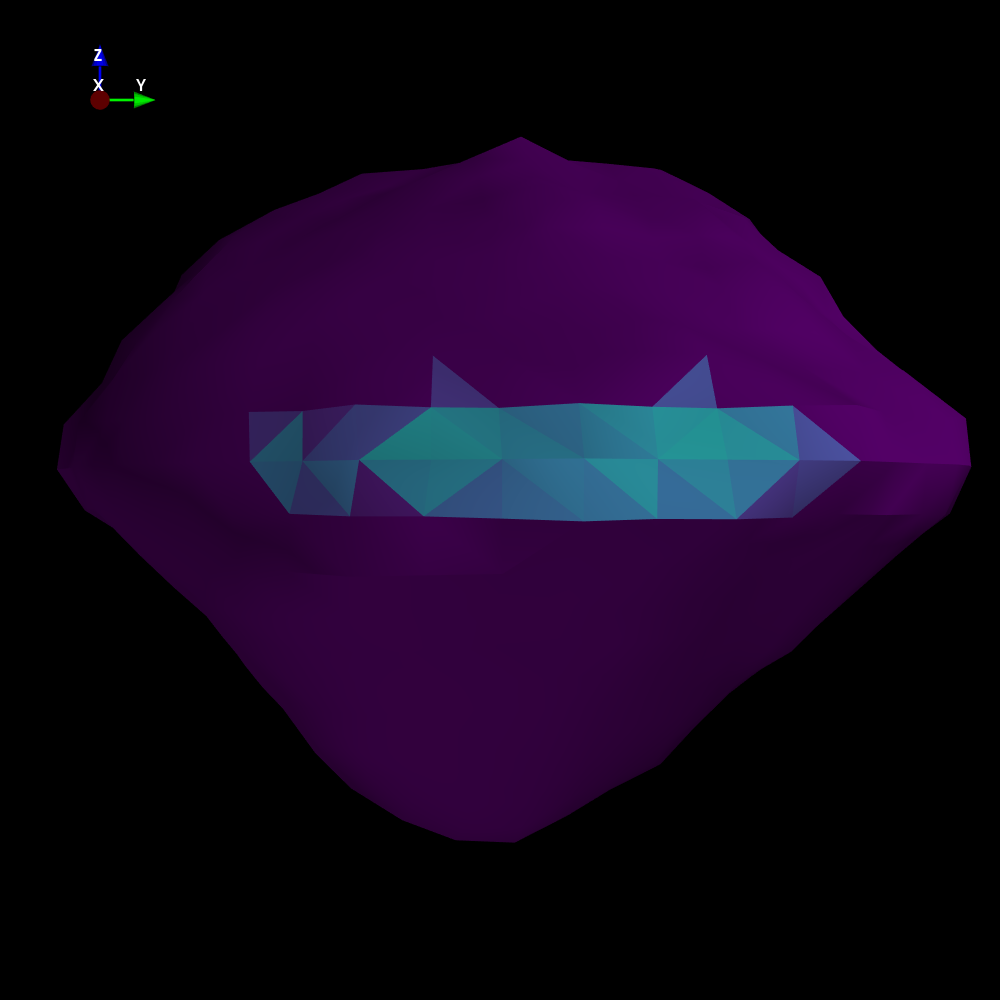} &
   \includegraphics[width=0.3\textwidth ]{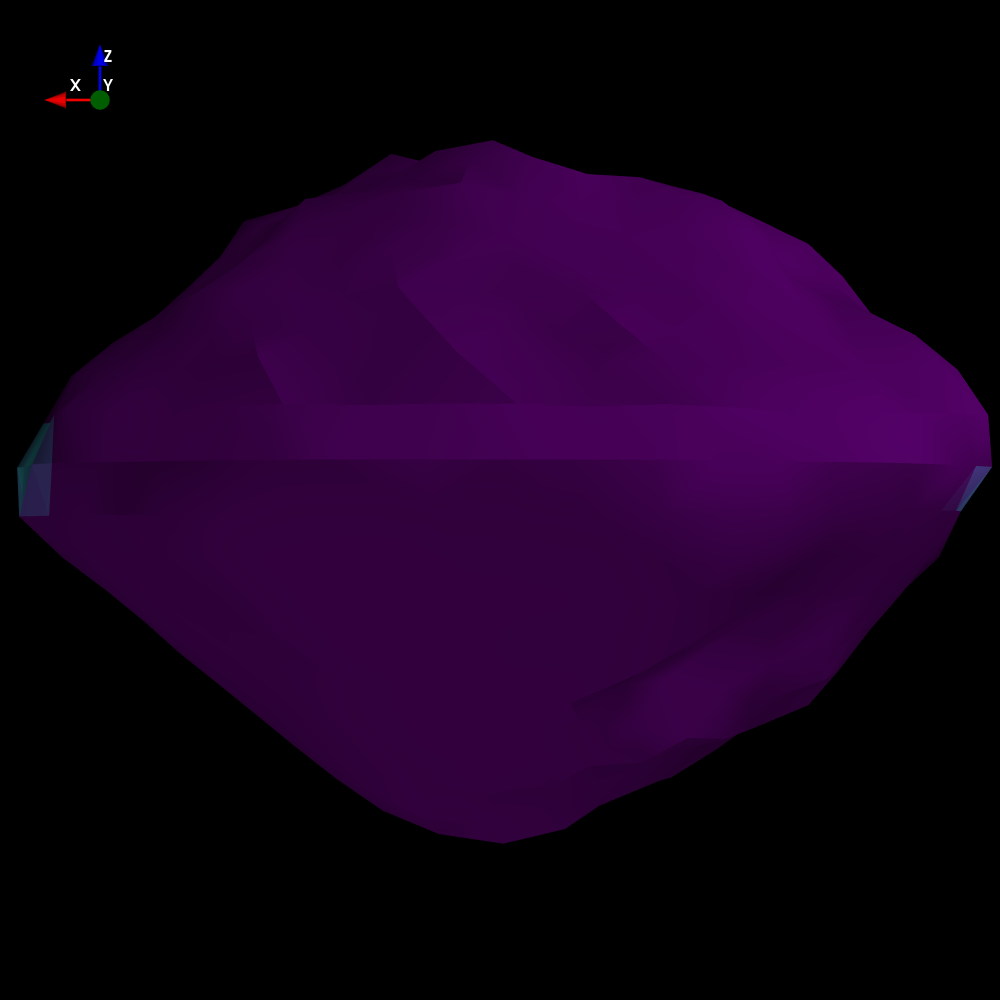}\\
   \includegraphics[width=0.3\textwidth ]{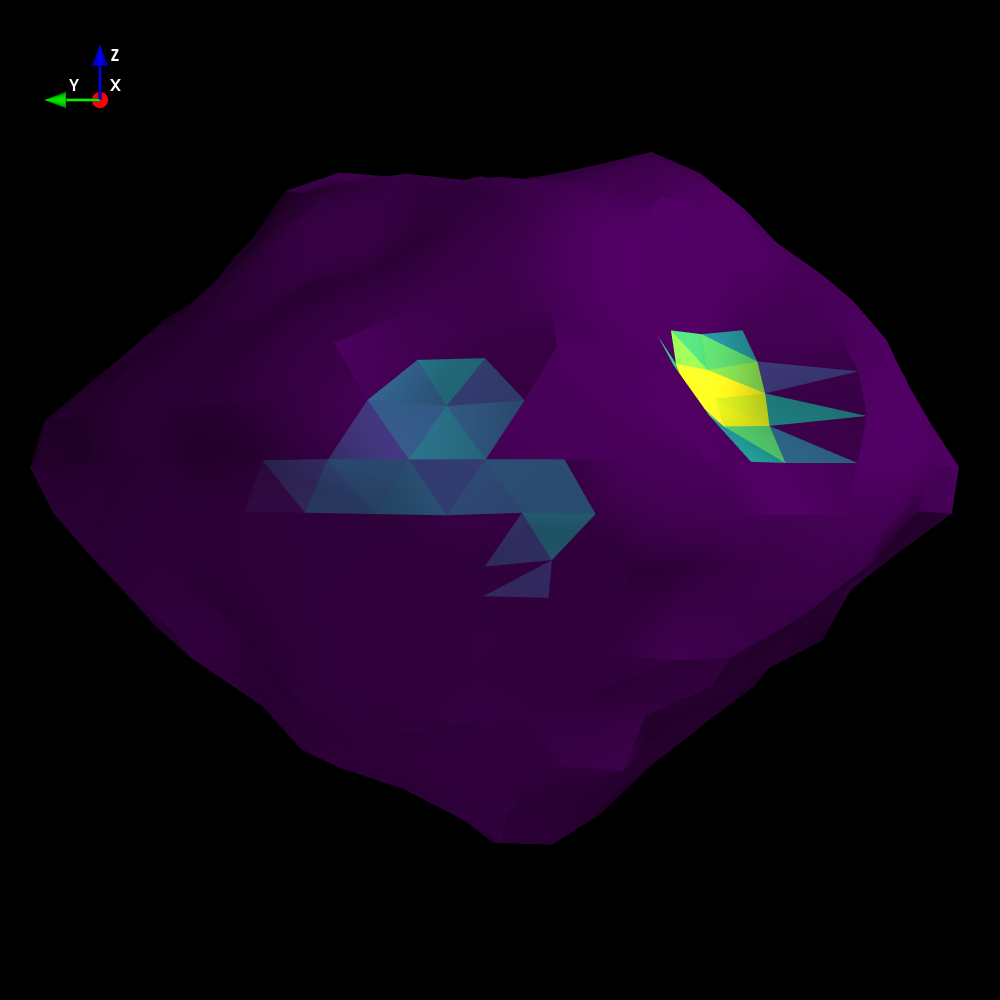} & 
   \includegraphics[width=0.3\textwidth ]{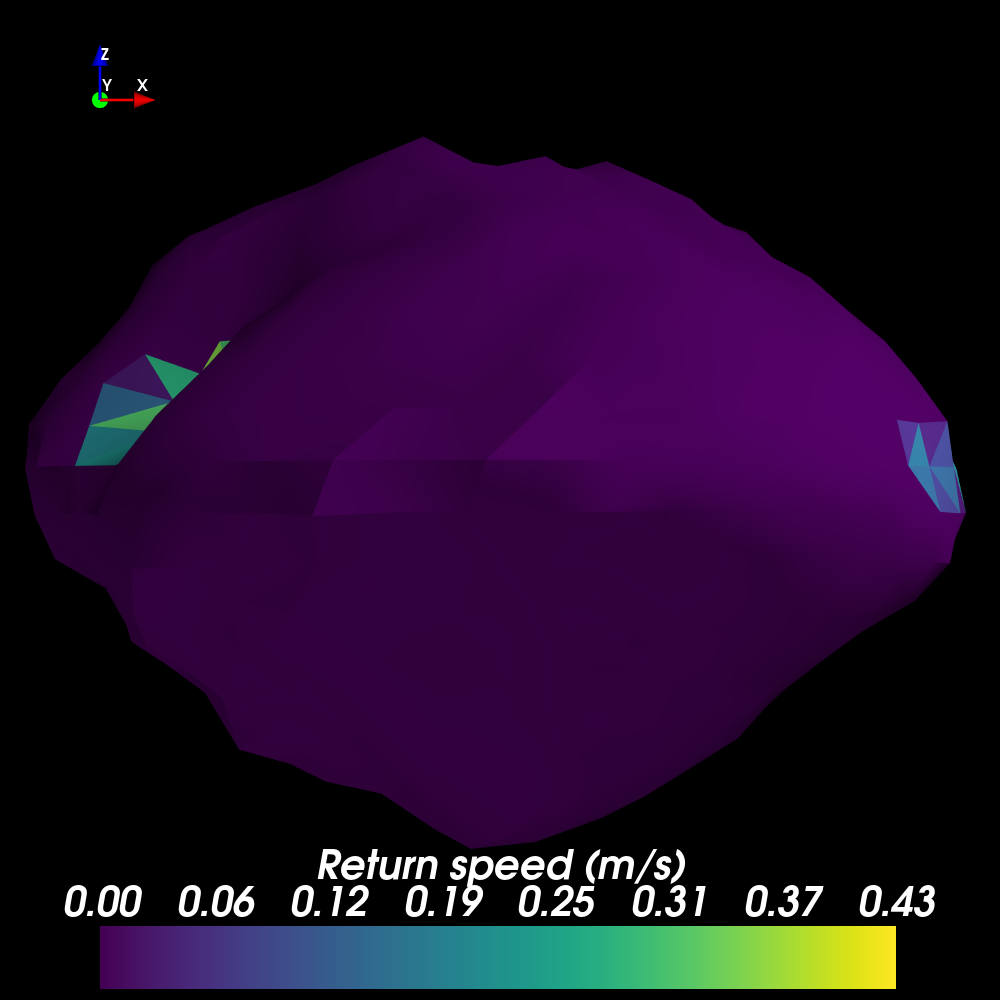} &
   \includegraphics[width=0.3\textwidth ]{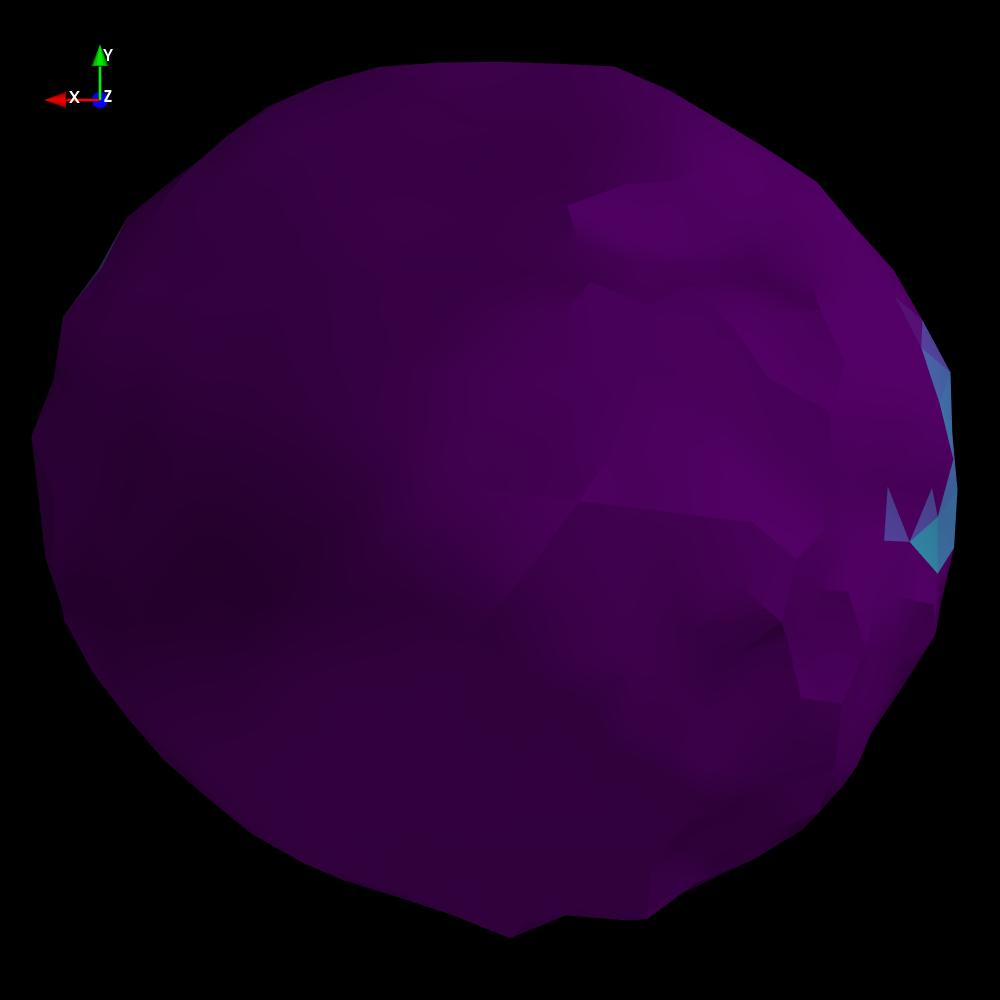}\\
  \end{tabular}
 \caption{Map of the return speeds on Phaethon's surface.
 \label{fig:return}}
\end{figure*}

The 3D surface corresponding to the minimum geopotential EP outside of Phaethon is referred to as the "rotational Roche lobe" \citep{2012omsp.book.....S}. The Roche lobe intersection marks the surface regions where the potential energy is theoretically sufficient for an object to be launched into space. Conversely, the regions enclosed by the Roche lobe indicate where the surface object requires additional energy to achieve escape. This extra energy corresponds to the return speed. 

If an object's initial velocity is below the return speed, it is guaranteed to fall back to the surface. The return speed is given as:
\begin{equation} \label{eq:v_return}
 v_\mathrm{ret}(\boldsymbol{r}) = \begin{cases}
			\sqrt{2[V(\boldsymbol{r})-C^*]} , & \text{if $V \geq C^*$}\\
            0, & \text{otherwise.}
		 \end{cases}
\end{equation}
where $C^*$ corresponds to the minimum geopotential EP outside of Phaethon \citep{2012omsp.book.....S, 2016Icar..276..116S}. According to Table \ref{tab:equil_coord}, this value is -3.58 m\textsuperscript{2}/s\textsuperscript{2} at point E5. The results are illustrated in Fig. \ref{fig:return}. If the initial velocity is above the return speed but less than the escape speed, the fate of the object depends on its interaction with the EPs. This interaction can result in outcomes ranging from fallback to long-term orbits, or even escape.

\subsubsection{Jacobi speed} \label{subsec:jacobi}

When an object located at the highest geopotential point on the surface begins to move downslope, it will naturally gain kinetic energy.  Assuming no energy was lost due to resisting forces like friction, this velocity corresponds to the Jacobi speed \citep{2016Icar..276..116S}. In other words, the Jacobi speed can be interpreted as the freefall speed from the highest geopotential point to a reference point. It is defined as:
\begin{equation} \label{eq:v_j}
 v_J(\boldsymbol{r}) =  \sqrt{2 [V_\mathrm{max}-V(\boldsymbol{r}) ]},
\end{equation}
where $V_\mathrm{max}$ is the maximum geopotential on the surface. The result is shown in Fig. \ref{fig:jacobi}. In realistic conditions, the kinetic energy of an object moving across an asteroid's surface would be affected by factors such as friction, collisions, etc. Nevertheless, the Jacobi speed can act as a useful basis for analyzing the particles' capability to be lifted off the surface. For instance, the Jacobi speed calculation indicates that an object migrating from the pole to the equator without energy loss can achieve escape speed. Combined with the return speed map in Fig. \ref{fig:return}, the Jacobi speed suggests that the most realistic and likely scenario would be for the particles to exceed the return speed and be lifted off the surface after a landslide event. As noted in Sect. \ref{subsec:return}, most of Phaethon's surface exhibits a return speed of zero, meaning that a particle displaced from a high geopotential point could theoretically escape shortly thereafter. 

\subsubsection{Trajectory of particles outside Phaethon} \label{subsec:trajectory}

To summarize, any object with a velocity below the return speed is guaranteed to fall back to the surface, while if the normal velocity is above the escape speed, it is guaranteed to escape the asteroid. However, for velocities that fall between these two thresholds, the object's trajectory can vary widely, depending on its initial conditions and interactions with EPs \citep{2016Icar..276..116S}. To investigate the dynamical properties of particles once they are lifted off the surface of Phaethon, we conducted a simple simulation in PKDGRAV.

First, we selected 130 outermost surface particles in the equatorial regions, which served as potential launch sites. We then placed 20-m-radius particles with a one-meter radial offset from each launch site. To minimize their gravitational influence, these particles were given a small density of 0.1 g/cm\textsuperscript{3}. Lastly, each particle was given the same initial velocity in the radially outward direction in addition to the rotational velocity of the corresponding launch site particles. We tested three initial radial velocities: 0.1, 0.5, and 1 m/s. The simulation timestep and parameters are identical to those listed in Table \ref{tab:param}. The locations and velocities of the particles were recorded every 1000 timesteps ($\sim 50.2$ s). The simulation ran for 26 million timesteps, which corresponds to just over 10 Phaethon rotations.

 Although the initial speed of 0.1 m/s is an order of magnitude lower than the escape speed, most areas of Phaethon have zero return speed (see Fig. \ref{fig:return}), which is theoretically enough to reach the lowest geopotential EP (E5). However, in our simulation result, all particles could not reach E5 before falling back to the surface. In other words, despite most of Phaethon's surface having a zero return speed, the particles from these regions can only escape if they can reach E5, which was difficult for most particles in this simulation, as the initial launch direction was radial. To effectively utilize E5 as an escape route even with low velocity, particles far from E5 must be launched with specific initial trajectories aiming towards E5 following the path of least resistance. In the case of particles whose initial launch locations were close to E5, their return speeds exceeded 0.1 m/s. As a result, all simulated particles fell back shortly after launch and only showed brief movements before settling into a stable position.

\begin{figure*}
 \resizebox{\hsize}{!}{\includegraphics{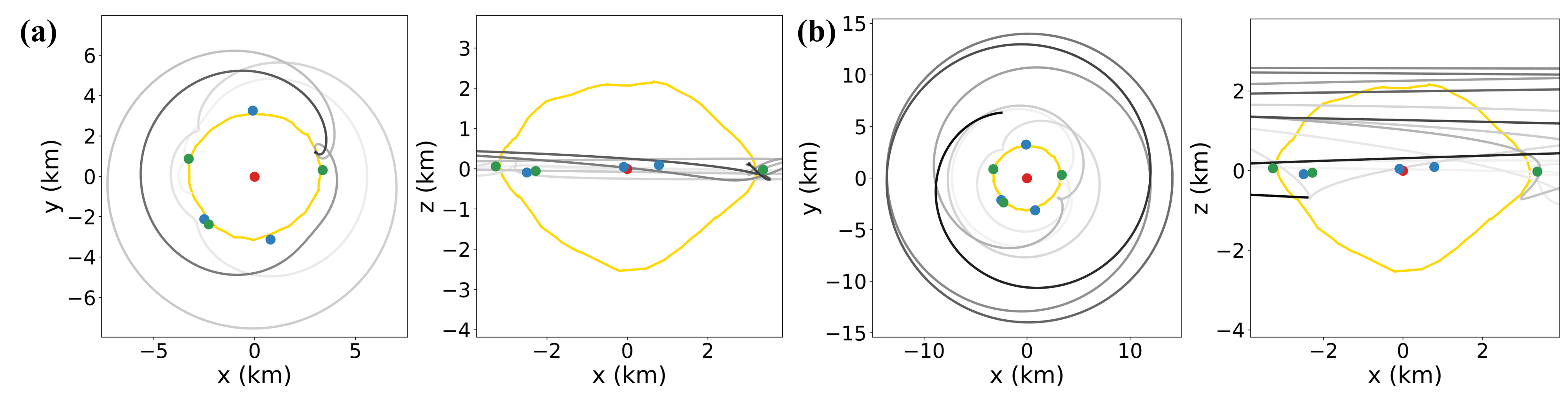}}
 \caption{Trajectory of a particle launched from the surface with (a) 0.5 m/s and (b) 1 m/s initial radial velocity. The darkness of the line shows the passage of time, from lighter to darker. The left and right panels are the projections of the plane $x-y$ and $x-z$, respectively. 
 \label{fig:traj_0.5_and_1}}
\end{figure*}

\begin{figure}
  \resizebox{\hsize}{!}{\includegraphics{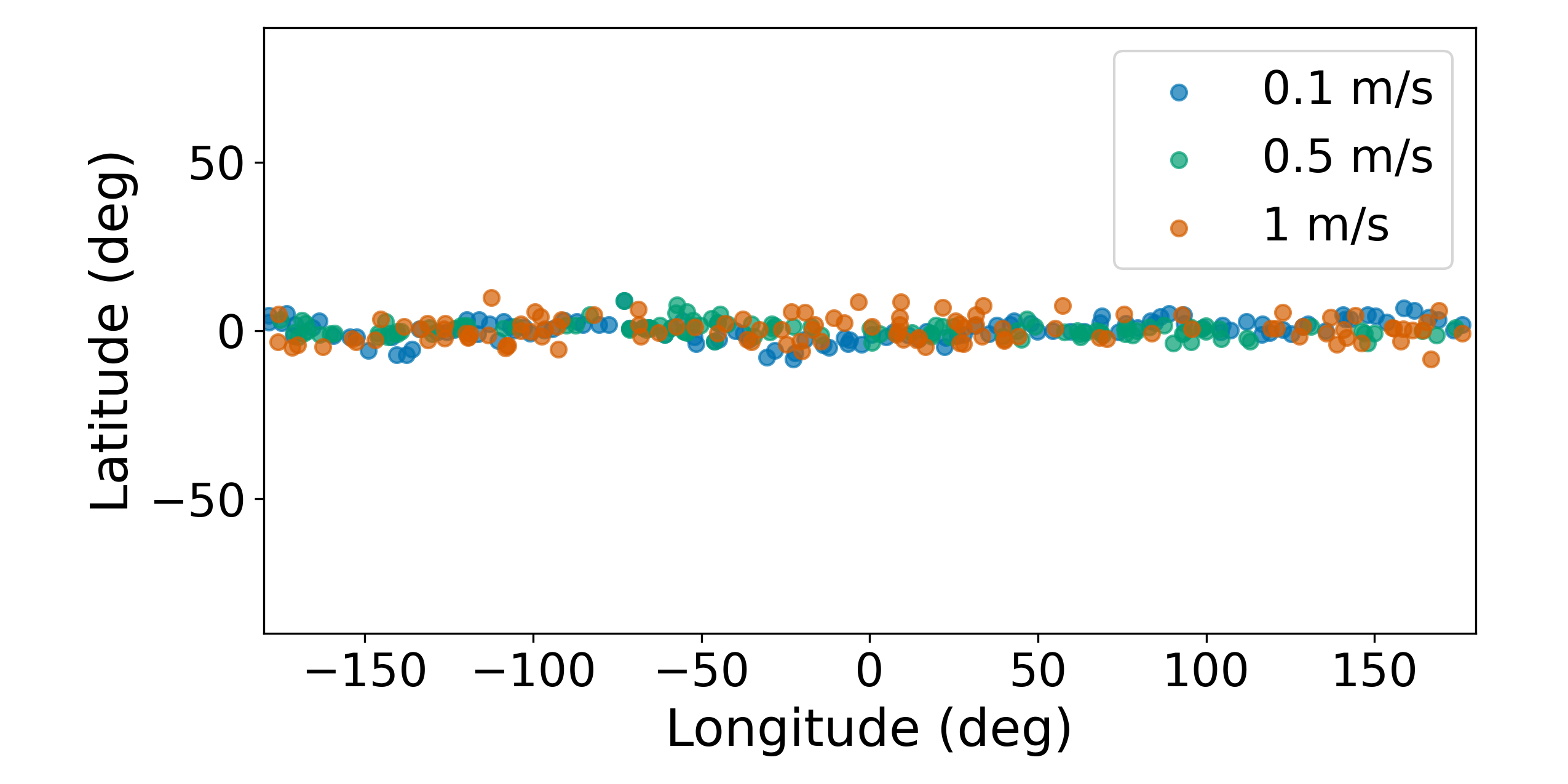}}
  \caption{Final positions of the launched particles after they returned to the surface. The different colors denote the different initial velocities of the particles. Particles that did not return to the surface by the end of the simulation are not shown.
  \label{fig:final_position}}
 \end{figure}

The initial speed of 0.5 m/s is below the escape speed but exceeds the maximum return speed of Phaethon. At this velocity, the particles have more opportunities to interact with the EPs. The top panel of Fig. \ref{fig:traj_0.5_and_1}  exemplifies this interaction; the particle exhibits a complex trajectory after being launched near E6 and appears to engage with E2 and E3 after rebounding. Additionally, the particles gain latitudinal ($z$-component) motion as well. When the initial velocity increases to 1 m/s, which is just below the escape speed in the equatorial region, this latitudinal motion becomes more pronounced, with some particles reaching beyond Phaethon's pole, as shown in the case of the bottom panel of Fig. \ref{fig:traj_0.5_and_1}. However, based on Fig. \ref{fig:final_position}, when the particles eventually fall back to the surface, they are still concentrated in the low-latitude region within 20\degr. However, we note that this is the result of a simplified simulation setup with limited initial locations and directions. Future studies could explore more detailed scenarios by varying the launch locations and directions, as well as incorporating the effects of solar radiation pressure on small dust particles \citep{2021JGRE..12606272A}.

\begin{figure}
  \resizebox{\hsize}{!}{\includegraphics{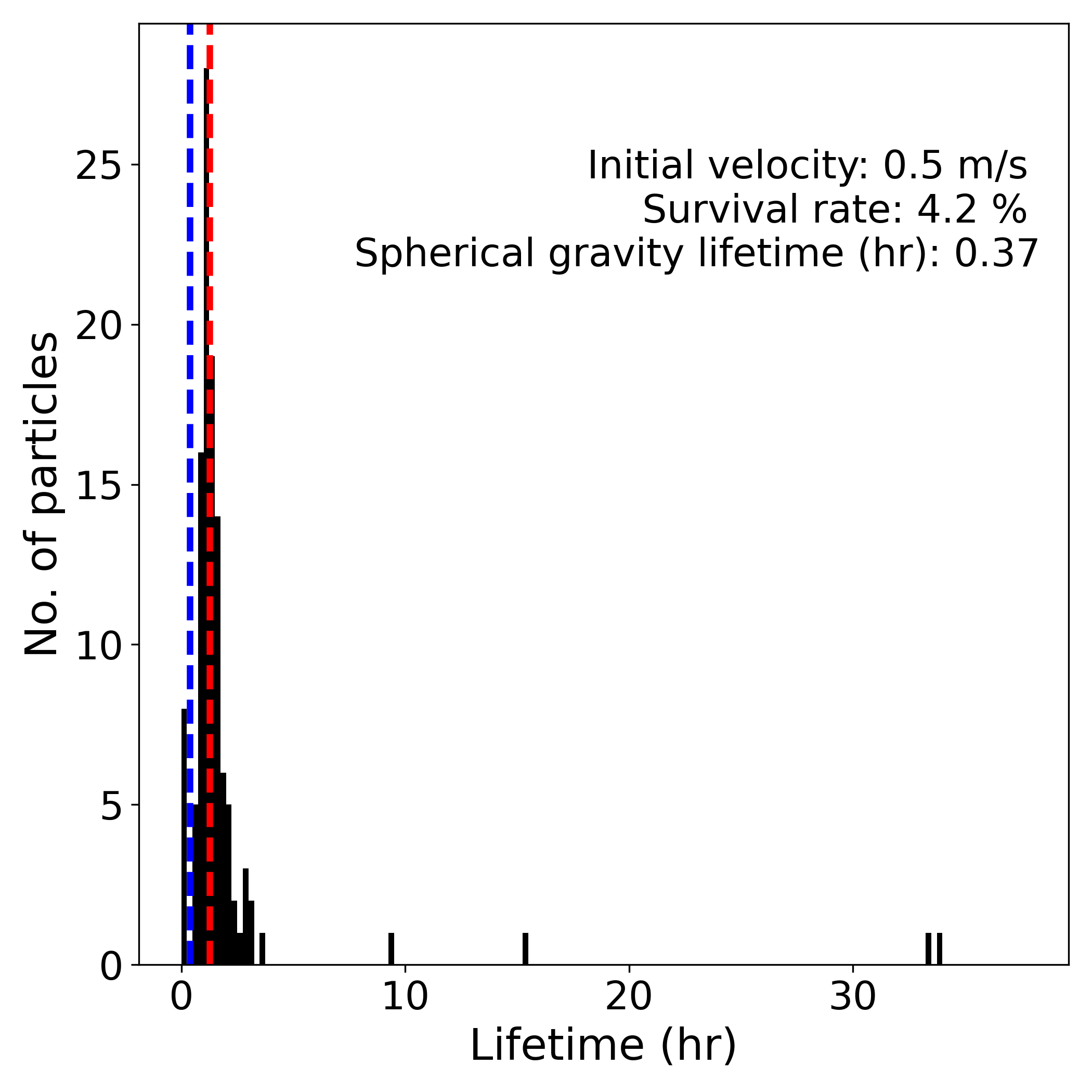}}
  \resizebox{\hsize}{!}{\includegraphics{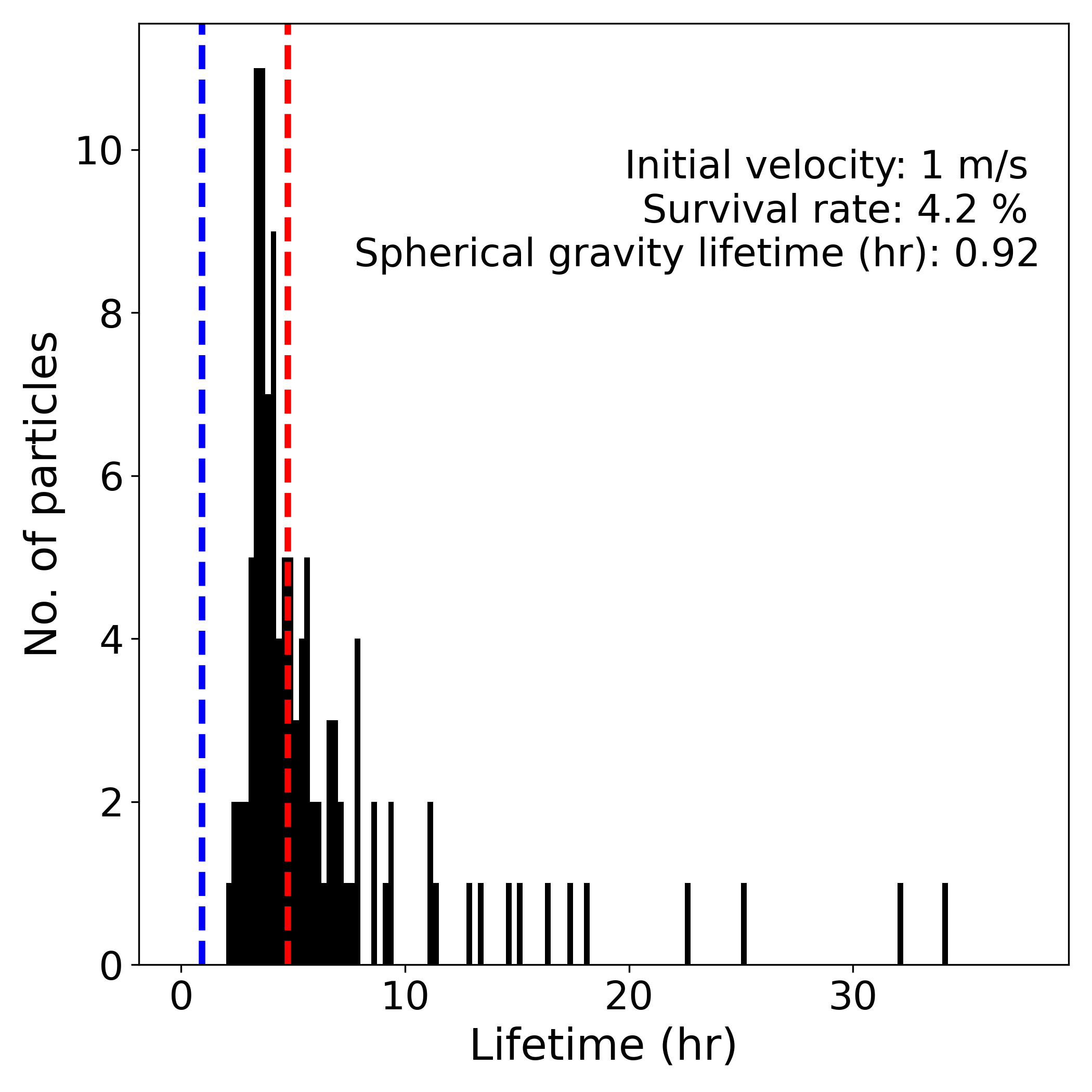}}
  \caption{Histograms of the launched particles' lifetimes for initial velocities of 0.5 m/s (upper) and 1 m/s (lower). The blue dotted line marks the expected lifetime, assuming Phaethon as a spherical body, while the red dotted line is the mean lifetime of the particles.
  \label{fig:lifetime}}
 \end{figure}

Although our simulations did not yield any particles that reach escape velocity through EP interactions, they demonstrated the influences of EPs on the particle trajectories. Fig. \ref{fig:lifetime} summarizes the lifetime of the particles, defined as the time taken for each particle to fall back and rest on the surface. Since EPs are products of Phaethon's irregular shape, we also analyzed a spherical body with the same mass and equivalent radius as Phaethon for comparison. The expected lifetime of this spherical Phaethon is marked with the blue dotted line in Fig. \ref{fig:lifetime}. The lifetime around a spherical body is noticeably shorter than that of most particles in our calculation for Phaethon. Moreover, a few particles survive beyond the simulation duration, although they are expected to eventually fall back to the surface, based on their velocities.

\section{Discussion} \label{sec:discussion}

In this section, we first outline the uncertainties involved in our method. We then review potential activity mechanisms on Phaethon and their possible effects on surface material in light of our findings. Next, we highlight Phaethon's unique characteristics compared to previously explored near-Earth asteroids (NEAs). Finally, we discuss the implications of this study for DESTINY\textsuperscript{+}'s upcoming in-situ observations.

\subsection{Model uncertainties} \label{subsec:uncertainty}\subsubsection{Mass density} \label{subsec:density_uncertainty}

Although this study uses the nominal mass density 1.58 g/cm\textsuperscript{3} by \citet{2022Icar..38815226M}, the study also reports an uncertainty of 0.45 g/cm\textsuperscript{3}, or about 28\%.
\begin{figure*}
 \resizebox{\hsize}{!}{\includegraphics{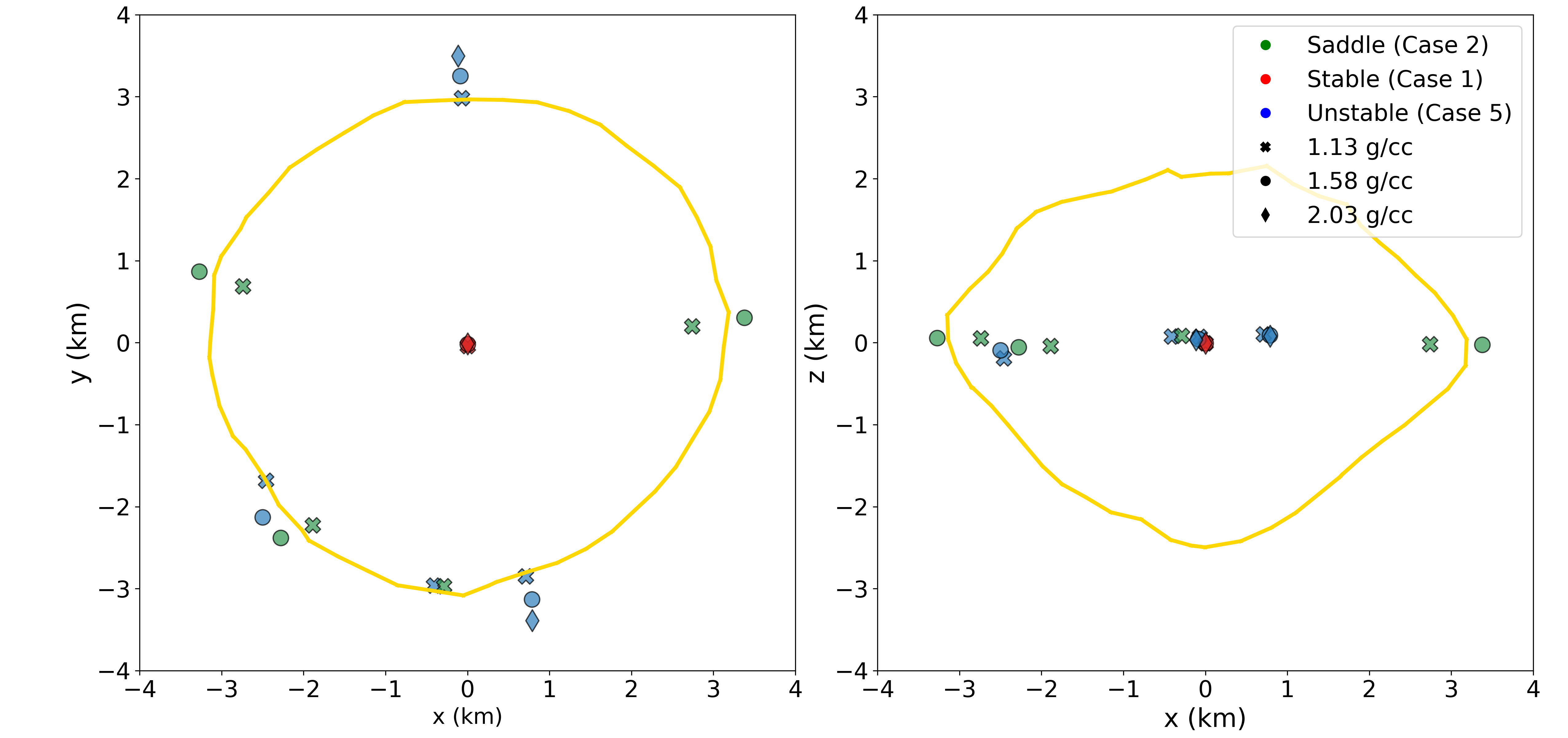}}
 \caption{The equilibrium points of Phaethon assuming different densities, marked with different markers.
 \label{fig:equilibrium_multiden}}
\end{figure*}
Fig. \ref{fig:equilibrium_multiden} shows how much the EPs can change within the $1\sigma$ uncertainty range (1.13 - 2.03 g/cm\textsuperscript{3}). First, we highlight the number of EPs decreasing for higher density. At the lowest density, we found nine EPs, while at the highest density, there are only three. A similar trend was found on Bennu \citep{2016Icar..276..116S}. However, \citet{2018P&SS..161..107Y} found that increasing the rotation speed resulted in the annihilation of EPs for eight of the ten small bodies they analyzed, including Bennu. Meanwhile, \citet{2020MNRAS.496.4154A} found that lower density leads to fewer EPs for contact binary object Arrokoth. Within this context, we conclude that the number of EPs is sensitive to mass density and shape model. 

The change of location for different density assumptions is also an important trend. As the density increases, so does the gravitational attraction of the body. Since the EP is where the centrifugal and gravitational acceleration is balanced, if the gravitational acceleration increases with density, the EPs require stronger centrifugal accelerations. Assuming that the angular velocity is fixed, this leads to the EPs being pushed further outwards. Here, we point out that for the lowest density, all the unstable points are near or inside the surface of Phaethon. \citet{2014ApJ...780..160H} defined this condition as the "first surface shedding". In other words, at the lower end of the $1\sigma$ density range, we can expect Phaethon to be undergoing global deformation and mass shedding. To summarize, within the currently estimated density range of Phaethon, the number and locations of the EPs can vary widely and have substantial consequences for its surface and near-surface environment.

Our assumption of the core-envelope structure is another possible source of uncertainty. As mentioned in Sec. \ref{sec:method}, this structure was used to account for the possibility of the Geminids being formed by a mass shedding event caused by rotational instability \citep{2024A&A...683A..68J}. However, a denser core is not required for mass shedding to occur if the interior is mechanically stronger than its envelope (see Sec. 1.2, \citet{2024PSJ.....5...54A}, and references therein). A prime example is Bennu, whose surface features suggest a high-cohesion interior \citep{2022NatCo..13.4589Z}, despite also being found to have an underdense core \citep{2020SciA....6.3350S}. Nevertheless, given that Phaethon is a larger and denser body than Bennu, we cannot rule out the possibility that their internal structures are different. Furthermore, \citet{2022Icar..37814914F} investigated various cases of core-envelope configurations and concluded that density inhomogeneity between the core and envelope has little effect on structural and surface stability, provided the bulk density remains unchanged.  

\subsubsection{Shape model} \label{subsec:shape_uncertainty}

The shape model has two types of uncertainty that are relevant to this work: size and smoothness. The relative uncertainties of Phaethon's lengths along its $x$--, $y$-- and $z$--axes are 3\%, 3\% and 6\%, respectively. Additionally, for radar-based shape modeling, there is a subjective weight that controls how smooth the model will be \citep{2007Icar..186..152M}. Increasing this weight discourages small-scale topographical roughness at the expense of worse fits to the observational data. The nominal model used in this study is the "rough" model.

\subsubsection{Combined effect of the uncertainties} \label{subsec:combined_uncertainty}

To evaluate the impact of uncertainties in the mass density and shape model on the surface analysis of Phaethon, we conducted additional calculations by varying three parameters: density, model size, and model smoothness. For density, we adjusted the PKDGRAV particle masses to achieve bulk densities of 1.13 and 2.03 g/cm\textsuperscript{3}, corresponding to $\pm1\sigma$ deviations from the nominal density. Similarly, we varied the model size by adjusting the $x$-- and $y$--axes together by $\pm 3\%$ and the $z$--axis by $\pm 6\%$ independently. Three levels of smoothness were also tested. Among these parameters, the mass density had the most significant impact, likely due to its larger uncertainty. When the mass density was increased, the unstable areas were considerably reduced, as increased gravity contributed to greater stability. We exhibit the slopes calculated after varying the model parameters in the Appendix \ref{app:model_slopes}. As our nominal results indicate that unstable regions exist on Phaethon, we focused on the most "stable" scenario to determine whether these regions keep their instability amid these uncertainties. The most "stable" case, with the fewest regions showing non-zero net surface acceleration and slopes exceeding the angle of friction, occurred with high mass density, larger size, and minimal smoothness. In this scenario, the net surface acceleration was zero across all facets. Thus, within the current uncertainty range of the parameters, we cannot rule out the possibility that Phaethon may actually be quite stable.

\subsection{Activity and ejection mechanisms} \label{subsec:activity}
\subsubsection{Possible present-day activity} \label{subsec:present_activity}

From Sect. \ref{subsec:escape}, we note that the escape speed in the equatorial region is $\sim 1$ m/s. This is comparable to the velocity of ejected particles detected on Bennu by OSIRIS-REx \citep{2019Sci...366.3544L}. Thermal fatigue is a possible cause of Bennu ejection events \citep{2020JGRE..12506325M}. While the diurnal and annual temperature variation on Bennu is $\sim 100$ K \citep{2020JGRE..12506323R}, Phaethon is expected to experience variations of several factors higher \citep{2019MNRAS.482.4243Y,2019P&SS..165..296L,2021A&A...654A.113B,2024NatAs...8...60M}. Hence, it is not unreasonable to expect Phaethonian particles to receive at the very least a similar amount of kinetic energy from such an event. Additionally, the high albedo and thermal inertia of Phaethon \citep{2018AA...620L...8H} could indicate a history of thermal fractures.

Another candidate for Bennu's ejection is meteoroid impact \citep{2020JGRE..12506282B}. Compared to Bennu, due to the small perihelion distance, Phaethon ventures deeper into the inner Solar System, where the meteoroid population is denser. This suggests that meteoroid impacts on Phaethon could be more frequent. Additionally, such impacts would be more energetic on Phaethon as its highly eccentric orbit leads to high relative velocities of interplanetary dust particles \citep{2019P&SS..165..194S}. Especially near the perihelion, an impact by cm-sized particles can lead to an ejection of thousands of kilograms of material \citep{2022JQSRT.28608224Z}. In short, if the same events that cause particle ejections on Bennu occur on Phaethon, it is possible for particles to reach escape speed, particularly at lower latitudes.

\citet{2023PSJ.....4...70Z} suggested that the present-day brightening of Phaethon in the perihelion passage is due to sodium volatilization. They estimated that only small dust particles of micron-scale or less would be able to escape from Phaethon from this mechanism. Additionally, recent thermal modeling suggests that thermal decomposition can also lift dust particles from Phaethon \citep{2024NatAs...8...60M}.

In short, present-day dust activity on Phaethon can be driven by thermal fracture, meteoroid impact, and/or sodium volatilization. While it would be challenging for DESTINY\textsuperscript{+} to directly observe such small particles, the takeaway should be that regolith particles are disturbed by such activity every perihelion passage. If the force exerted by any of these mechanisms is enough to overcome friction and displace the regolith material, it could trigger downslope mass movements, such as landslides where boulder-sized objects can be affected as well. From Fig. \ref{fig:jacobi}, one can infer that if sodium activity displaces some mass at mid to high latitude (high geopotential region), by the time this mass reaches the equatorial region, it may reach a velocity comparable to the escape speed.

\subsubsection{Geminid formation event} \label{subsec:geminid}

Whether these mechanisms caused the Geminid meteor stream in the past is a separate question. Key constraints from the Geminid stream are its estimated mass range ($\sim 10^{12} - 10^{17}$ g, \citet{1983Natur.306..116H,1994A&A...287..990J,2017P&SS..143...83B,2017P&SS..143..125R,2024A&A...683A..68J}), age range ($\sim 2 - 18$ kyr, \citet{1999SoSyR..33..224R,2007MNRAS.375.1371R,2024A&A...683A..68J}) and the existence of large particles of $\sim 1 - 10$ cm in size \citep{2017P&SS..143...83B}. Based on Bennu's ejection, thermal fracture and meteoroid impacts can produce cm-sized ejecta. However, for all the mechanisms mentioned above, it is unclear if they are able to produce the Geminid stream mass. It is also possible that the Geminid formation mechanism is currently inactive. Historically, comet-like ejection by ice sublimation was the first proposed activity mechanism \citep{1989A&A...225..533G}. However, during its millions of years in the inner Solar system, it is unlikely for water ice in Phaethon to survive long enough to satisfy the Geminid age range \citep{2010AJ....140.1519J, 2019MNRAS.482.4243Y, 2022Icar..38815226M}. On the other hand, Geminid formation by YORP-induced rotational instability was suggested by \citet{2020ApJ...892L..22N}. \citet{2024A&A...683A..68J} performed dynamical simulations and found that particle ejection by rotational instability that occurred $\sim 18$ kyr ago can create the Geminid meteor shower. The results of our study indicate that rotational instability is not likely to be occurring at present on Phaethon. However, we cannot rule out the possibility that Phaethon experienced rotational instability in the past, undergoing global mass migration and even mass ejection events before decelerating to its current rotation \citep{2020ApJ...892L..22N}. If this event is the origin of the Geminid stream, most of Phaethon's surface would consist of freshly exposed material. We also note that three decades of lightcurve data indicate that Phaethon's rotation is currently accelerating \citep{2022DPS....5451407M}. The YORP effect predicts a slower acceleration, but the observed acceleration could be explained by tangential YORP, which depends on small-scale morphology and cannot be calculated accurately from the current shape model \citep{2024Senshu}. Phaethon may undergo cycles of spin-up, spin-down, and spin-up again on timescales shorter than the YORP timescale.

\subsection{Comparison with other asteroids} \label{subsec:comparison}

\begin{figure*}
  \centering 
  \begin{tabular}{ccc}
   \includegraphics[width=0.3\textwidth ]{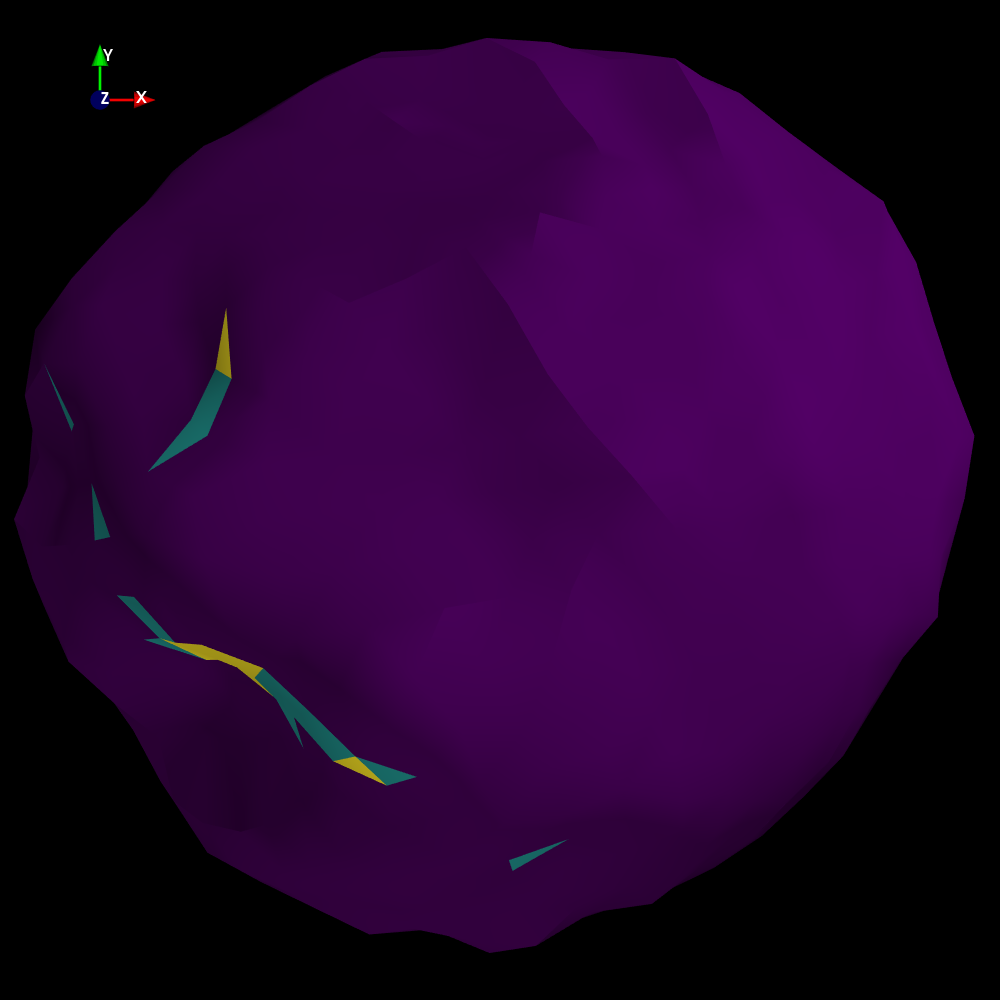} & 
   \includegraphics[width=0.3\textwidth ]{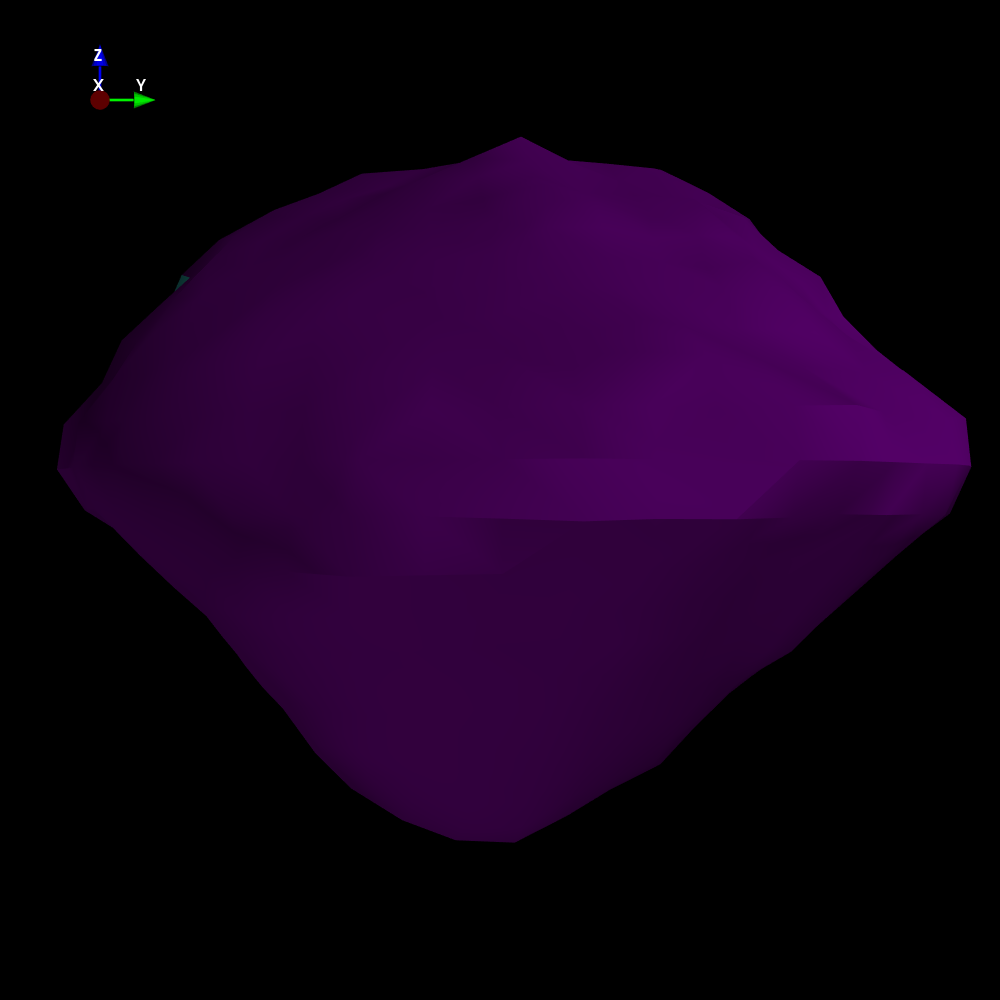} &
   \includegraphics[width=0.3\textwidth ]{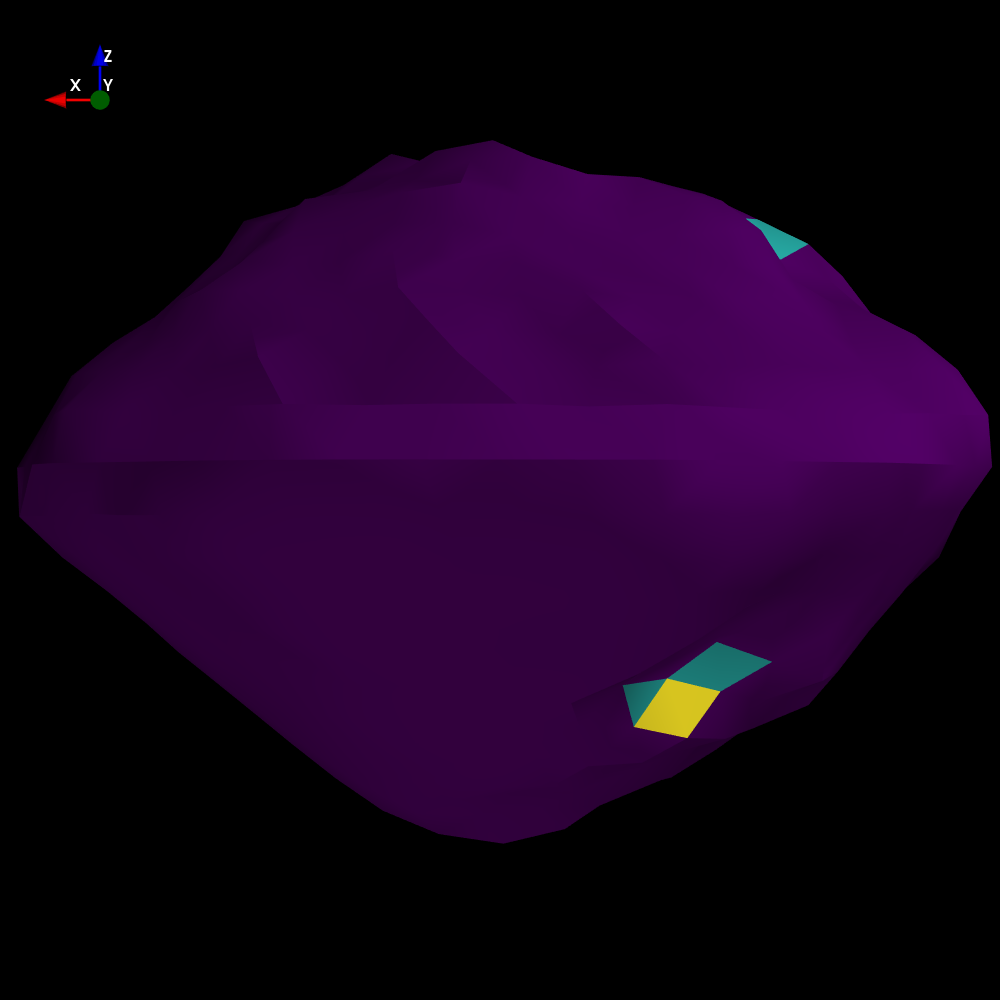}\\
   \includegraphics[width=0.3\textwidth ]{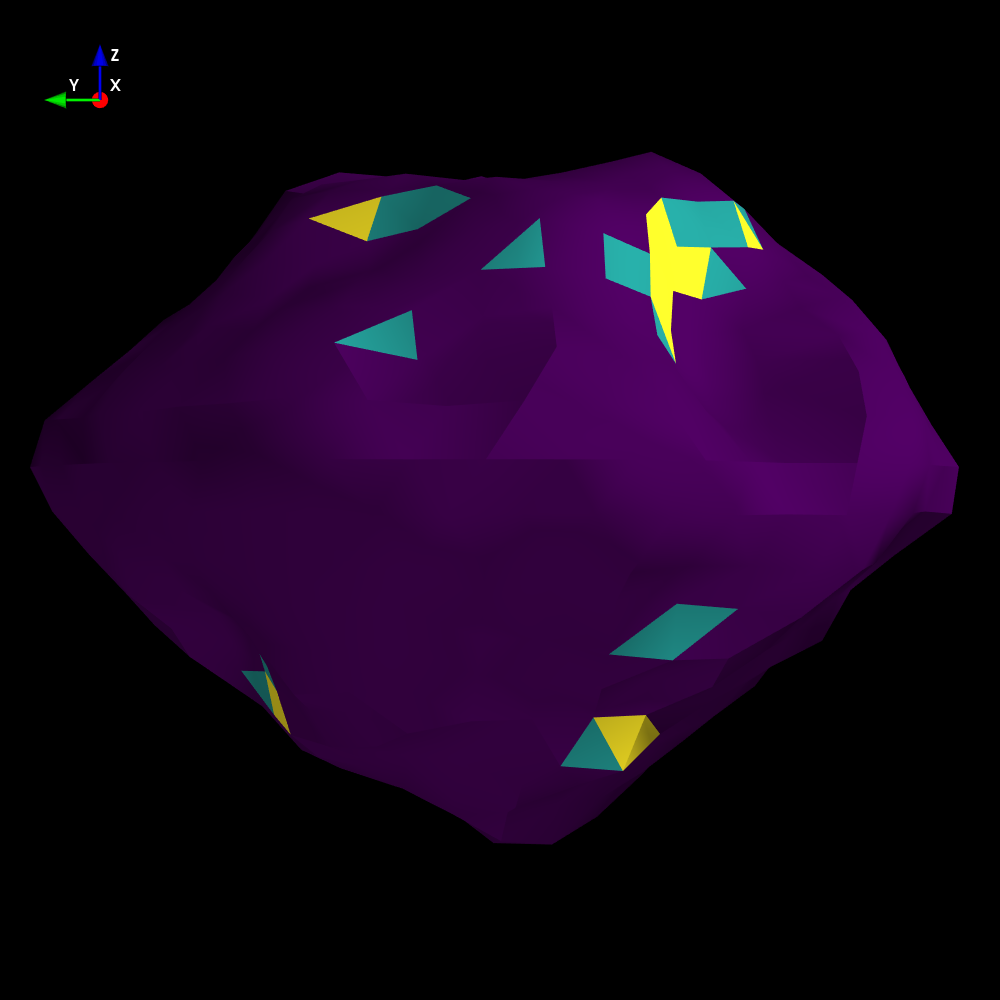} & 
   \includegraphics[width=0.3\textwidth ]{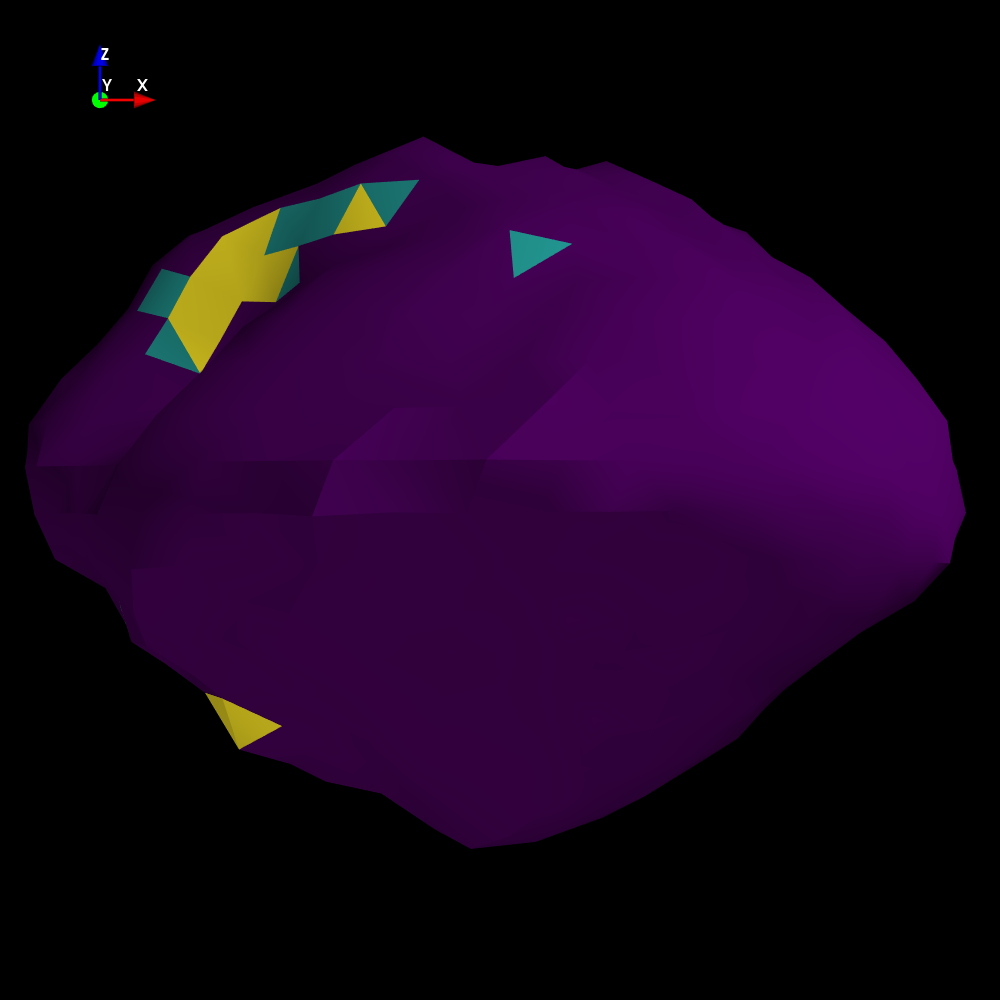} &
   \includegraphics[width=0.3\textwidth ]{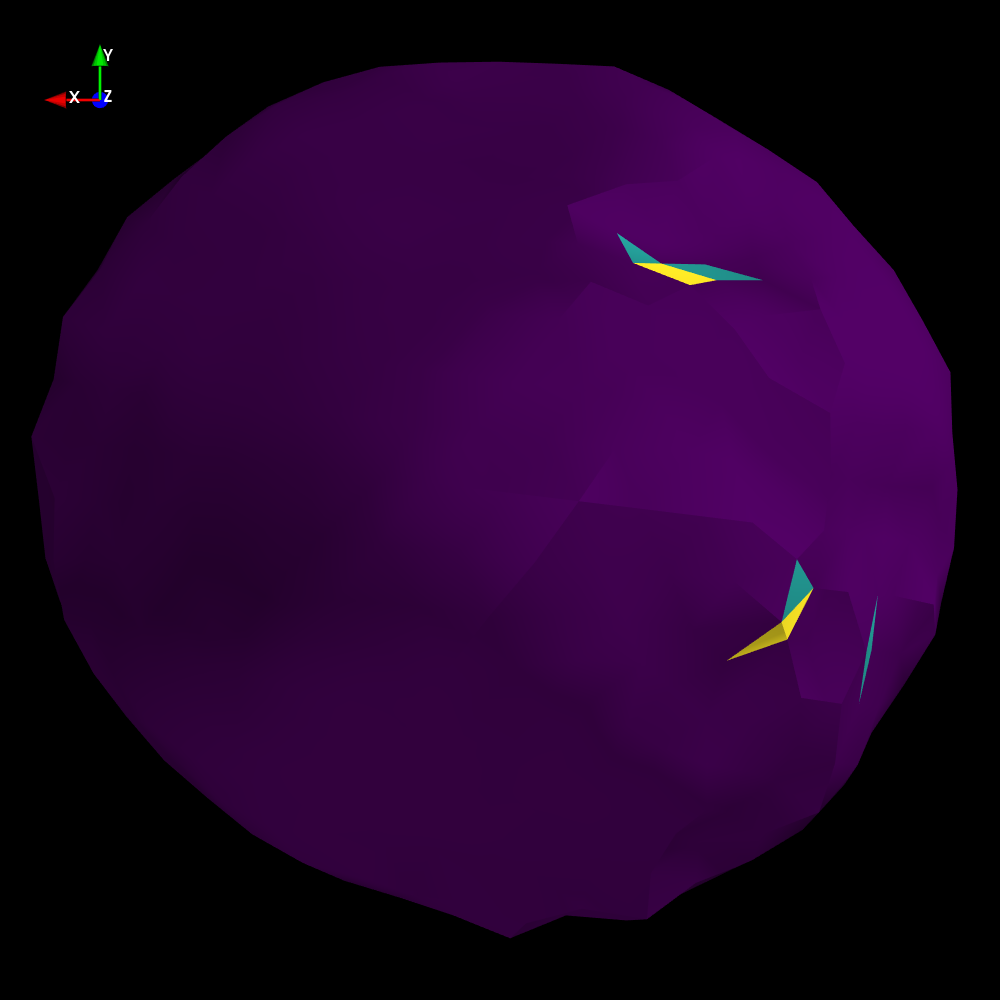}\\
  \end{tabular}
 \caption{Stability of Phaethon's surface. The conditions of instability are (i) non-zero net surface acceleration and (ii) slope larger than 40\degr. Purple, green and yellow represent none, one and both conditions satisfied, respectively, for each face.
 \label{fig:unstable}}
\end{figure*}

While the recent explorations of near-Earth asteroids (NEAs) Bennu, Ryugu, and Didymos have provided various new insights into the evolution and structure of asteroids, our results show that Phaethon is unique in this context. Firstly, Phaethon's large size yields high surface acceleration. The maximum surface acceleration found on Ryugu and Didymos is $\sim 1.5 \times 10^{-4}$ m/s, and $\sim 8 \times 10^{-5}$ m/s for Bennu \citep{2019Sci...364..268W, 2019MNRAS.482.4243Y,2019NatAs...3..352S}. On Phaethon, we estimate the maximum surface acceleration to be $6.4 \times 10^{-4}$ m/s, which is up to several factors higher than the aforementioned asteroids. In other words, the Phaethon's surface should theoretically be more mobile. 

In particular, there is one feature on Phaethon where the surface acceleration tends to be highest. Fig. \ref{fig:unstable} summarizes the results in Figs. \ref{fig:net_acc} and \ref{fig:slope} and identifies the most unstable regions. We note, as explained in Sec. \ref{subsec:uncertainty}, that the results are subject to uncertainties. Therefore, we emphasize that Fig. \ref{fig:unstable} serves as a visual summary of the nominal results and is intended to highlight regional variations in instability rather than provide definitive indicators of instability. Due to the low resolution of the model, regions where only one or two faces are marked unstable should be considered unreliable. However, the large depression at the mid-latitude of the northern hemisphere (marked with the white arrows in Fig. \ref{fig:unstable}) shows a concentration of unstable faces. This region occupies $\sim 2.3$ km\textsuperscript{2}, with $\sim 0.8 $ km\textsuperscript{2} of it being possibly unstable (yellow or green in Fig. \ref{fig:unstable}), and correspond to the feature (d) identified in the radar observations by \citet{2019P&SS..167....1T}. The slopes in these regions can reach 51.4\degr and have some of the highest net surface accelerations as well. Although these calculations strongly indicate that this region is unstable, this does not guarantee mass movement in this region. In fact, our PKDGRAV model remained stable for multiple Phaethon rotations, suggesting that despite the high slope, it is possible for Phaethon to maintain this cliff-like structure. However, we remind that our PKDGRAV particles are more than 10 m in radius. Thus, while boulder-sized objects can sustain this cliff, that might not be the case for grains and dust particles. We conjecture that the unstable structure consists of exposed boulders, while regolith particles moved to lower geopotential areas just south of this structure, forming a deposit. If this deposit was formed by a local landslide event, its shape would follow the angle of repose \citep{2023Icar..40015563C}. Considering that the angle of repose is typically $\sim 30-40$\degr, the slope at the southern rim of the depression feature is around this range, implying that this is the deposit section in this region. In short, we argue that Phaethon has a large depression consisting of cliffs and deposits, which is a unique geological feature that was not found on recently explored NEAs. The cliffs may have exposed subsurface boulders, which would potentially give us more insight into the interiors of rubble-pile asteroids, and from the shape of the deposits, it would be possible to constrain the regolith material properties to some degree. As DESTINY\textsuperscript{+} is a flyby mission, it will only be able to observe about half of Phaethon's surface and has to select the observation region by adjusting the timing of the closest approach. We propose that the northern hemisphere depression be considered as a potential focus during this selection process.

\subsection{Implications and expectations for DESTINY\textsuperscript{+}} \label{subsec:destinyplus}

When an airless body is exposed to solar winds and micrometeorite bombardments, its surface can undergo physical and/or chemical modifications that result in optical changes over time -- a process known as space weathering \citep{clark2002asteroid}. Two recently visited top-shaped asteroids, Bennu and Ryugu, both displayed latitudinal color variation, indicating different degrees of space weathering due to global mass movement \citep{2024Icar..42016204Y}. However, space weathering on C-complex asteroids is particularly complicated.

On Bennu, smaller (and presumably younger) craters appear redder than its surroundings, implying that Bennu's surface becomes bluer over time \citep{2021Icar..35714252D, 2022Sci...377..285L,2023Icar..40015563C}. However, Bennu's equatorial region, which is thought to be the oldest regions on Bennu due to its low geopotential \citep{2019NatAs...3..352S} and high crater density \citep{2022NatGe..15..440B}, exhibits redder and darker characteristics \citep{2021PSJ.....2..117L}. Based on these observations, \citet{2023Icar..40015563C} concluded that space weathering on Bennu leads to non-linear color variations, from red to blue and back to red. However, \citet{2020Sci...370.3660D} and \citet{2021PSJ.....2..117L} noted that the equatorial region appears bluer in the blue-to-visible wavelengths, potentially due to 0.55 $\mu$m enhancement from magnetite production. On Ryugu, latitudinal variations and the colors of young craters indicate that space weathering results in the reddening and darkening of surface materials \citep{2019Sci...364..252S,2020Sci...368..654M,2021NatCo..12.5837T}. \citet{2024Icar..42016204Y} cross-calibrated the images from Hayabusa2 and OSIRIS-REx to compare the space weathering trends of the two asteroids. They found that Ryugu and Bennu may have started with similar spectra but underwent different weathering processes, possibly due to variations in surface dust size and coating thickness.

Based on the space weathering trends found by these previous missions, we can conjecture what DESTINY\textsuperscript{+} may find on Phaethon. Fig. \ref{fig:surf_acc} and \ref{fig:slope} suggest that mass movement from mid-to-low latitude should be expected on Phaethon, similar to what was observed on Bennu. Using the Telescopic CAmera for Phaethon (TCAP) onboard DESTINY\textsuperscript{+}, boulder orientations from surface images will be a strong indicator of such global mass movement trend on Phaethon \citep{2020JGRE..12506475J}. Additionally, it is reasonable to infer that the equatorial region is the most static and, therefore, likely the most mature and space-weathered. In contrast, mid-latitude regions would be less space-weathered and thus fresher. This latitudinal difference in space weathering can be revealed through multiband observations with the Multiband CAmera for Phaethon (MCAP) on DESTINY\textsuperscript{+}. The space weathering timescale is typically thought to be $\lesssim 10^6$ years for main-belt S-type asteroids (e.g. \citet{2009Natur.458..993V}), and as short as $10^3$ years based on a recent analysis on the Hayabusa images of near-Earth asteroid Itokawa \citep{2022A&A...667A..93J}. For C-complex asteroids, \citet{2024AJ....167..224H} estimated the space weathering timescale for main-belt Ch/Cgh asteroids to be $\sim 10^{3.5} - 10^{4.5}$ years. If we scale this value to Phaethon's orbit, following $T_\mathrm{sw} \propto a^2 \sqrt{1-e}$ \citep{2001JGR...10610039H}, where $a$ and $e$ are the semimajor axis and eccentricity, the timescale decreases to $\sim 10^{2.3} - 10^{3.3}$ years, shorter than the estimated age of the Geminid stream, $\sim 10^3 - 10^4$ years \citep{2007MNRAS.375.1371R, 2017P&SS..143..125R, 2024A&A...683A..68J}.

On the opposite side of space weathering is resurfacing, which refers to the various processes, such as tidal interaction, impact-induced seismic shaking, thermal fatigue and YORP spin-up, that expose fresher material from below the surface \citep{2022A&A...667A..93J}. The potential past and present activity mechanisms, discussed in Sec. \ref{subsec:activity}, can all be considered as Phaethon's resurfacing mechanisms. Due to Phaethon's fast rotation, these mechanisms may trigger landslides and launch materials into space. The equatorial region is prone to potential fallback impacts from objects that were briefly launched into space (Fig. \ref{fig:final_position}). If Phaethon's equator is highly populated with boulders, it could indicate that such launch-and-fallback events, and large landslides that caused these events, happen frequently. 

Phaethon's unique status as the only km-sized asteroid with a very short perihelion distance implies that the tension between space weathering and resurfacing will be vigorous near the perihelion. The near-perihelion phase is where Phaethon is subjected to intense solar wind and denser regions of the zodiacal cloud, leading to stronger space weathering. Conversely, it is also where Phaethon experiences extreme temperature variations and more frequent meteoroid impacts, potentially making the resurfacing process more efficient on Phaethon than on Bennu. A further complication at perihelion is sintering. Polarimetric studies by \citet{2018NatCo...9.2486I} and \citet{2022MNRAS.516L..53G} suggest that Phaethon is populated with large grains, possibly produced by sintering near the perihelion, although its role in space weathering and resurfacing remains unclear.

In summary, on a global scale, since the space weathering timescale is shorter than the average age of the surface, which we assume to be equal to the Geminid age, Phaethon’s surface should be at or near saturation in terms of weathering, in the absence of resurfacing events. However, the near-Sun, high-eccentricity orbit of Phaethon makes it susceptible to multiple resurfacing mechanisms. Additionally, Fig. \ref{fig:surf_acc} indicates that the regolith movement might be more efficient on Phaethon compared to other visited NEAs due to its larger size, supporting the likelihood of latitude-dependent variations in space weathering. On a local scale, as discussed in Sect. \ref{subsec:comparison}, Phaethon features a unique depression geology at the northern hemisphere. The instability of this area implies that it is among the freshest regions on Phaethon, providing an additional reason for it to be strongly considered as part of the DESTINY\textsuperscript{+} observation region.

\section{Summary and Conclusions} \label{sec:conclusion}

We presented the first dynamical analysis of the DESTINY\textsuperscript{+} mission target (3200) Phaethon. Using the latest shape model generated from various observations, we computed Phaethon’s gravitational field with the mascon method, filling the model with particles through the SSDEM N-body code PKDGRAV. Next, we calculated the geopotential and derived surface properties such as surface acceleration, net surface acceleration, and slope. Additionally, we determined equilibrium points, escape speed, return speed, and Jacobi speed to evaluate conditions for material to escape from Phaethon's surface. Our findings indicate that even small triggers could induce landslides on Phaethon, potentially launching surface material. According to our numerical simulations, particles launched with velocities between the return speed and escape speed could remain suspended above the surface for extended periods due to interactions with equilibrium points (EPs) before eventually returning to the surface. We also highlighted a prominent depression in Phaethon’s northern hemisphere, hypothesizing a potential cliff deposit structure, suggesting this as one of the freshest regions. However, our results are influenced by uncertainties, particularly that of the mass density. We hope that the DESTINY\textsuperscript{+} mission will help to reduce these uncertainties and offer further insight into the space weathering trends of carbonaceous asteroids, as well as the resurfacing and activity mechanisms on near-Sun asteroids.

\begin{acknowledgements}
This research at Seoul National University was supported by a grant from the Korean National Research Foundation (NRF) (MEST) funded by the Korean government (No. 2023R1A2C1006180). SEM was supported by NASA grant number 80NSSC19K0523 awarded to University of Central Florida.
HJ and MI thank Yoonsoo Bach, Sunho Jin, Jooyeon Geem, Bumhoo Lim and Seungyoo Lee for their helpful comments and discussions, as well as  Prof. Hyung Mok Lee, Prof. Woong-Tae Kim and Dr. Hee Il Kim for allowing us the usage of the computational facility gmunu.

\end{acknowledgements}

%
%

\bibliographystyle{aa} 
\bibliography{phaethon_potential.bib} 

\begin{appendix} 
\onecolumn
\FloatBarrier
\section{Additional figures} \label{app:more_figures}
\begin{figure*}[ht]
  \centering 
  \begin{tabular}{ccc}
   \includegraphics[width=0.26\textwidth ]{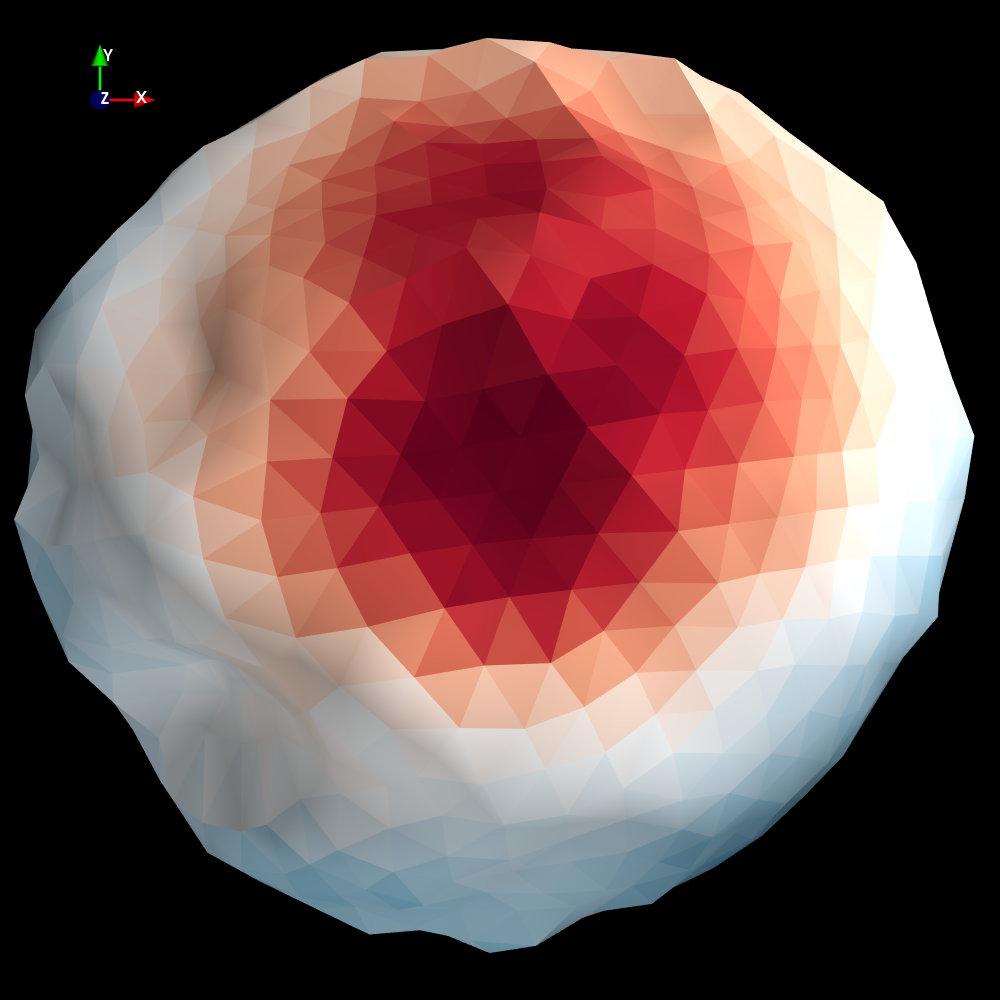} & 
   \includegraphics[width=0.26\textwidth ]{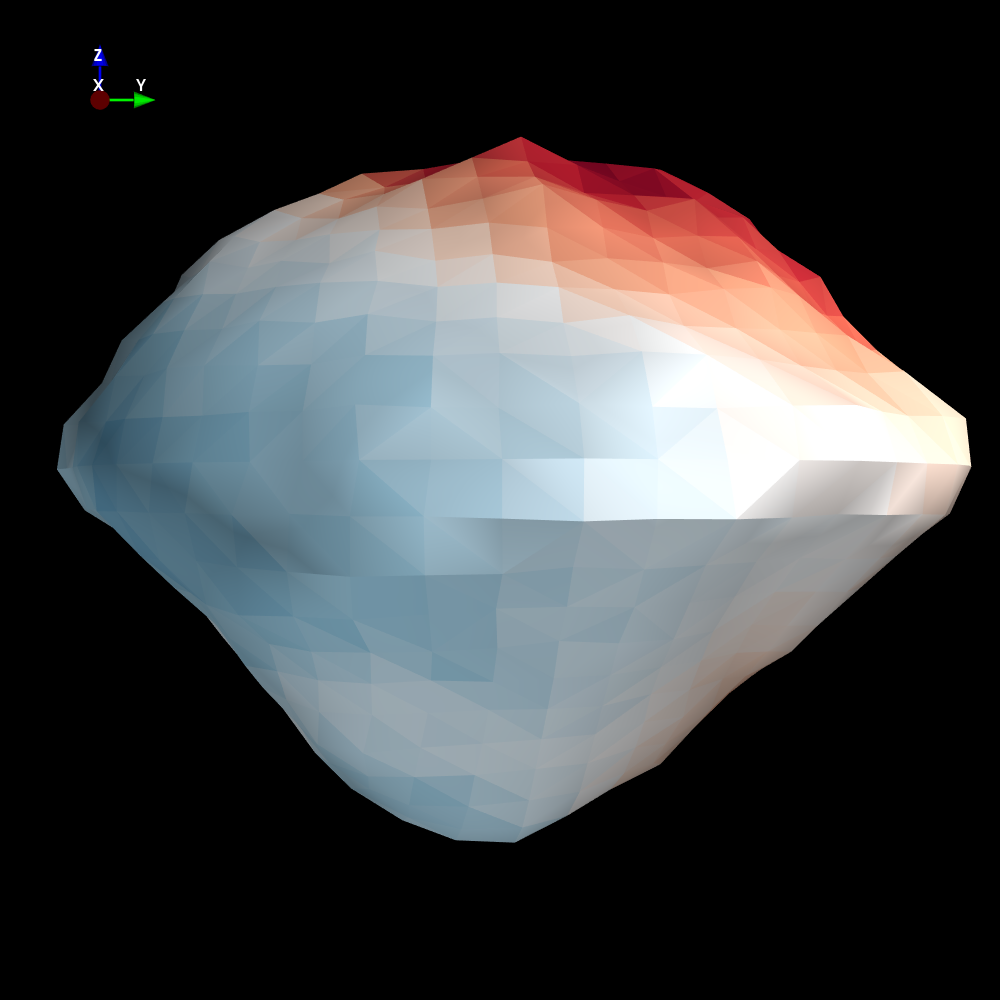} &
   \includegraphics[width=0.26\textwidth ]{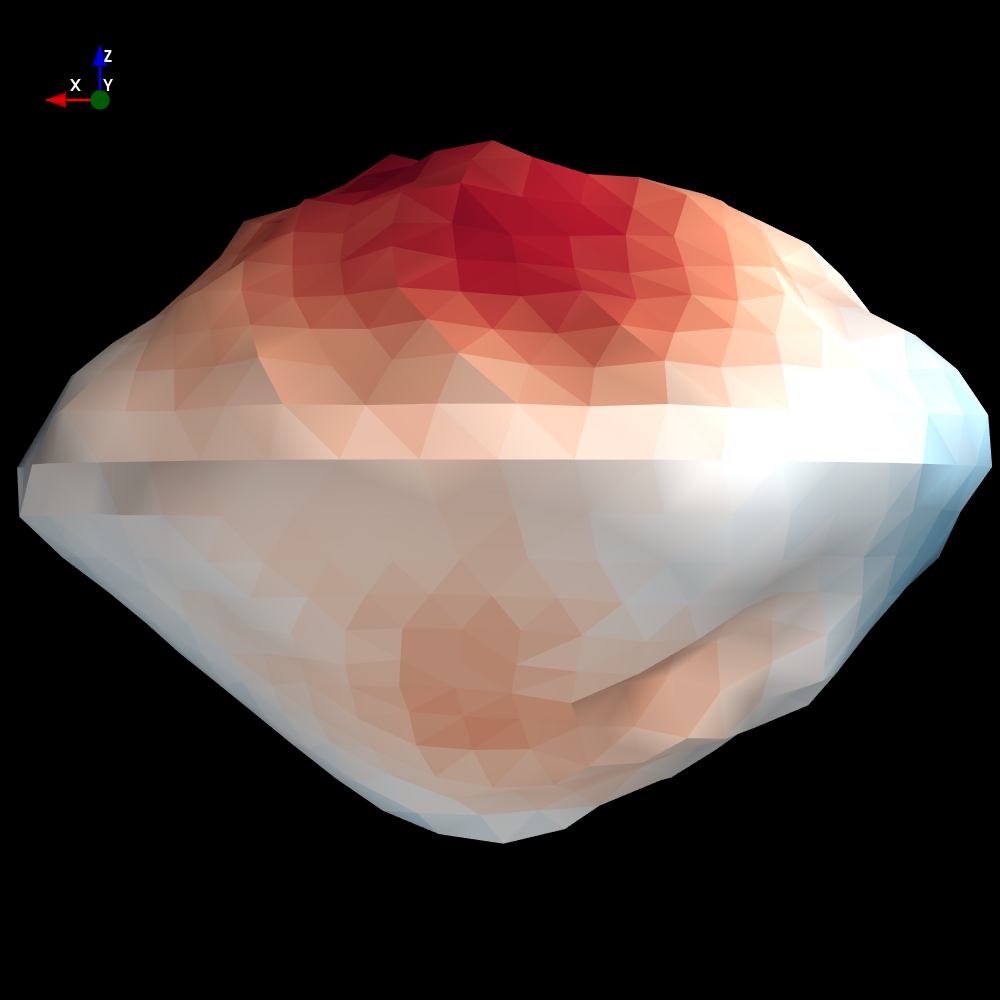}\\
   \includegraphics[width=0.26\textwidth ]{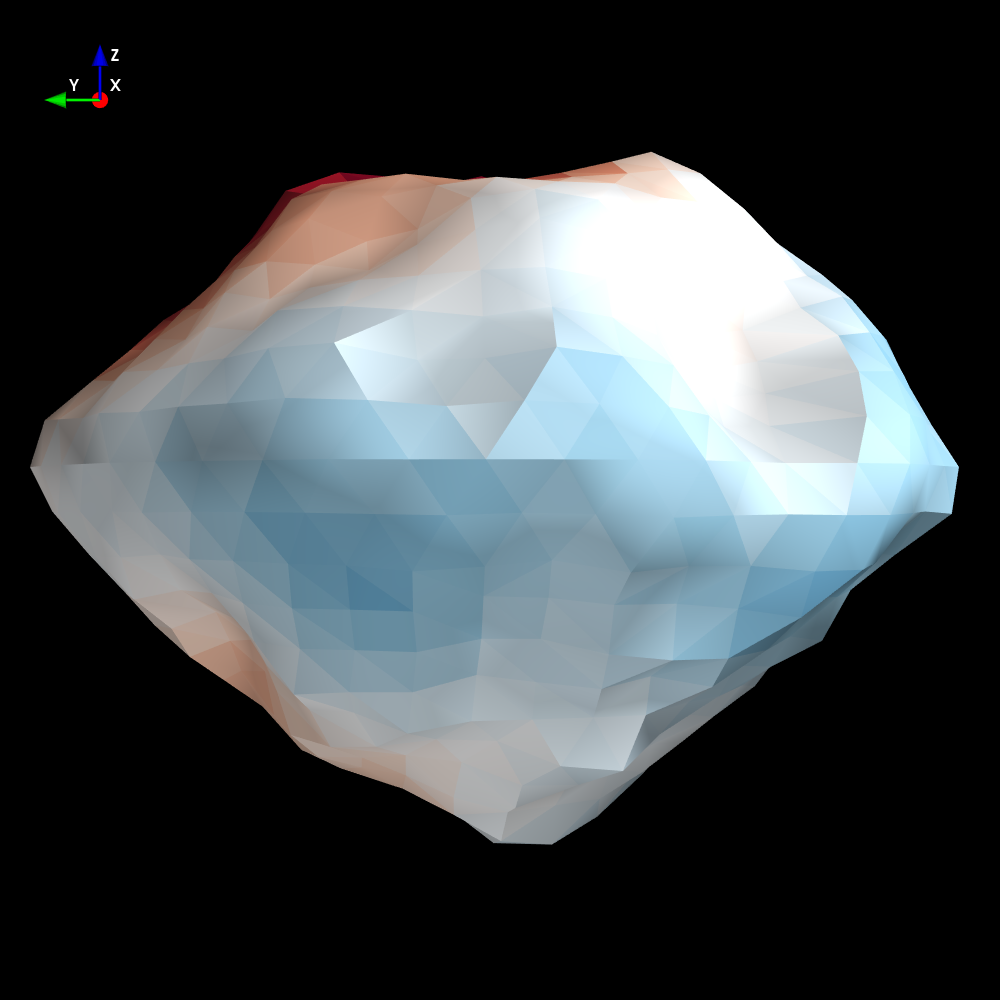} & 
   \includegraphics[width=0.26\textwidth ]{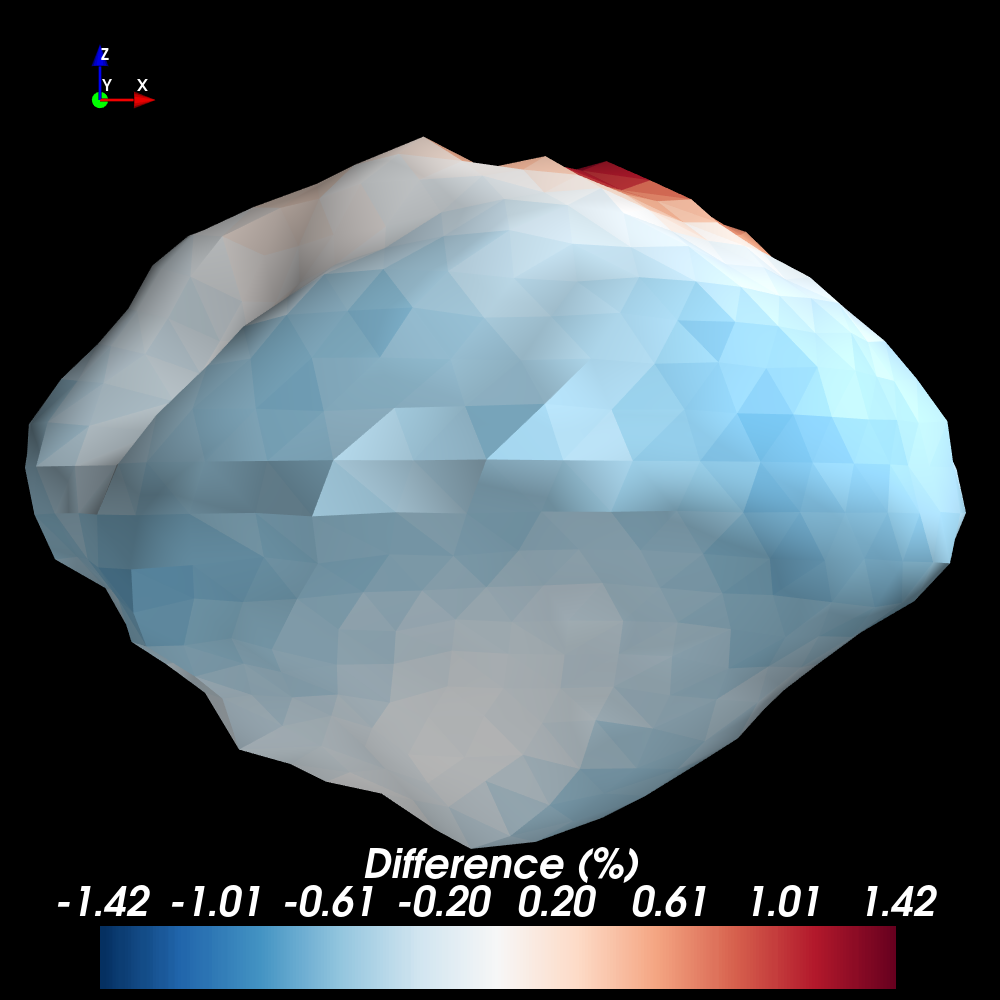} &
   \includegraphics[width=0.26\textwidth ]{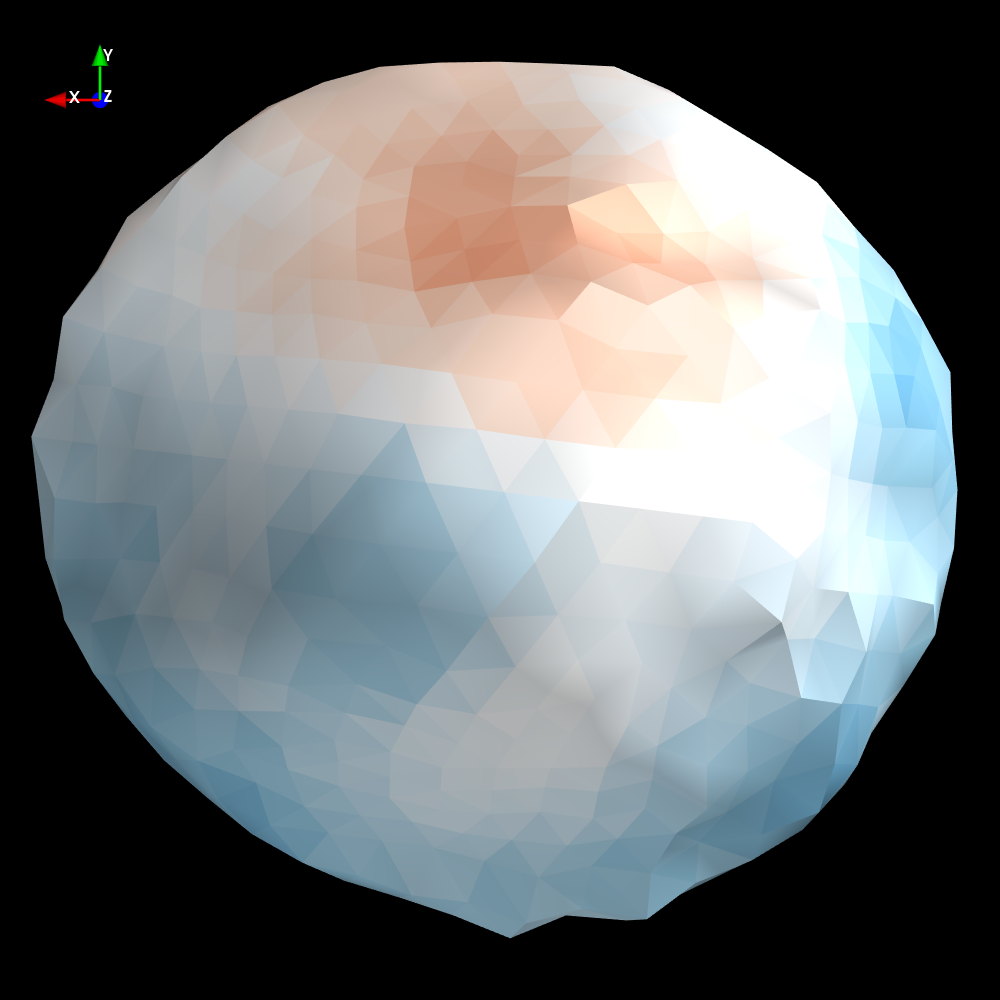}\\
  \end{tabular}
  \caption{Maps of relative difference in the surface geopotential calculated using the mascon method in comparison with the polyhedron method. We assume a uniform density for the polyhedron method calculation. The apparent directions in each panel are the same as in Fig. \ref{fig:geopotential}.
 \label{fig:pot_diff_simple}}
\end{figure*}

\begin{figure*}[ht]
  \centering 
  \begin{tabular}{ccc}
   \includegraphics[width=0.26\textwidth ]{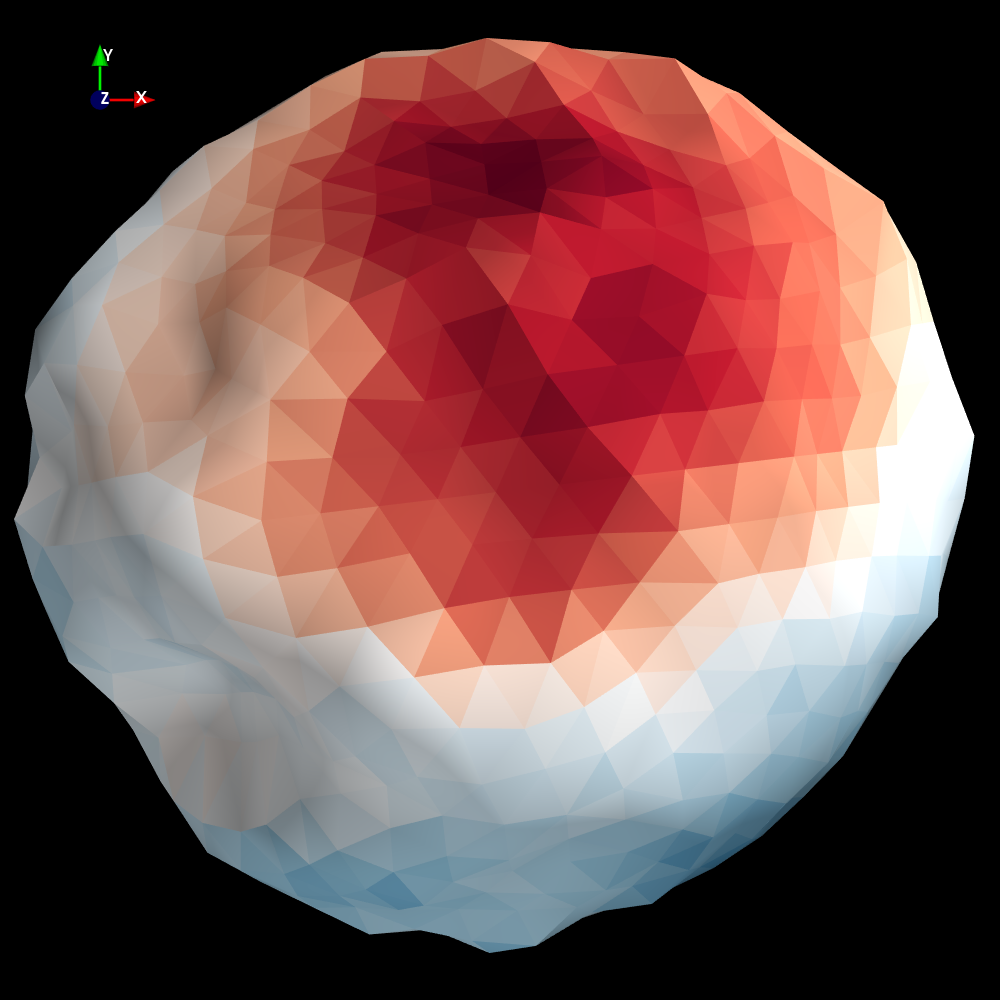} & 
   \includegraphics[width=0.26\textwidth ]{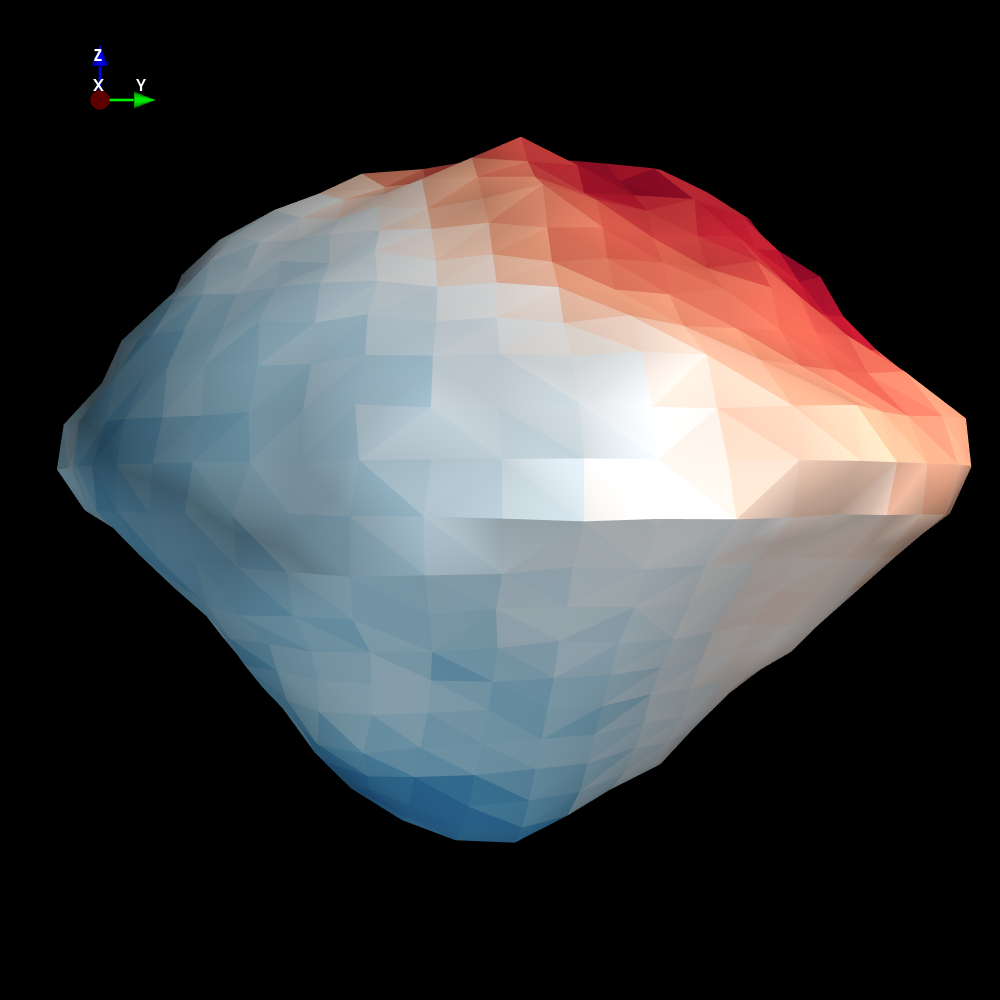} &
   \includegraphics[width=0.26\textwidth ]{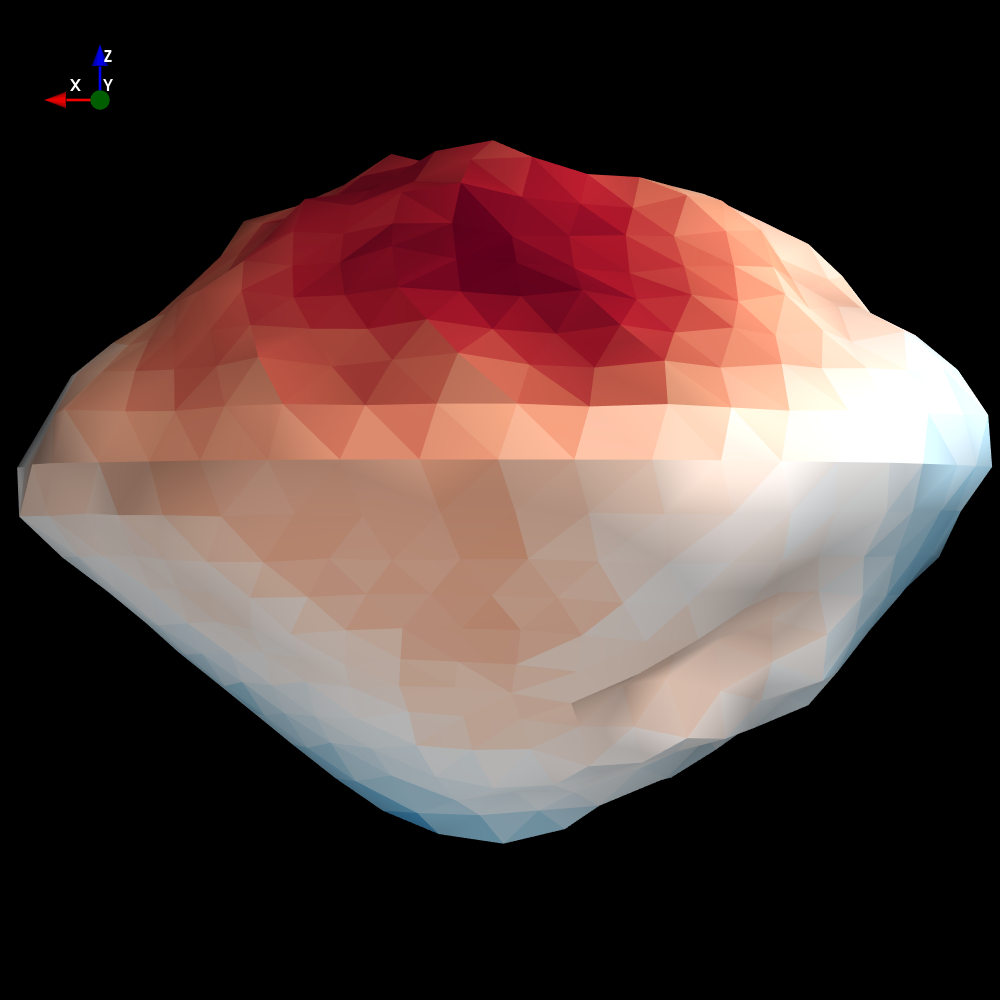}\\
   \includegraphics[width=0.26\textwidth ]{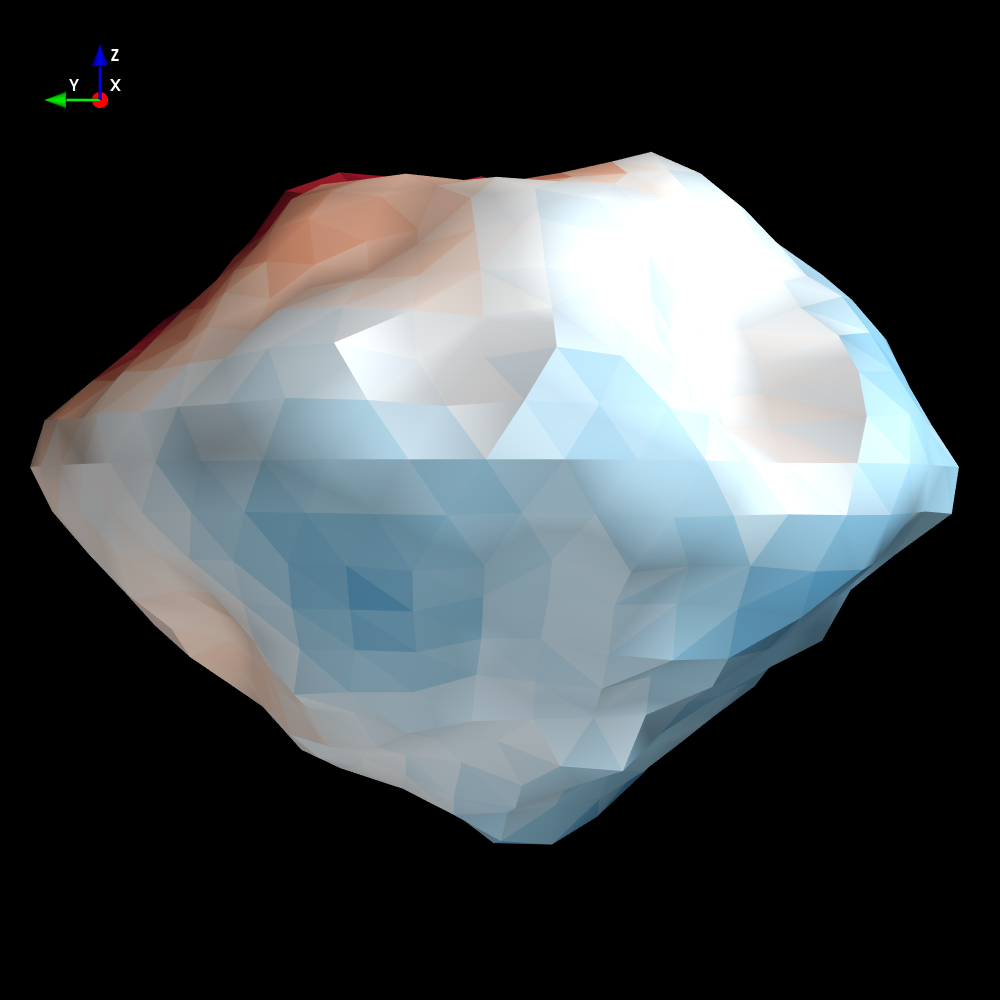} & 
   \includegraphics[width=0.26\textwidth ]{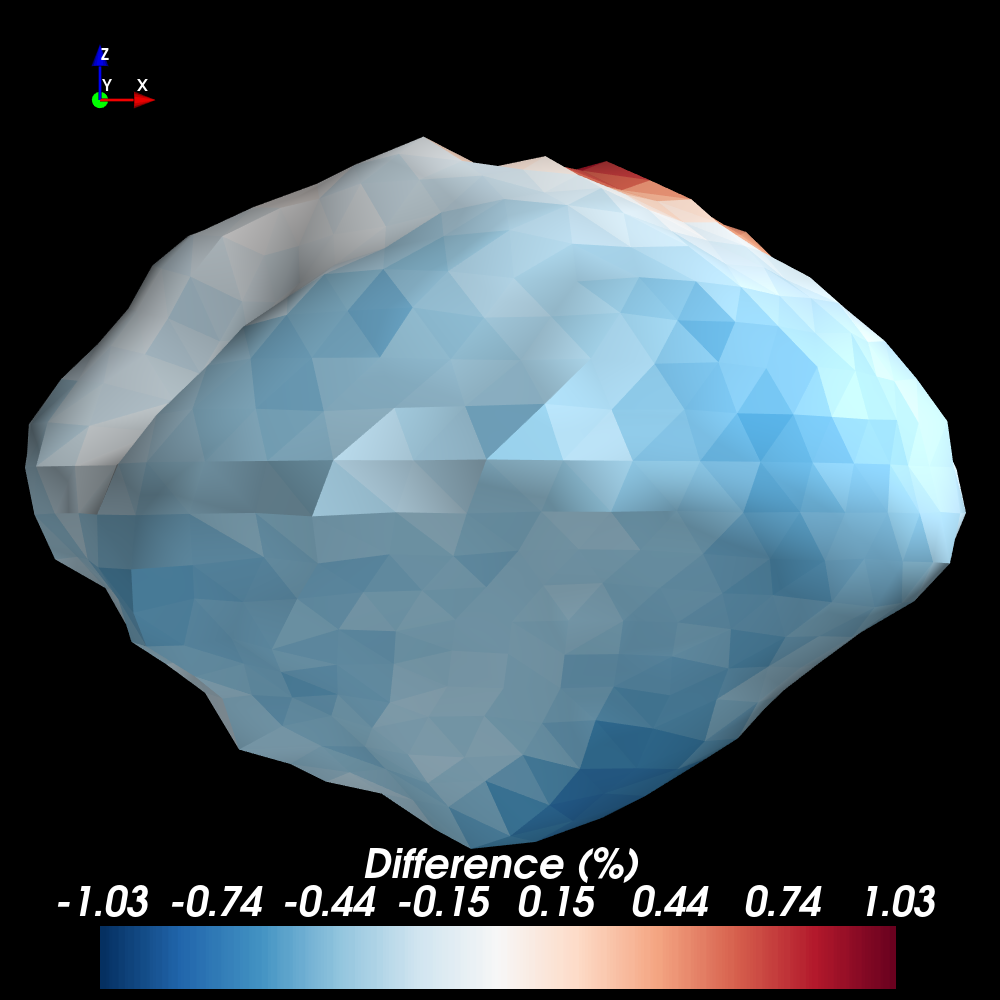} &
   \includegraphics[width=0.26\textwidth ]{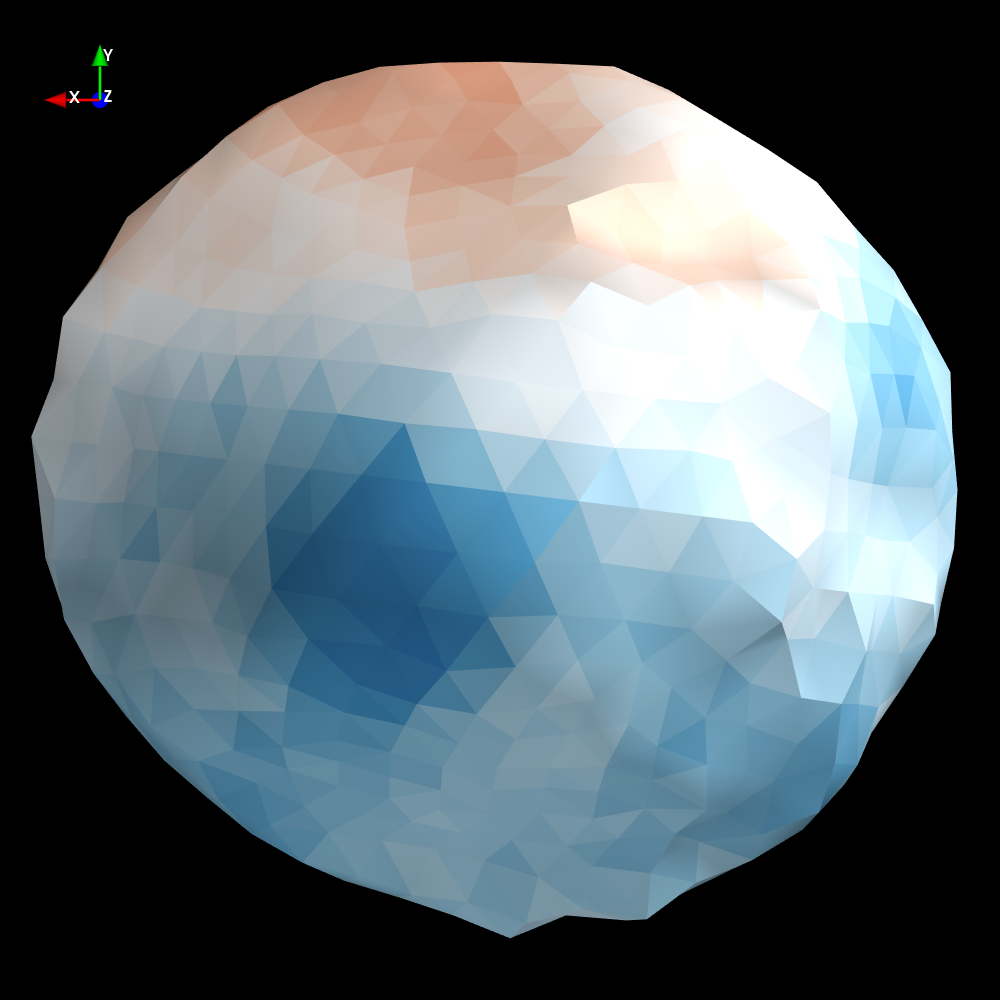}\\
  \end{tabular}
  \caption{Maps of relative difference in the surface geopotential calculated using the mascon method in comparison with the polyhedron method. Here, the polyhedron method with the core-envelope structure is considered. The apparent directions in each panel are the same as in Fig. \ref{fig:geopotential}.
 \label{fig:pot_diff_core}}
\end{figure*}

\begin{figure*}[ht]
  \centering 
  \begin{tabular}{ccc}
   \includegraphics[width=0.26\textwidth ]{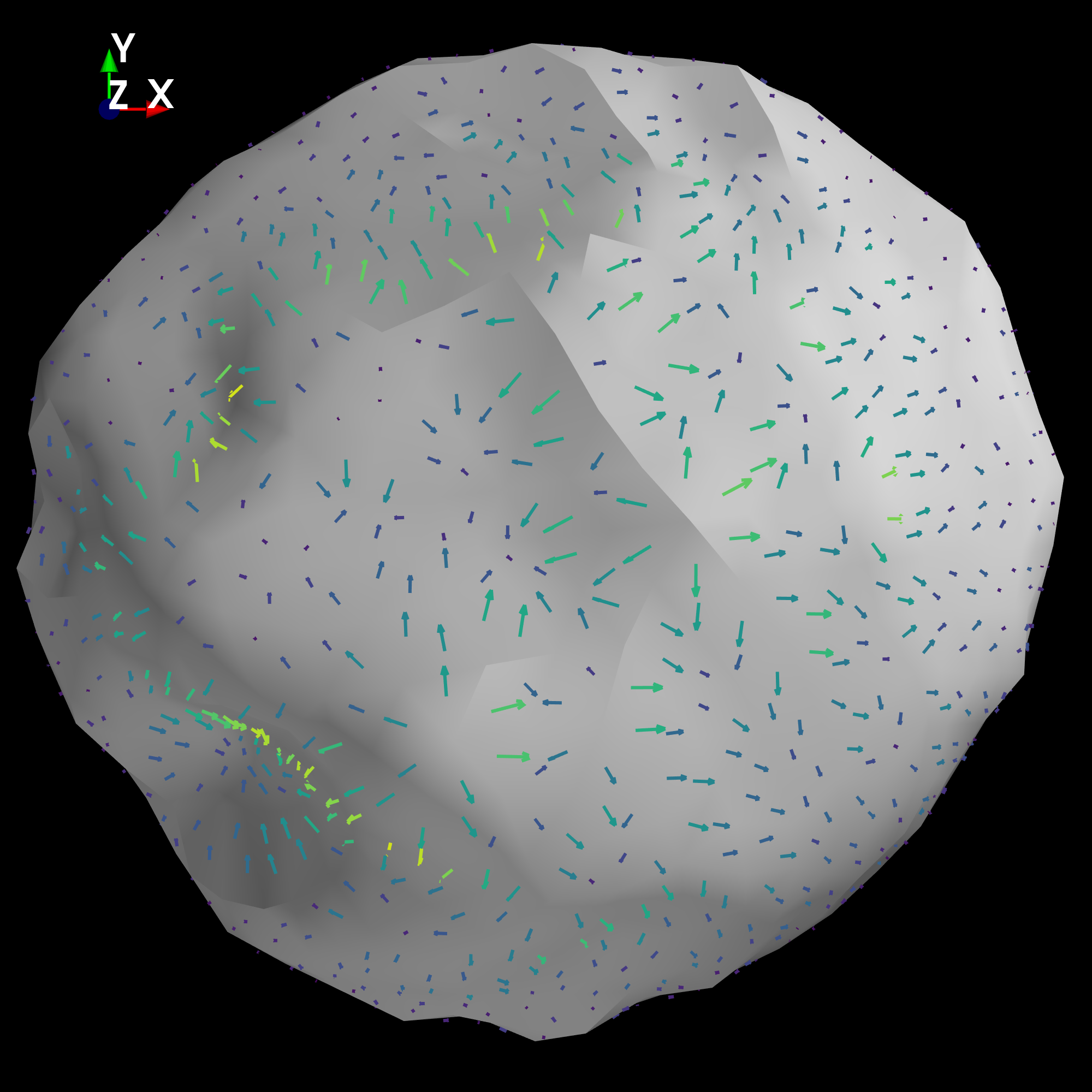} & 
   \includegraphics[width=0.26\textwidth ]{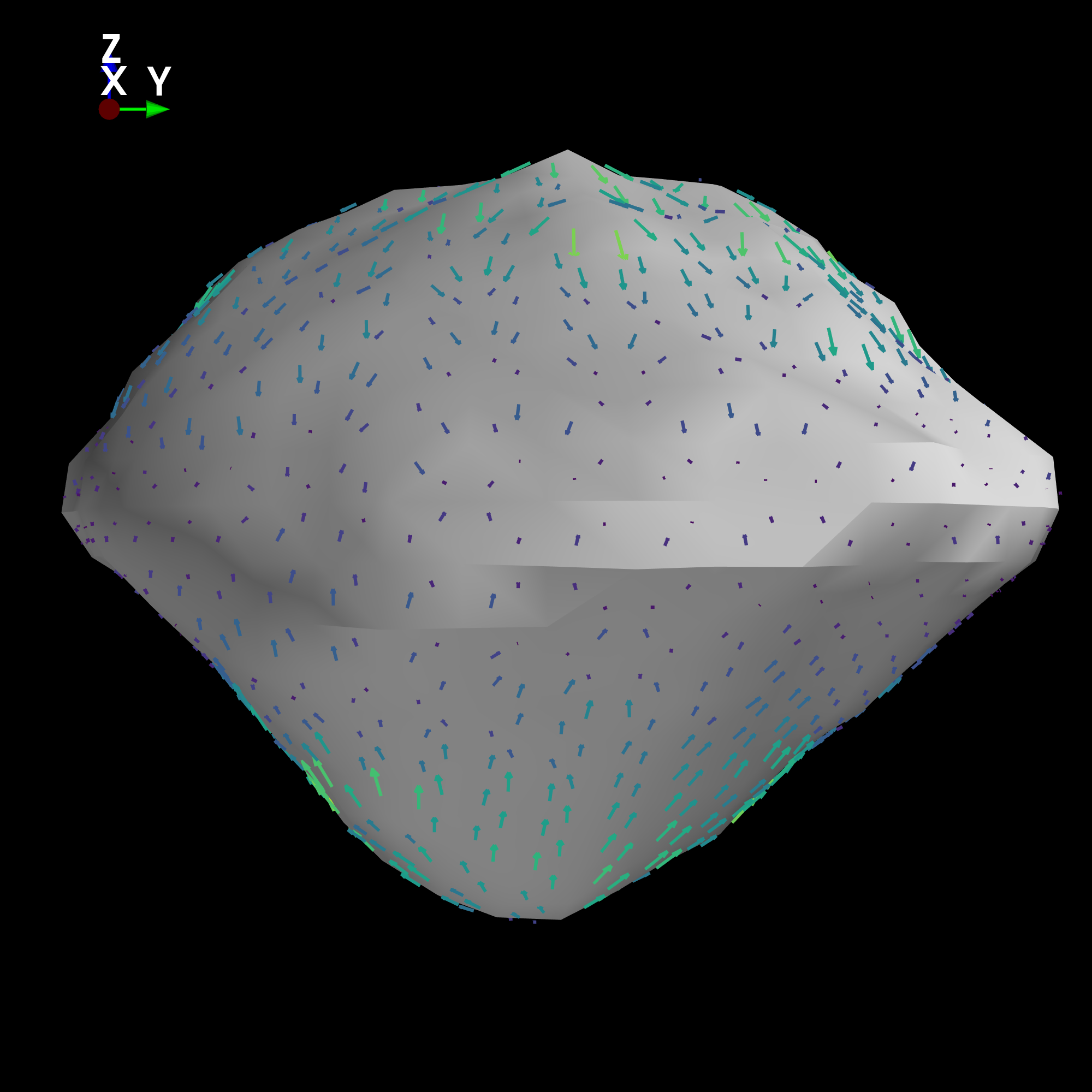} &
   \includegraphics[width=0.26\textwidth ]{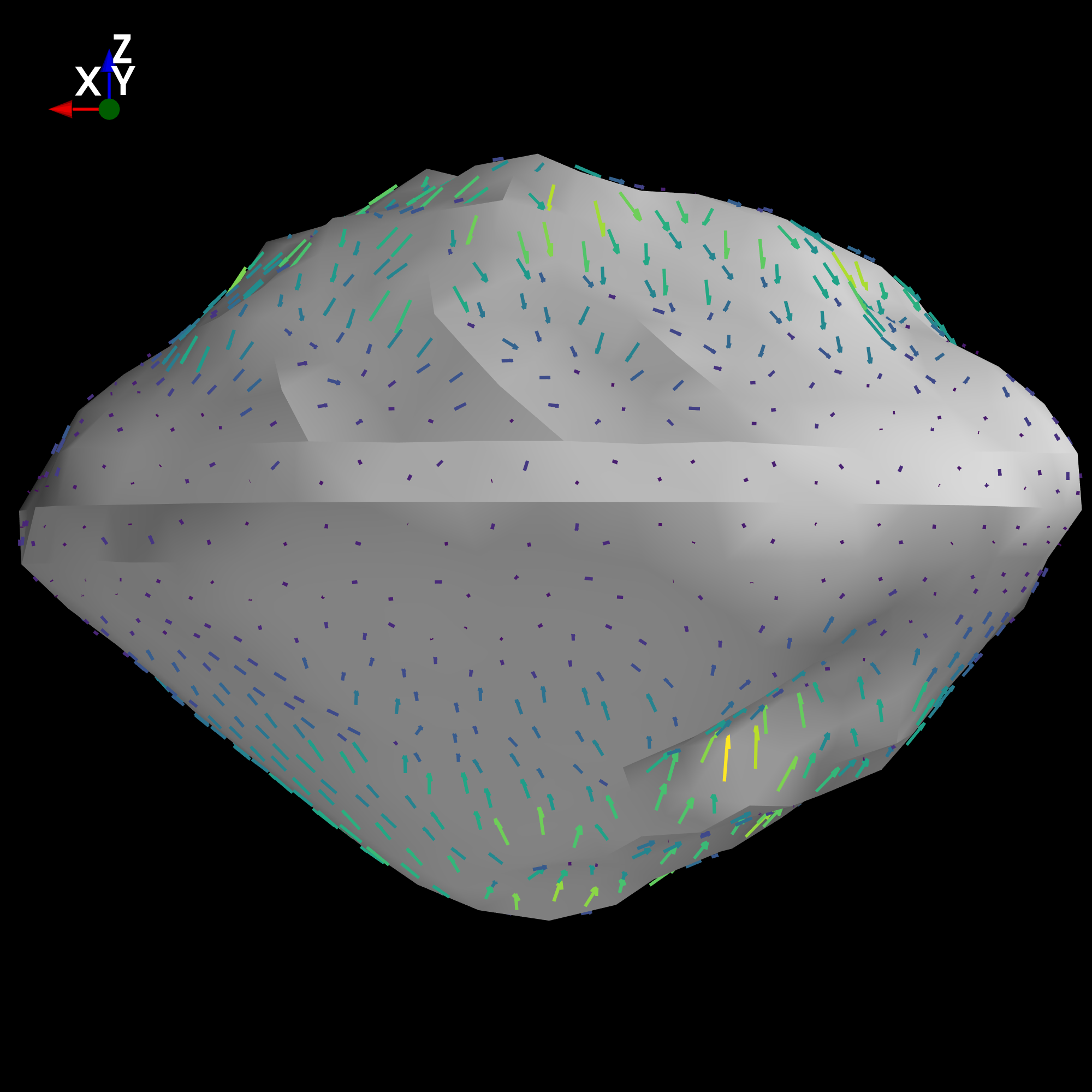}\\
   \includegraphics[width=0.26\textwidth ]{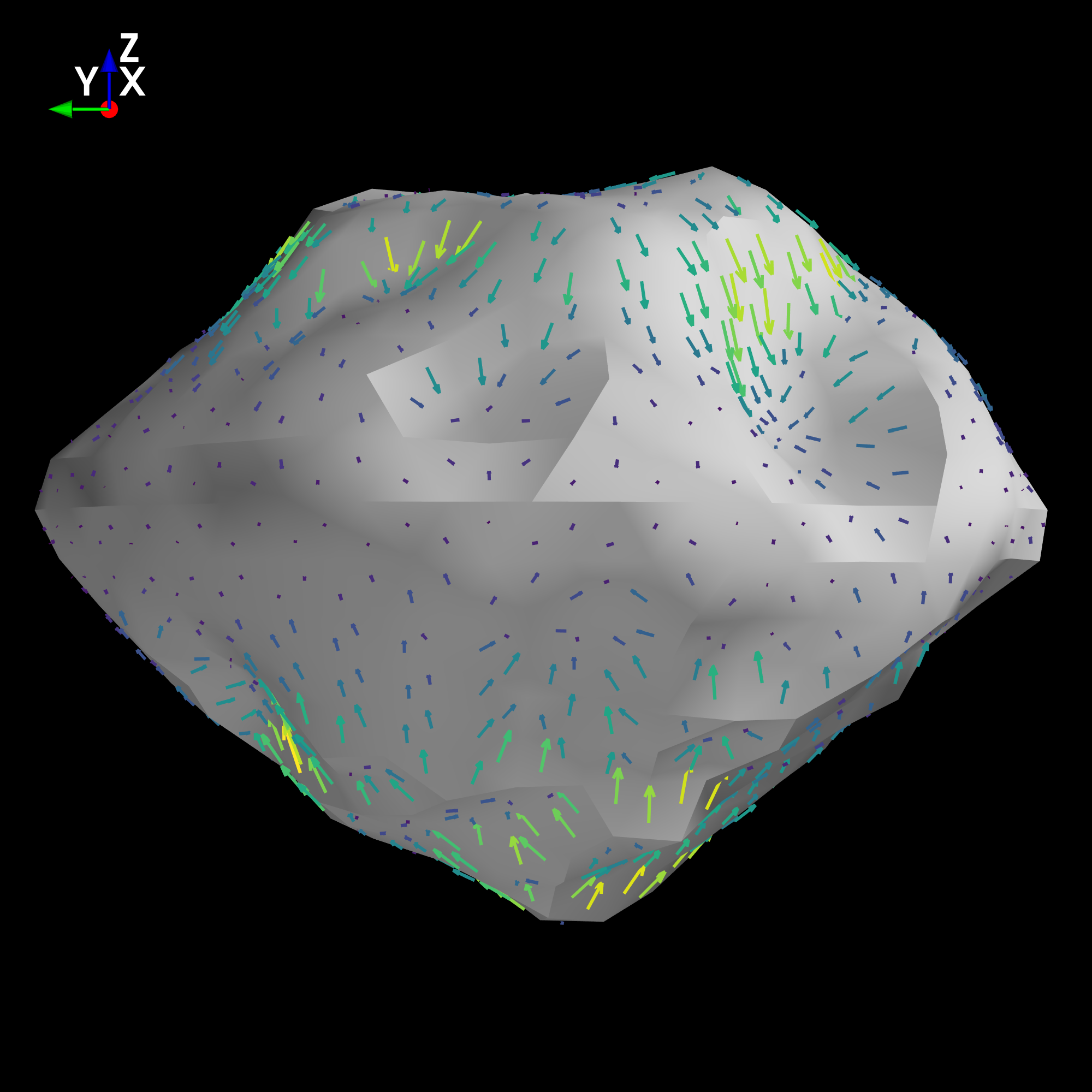} & 
   \includegraphics[width=0.26\textwidth ]{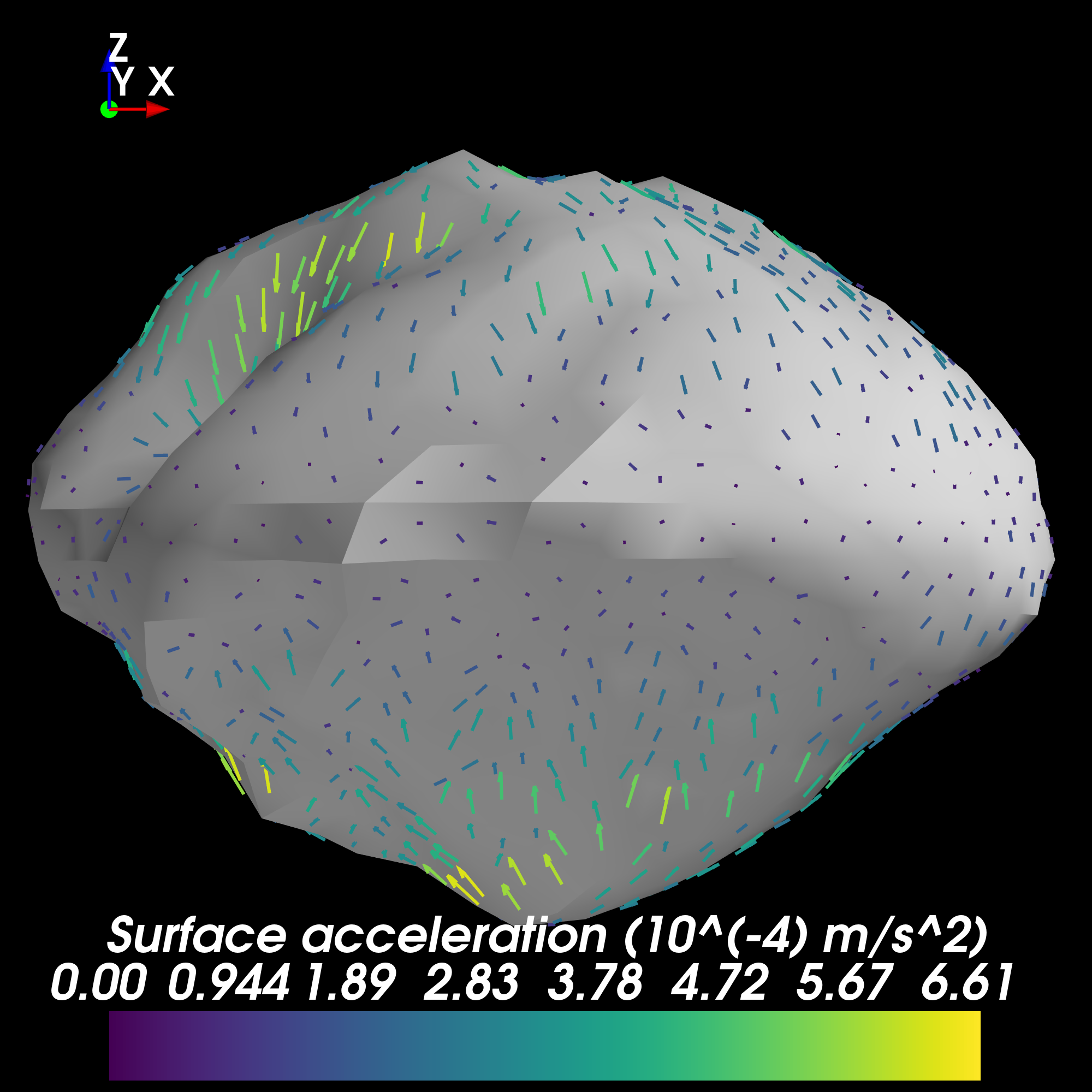} &
   \includegraphics[width=0.26\textwidth ]{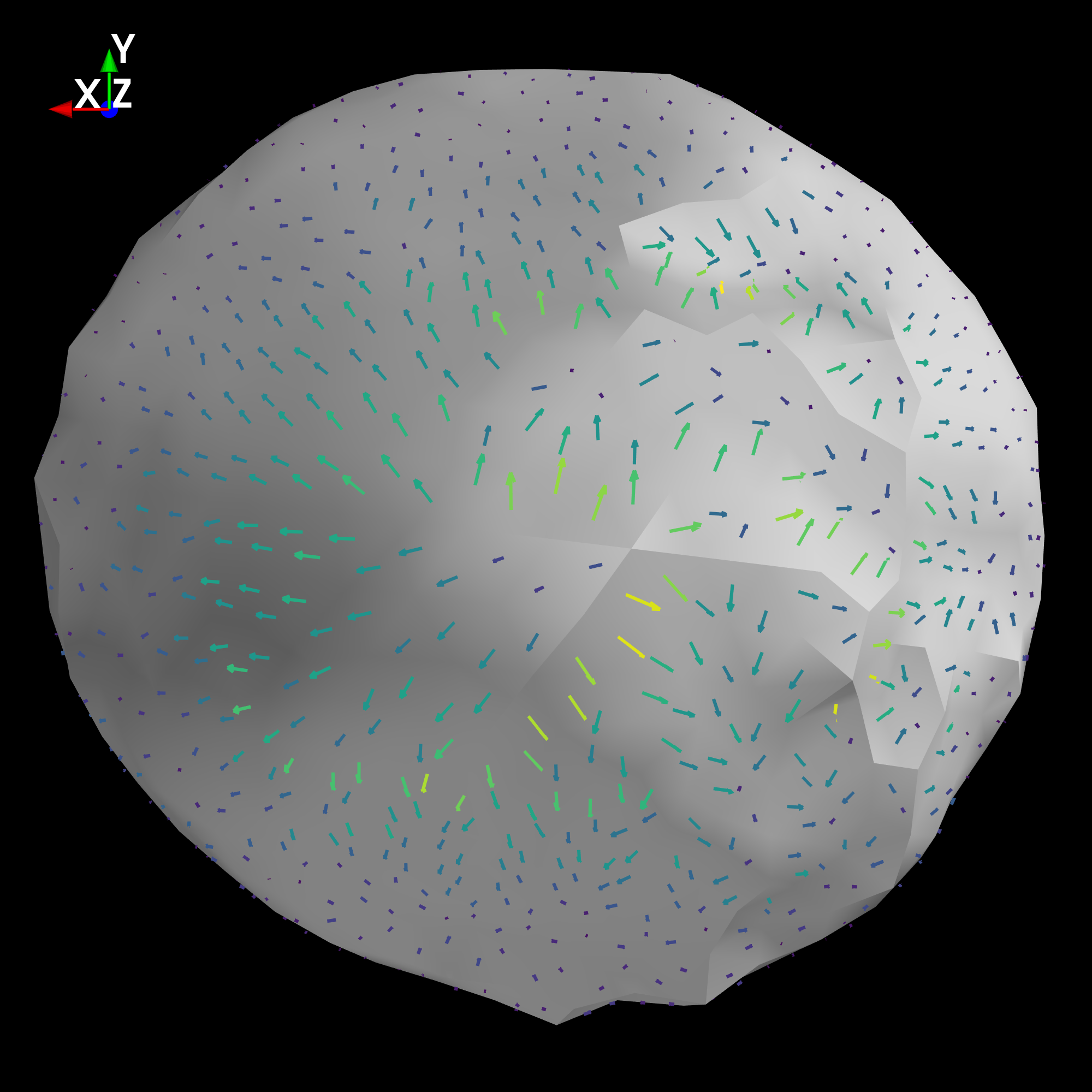}\\
  \end{tabular}
 \caption{Maps of the surface acceleration on the Phaethon surface.
 \label{fig:surf_acc}}
\end{figure*}

\begin{figure*}[ht]
  \centering 
  \begin{tabular}{ccc}
   \includegraphics[width=0.26\textwidth ]{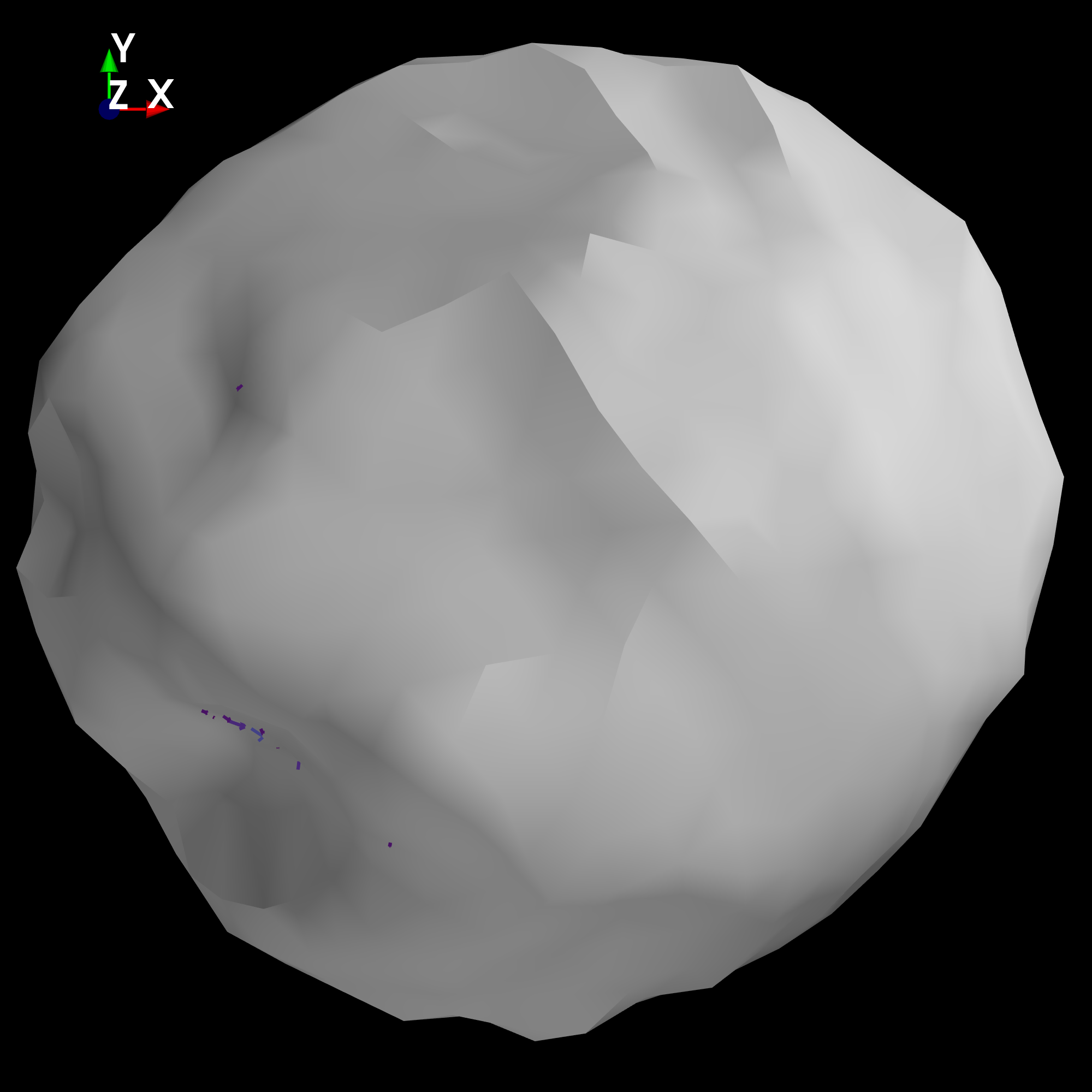} & 
   \includegraphics[width=0.26\textwidth ]{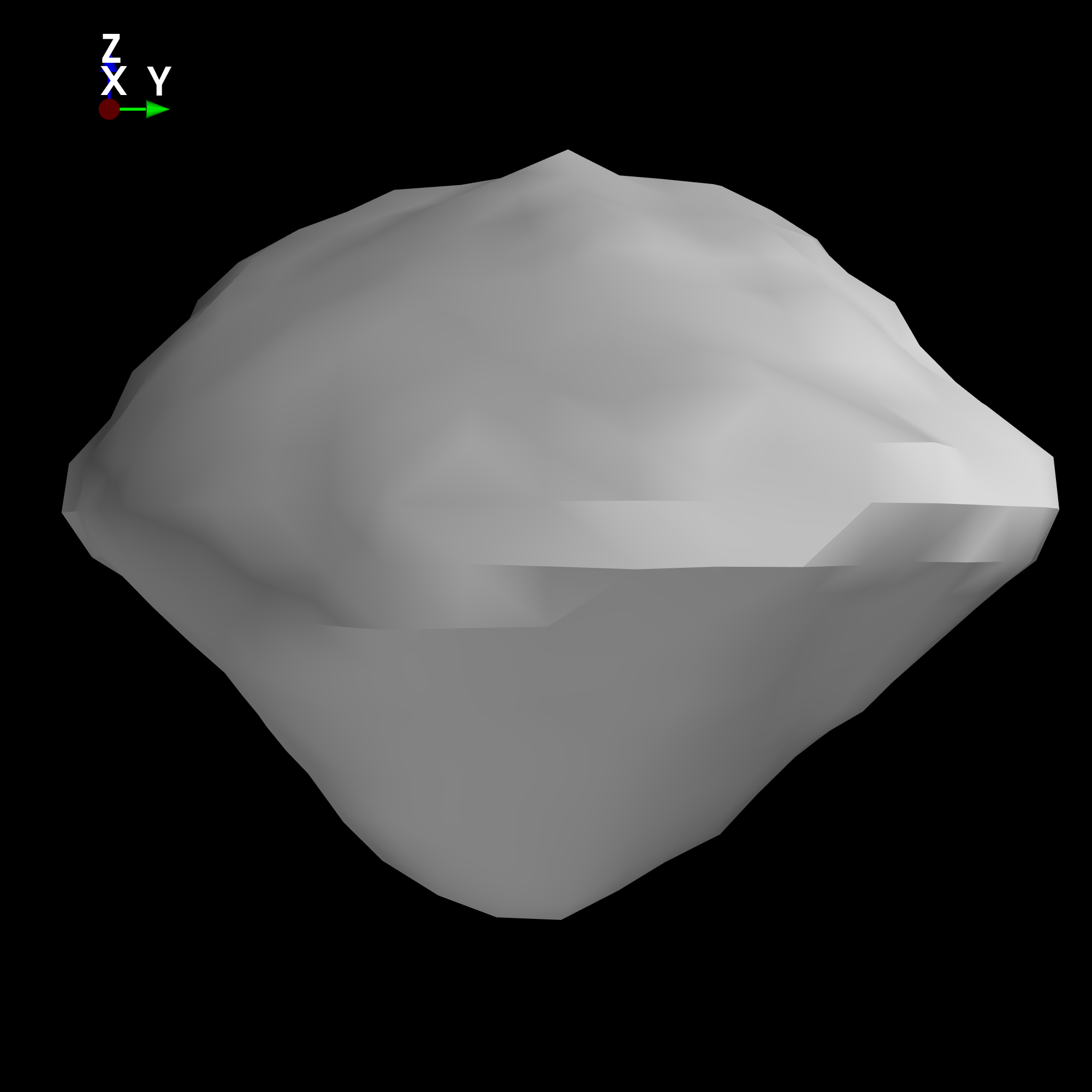} &
   \includegraphics[width=0.26\textwidth ]{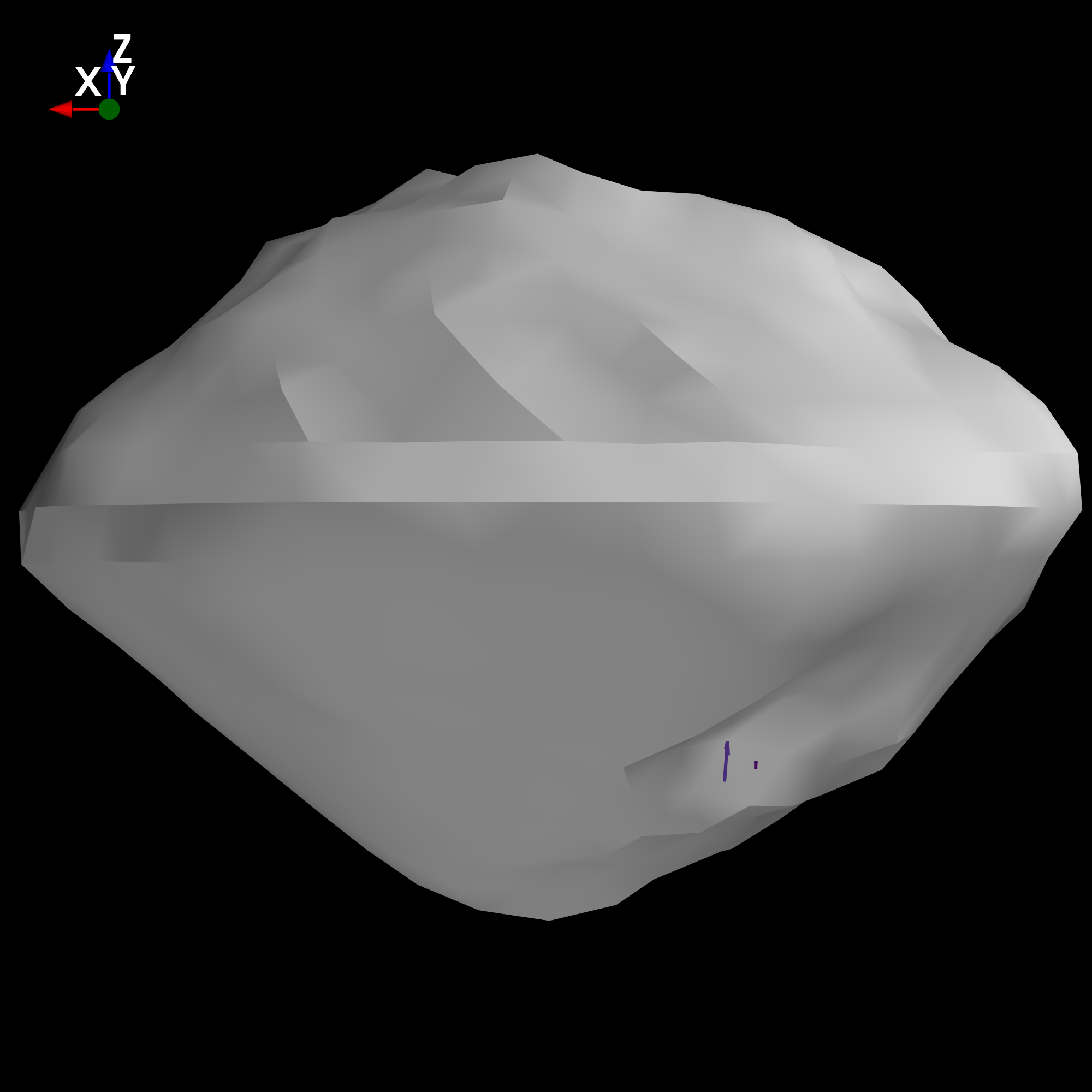}\\
   \includegraphics[width=0.26\textwidth ]{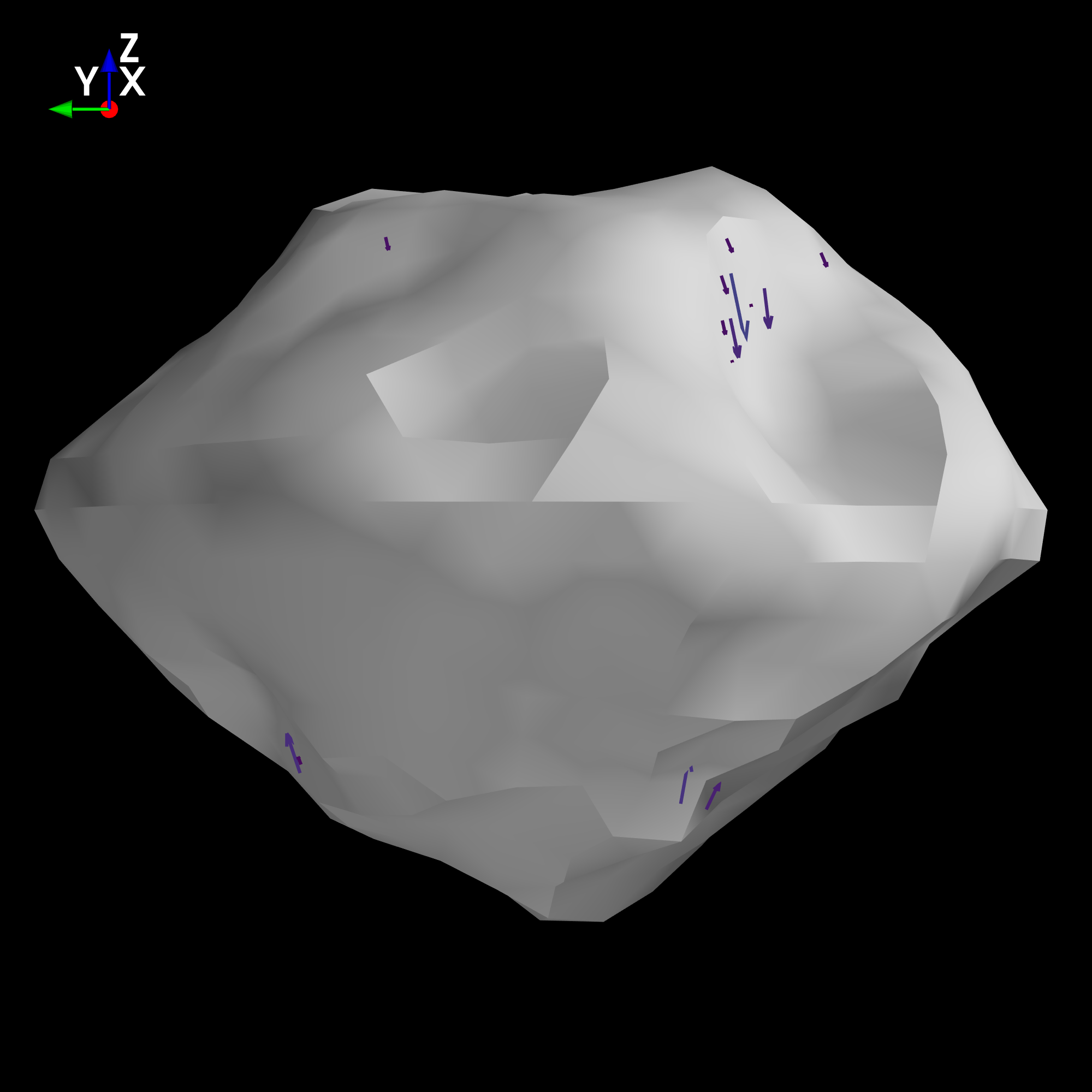} & 
   \includegraphics[width=0.26\textwidth ]{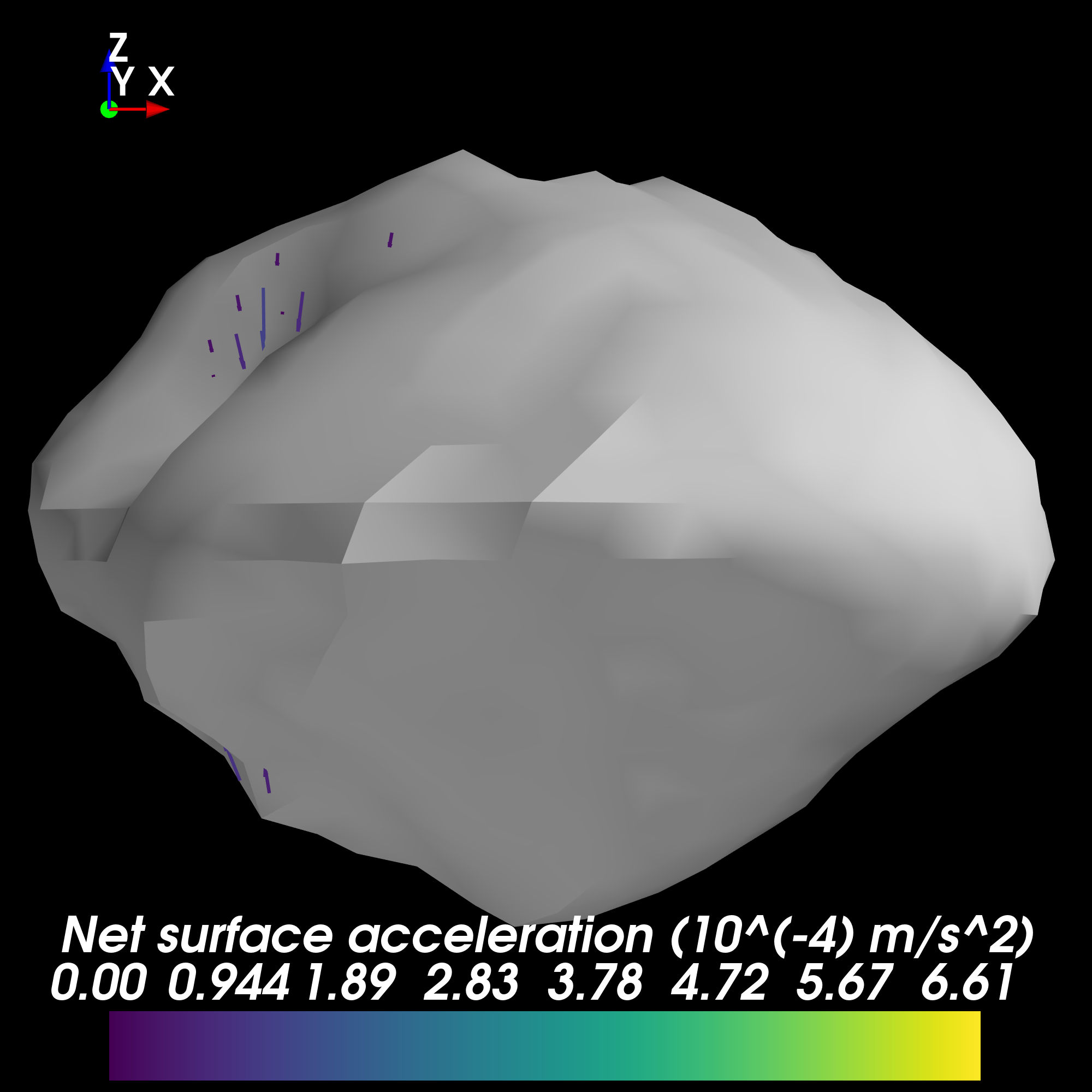} &
   \includegraphics[width=0.26\textwidth ]{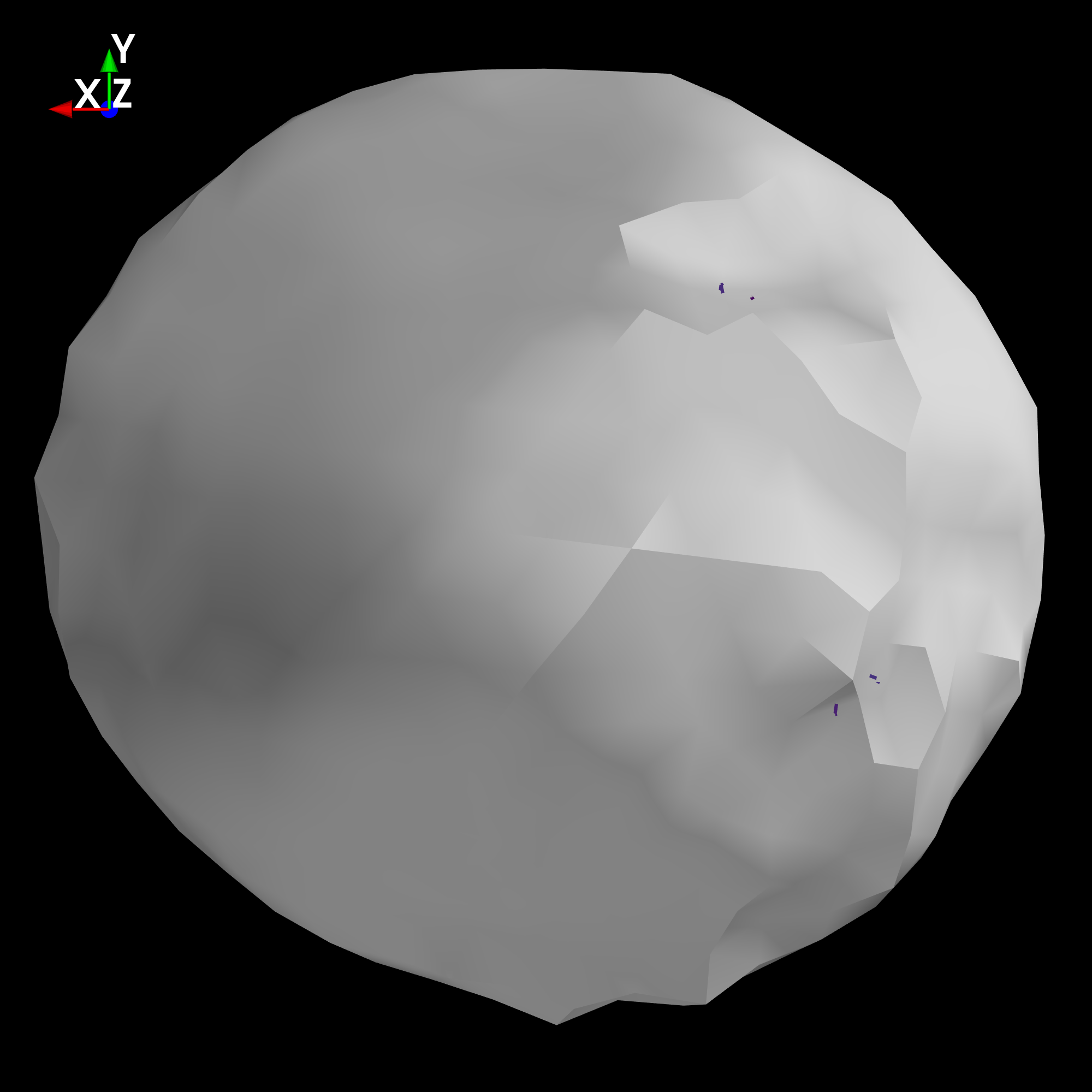}\\
  \end{tabular}
 \caption{Maps of the net acceleration on the Phaethon surface.
 \label{fig:net_acc}}
\end{figure*}

\begin{figure*}[ht]
  \centering 
  \begin{tabular}{ccc}
   \includegraphics[width=0.26\textwidth ]{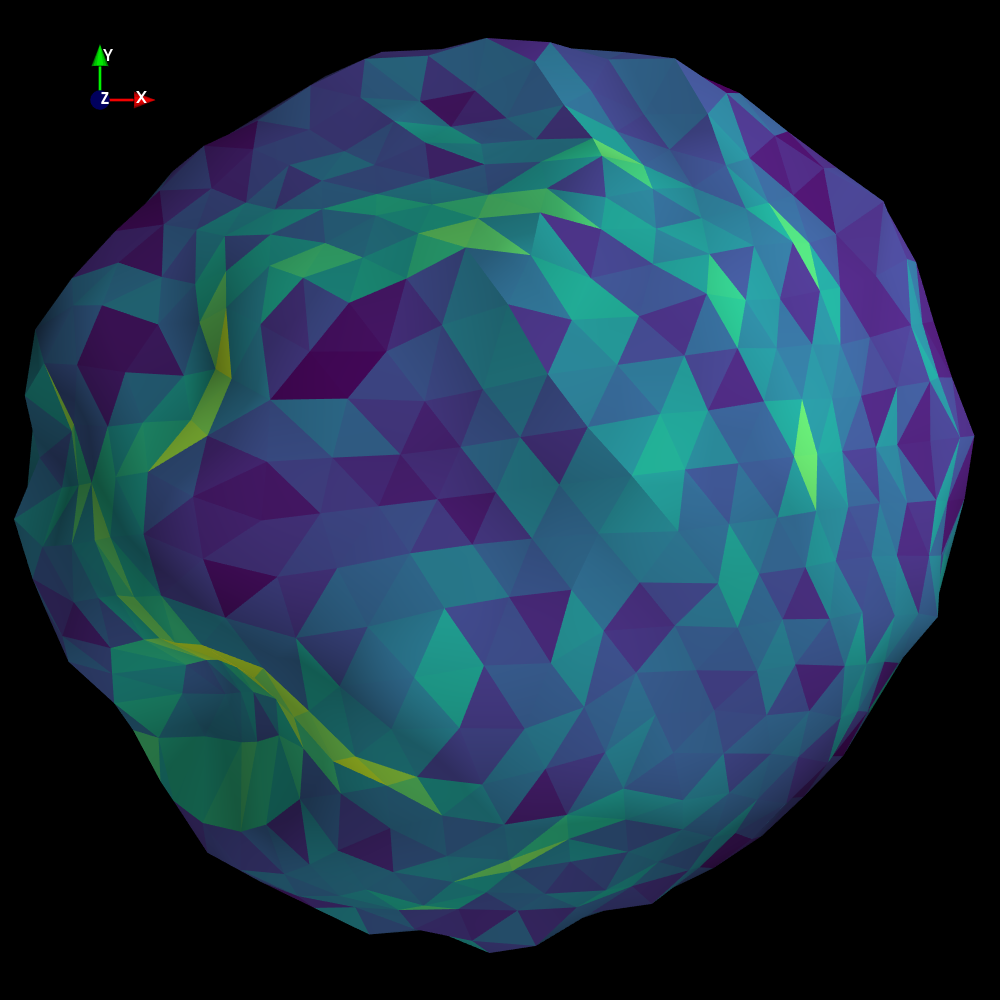} & 
   \includegraphics[width=0.26\textwidth ]{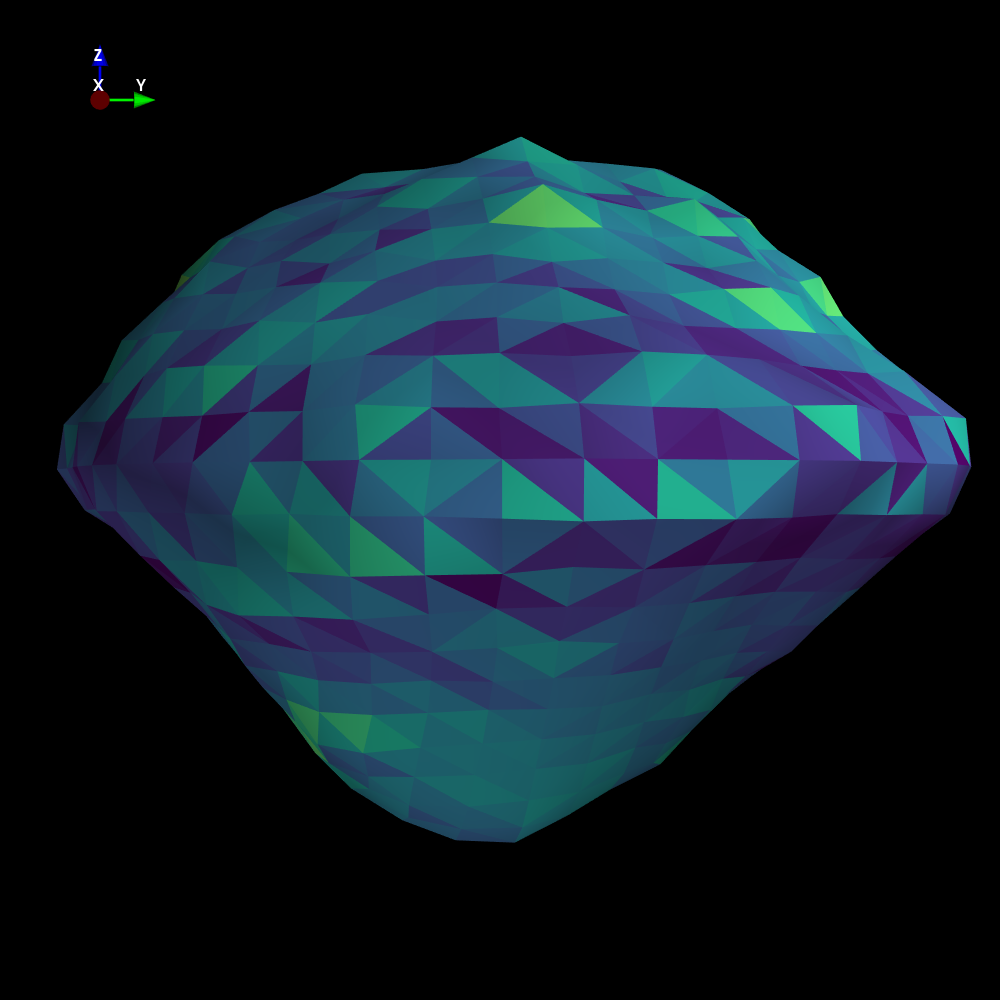} &
   \includegraphics[width=0.26\textwidth ]{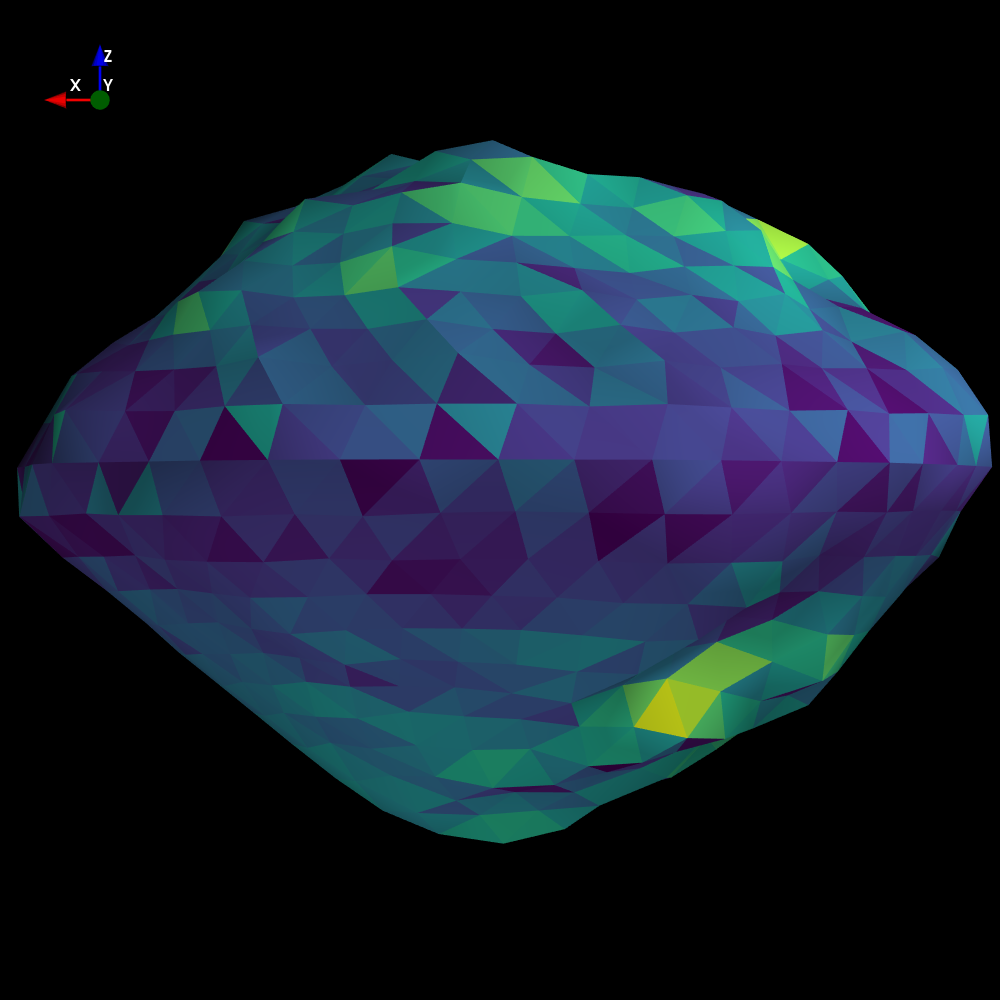}\\
   \includegraphics[width=0.26\textwidth ]{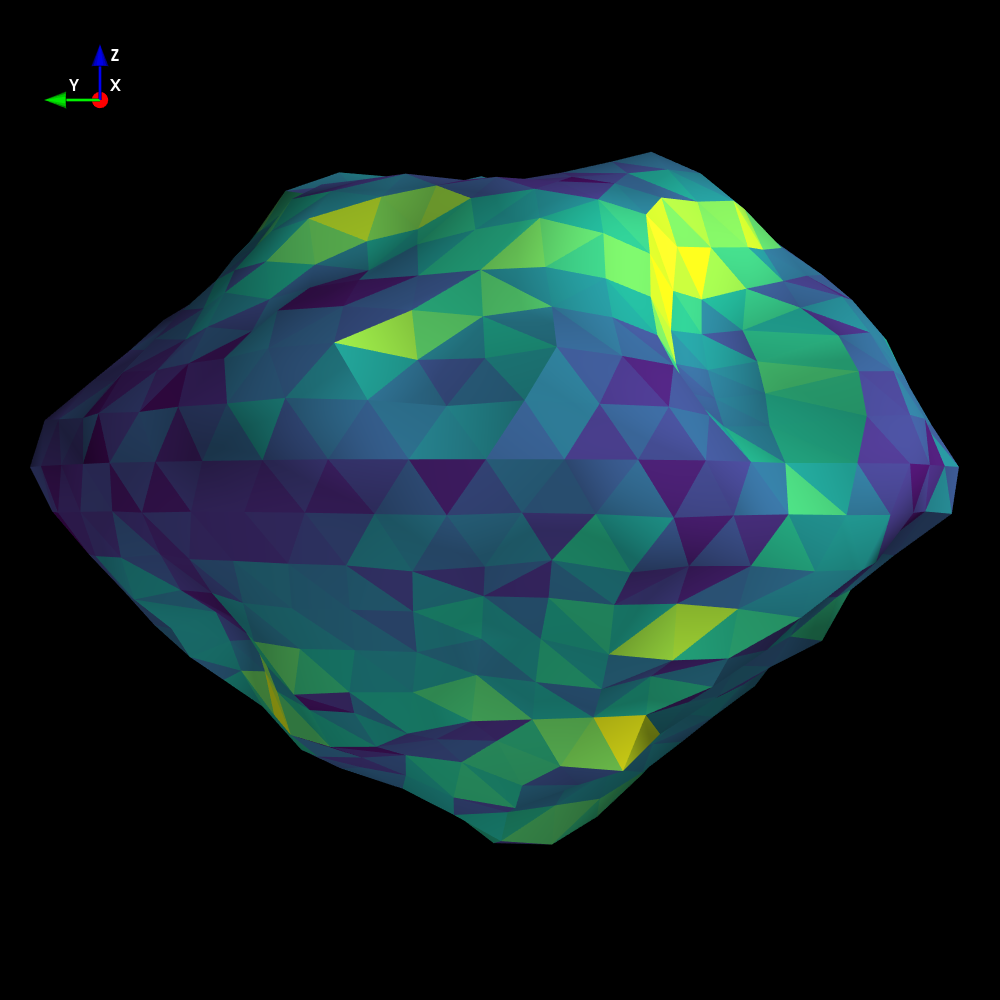} & 
   \includegraphics[width=0.26\textwidth ]{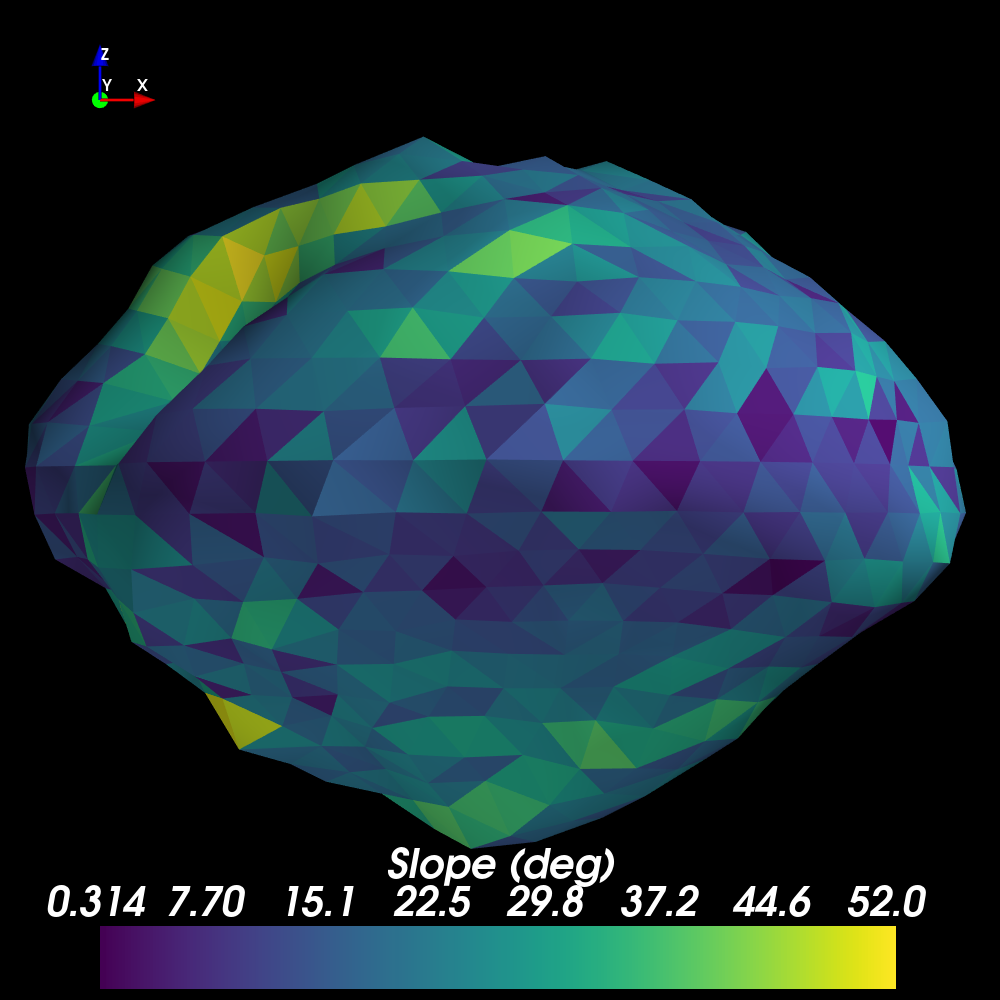} &
   \includegraphics[width=0.26\textwidth ]{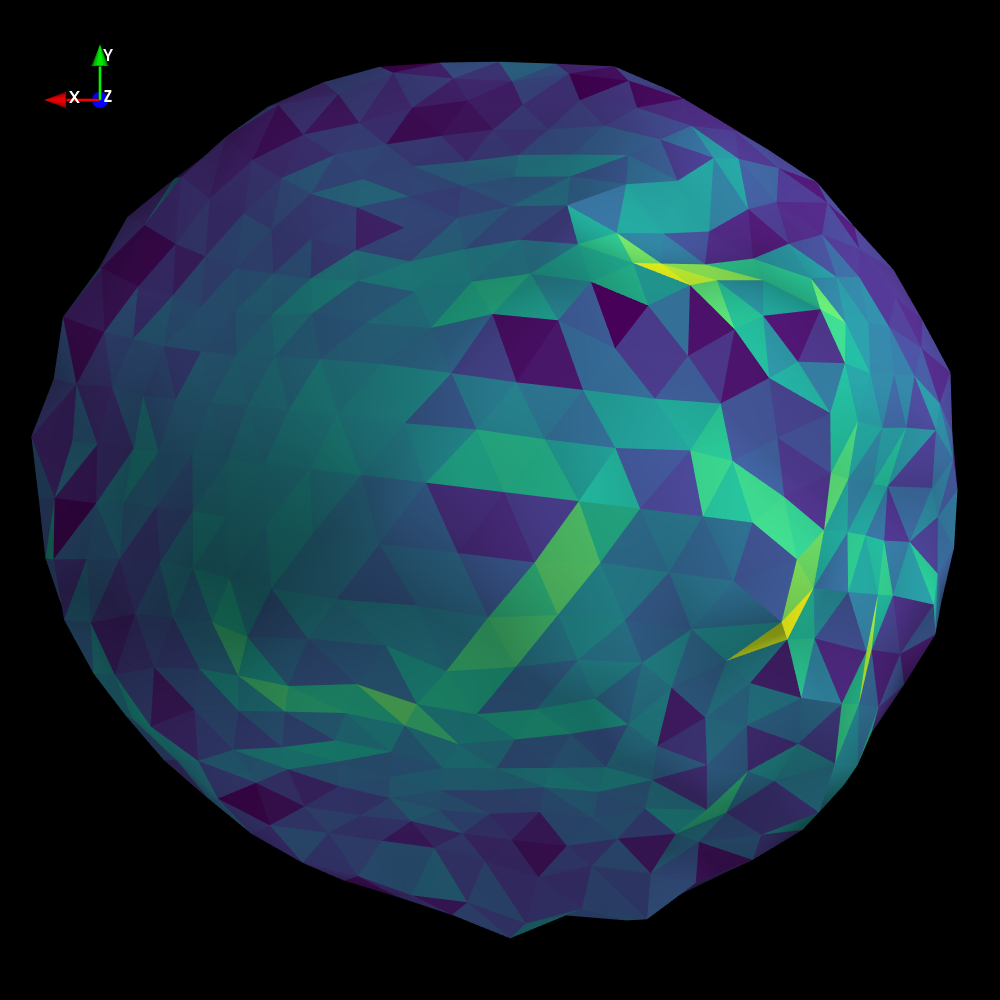}\\
  \end{tabular}
 \caption{Slope maps on the Phaethon surface}
 \label{fig:slope}
\end{figure*}

\begin{figure*}[ht]
  \centering 
  \begin{tabular}{ccc}
   \includegraphics[width=0.26\textwidth ]{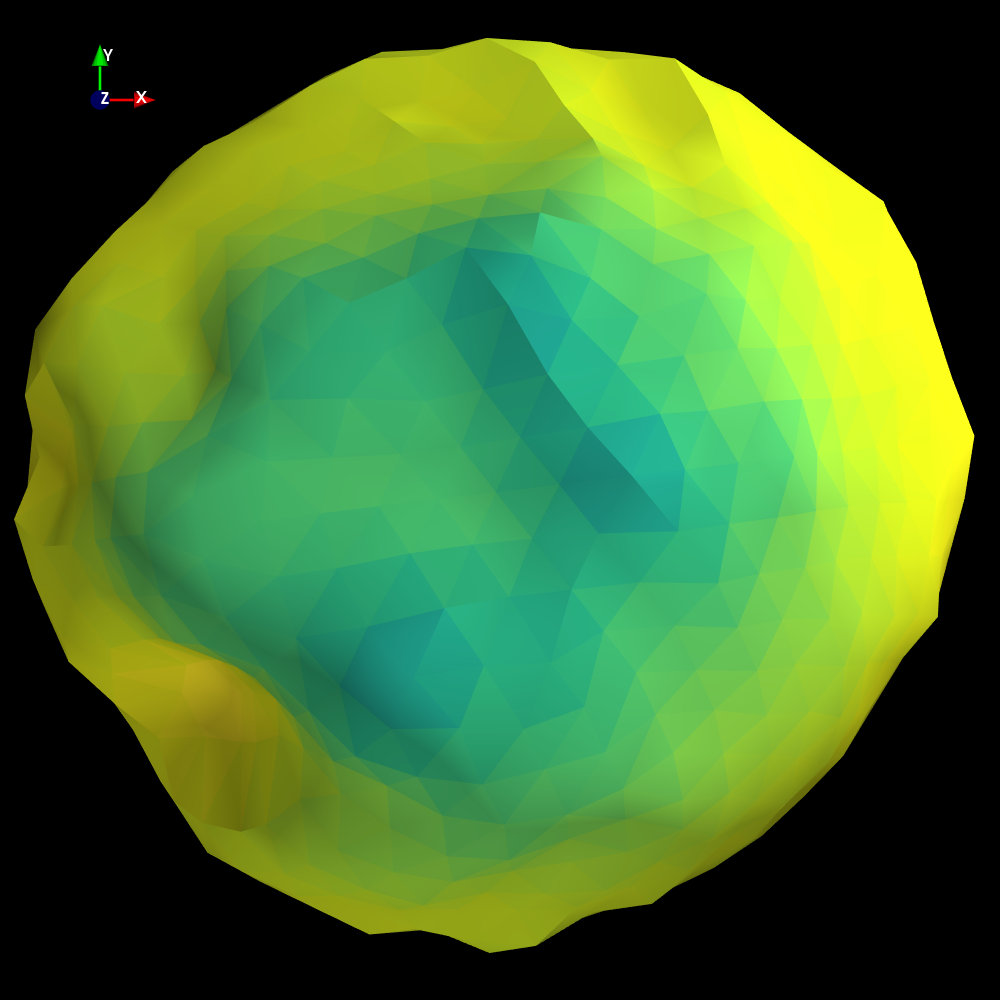} & 
   \includegraphics[width=0.26\textwidth ]{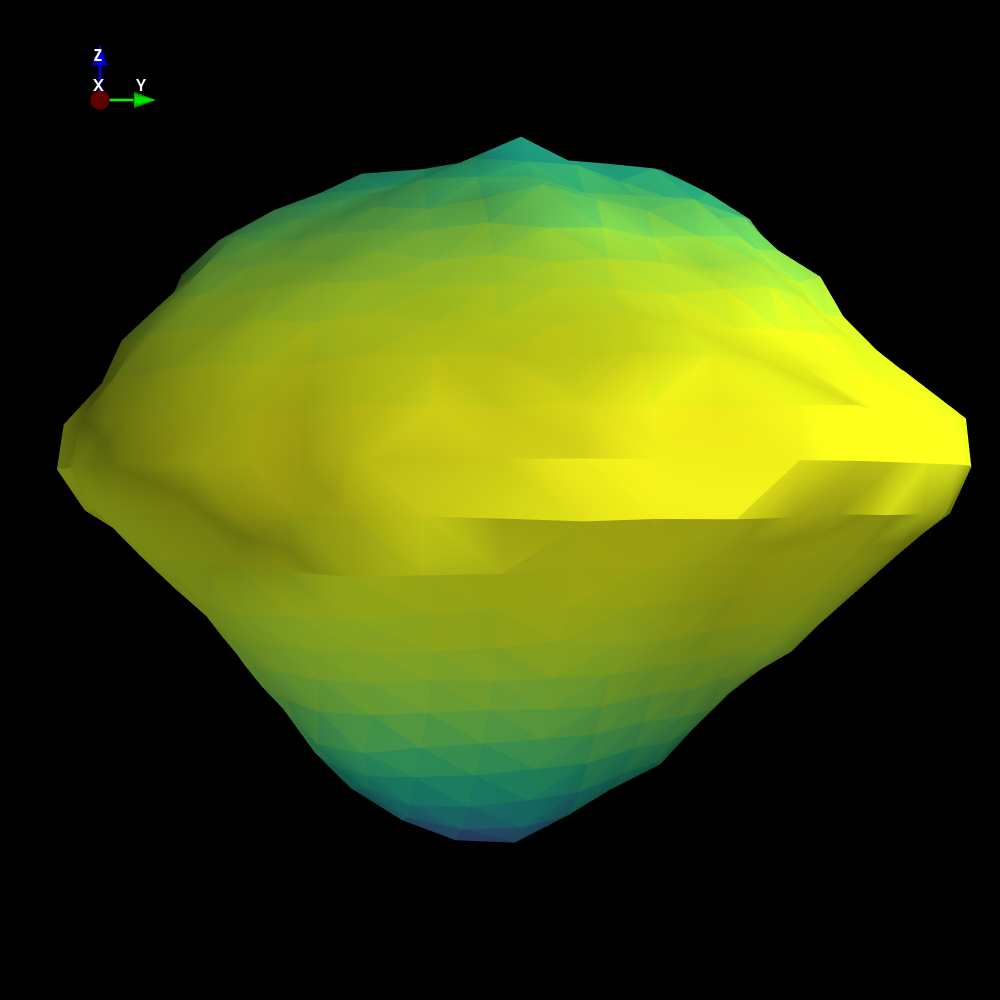} &
   \includegraphics[width=0.26\textwidth ]{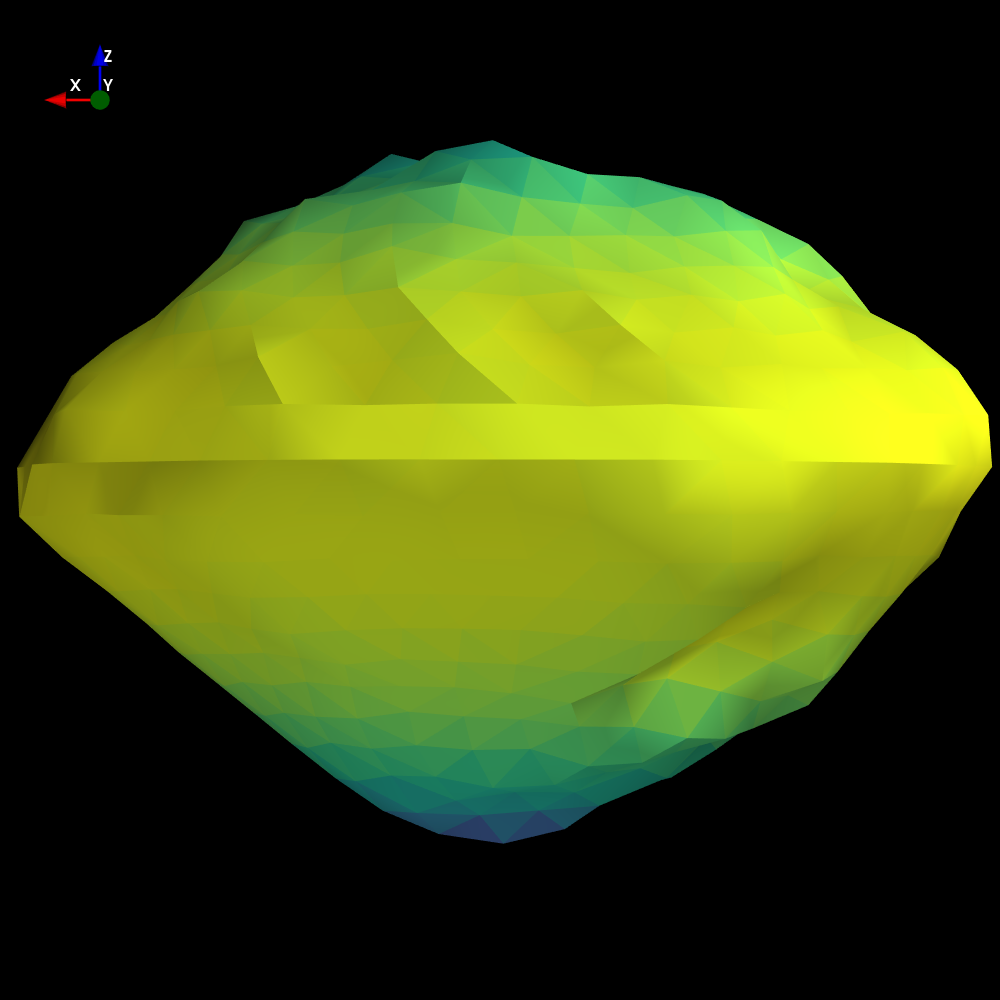}\\
   \includegraphics[width=0.26\textwidth ]{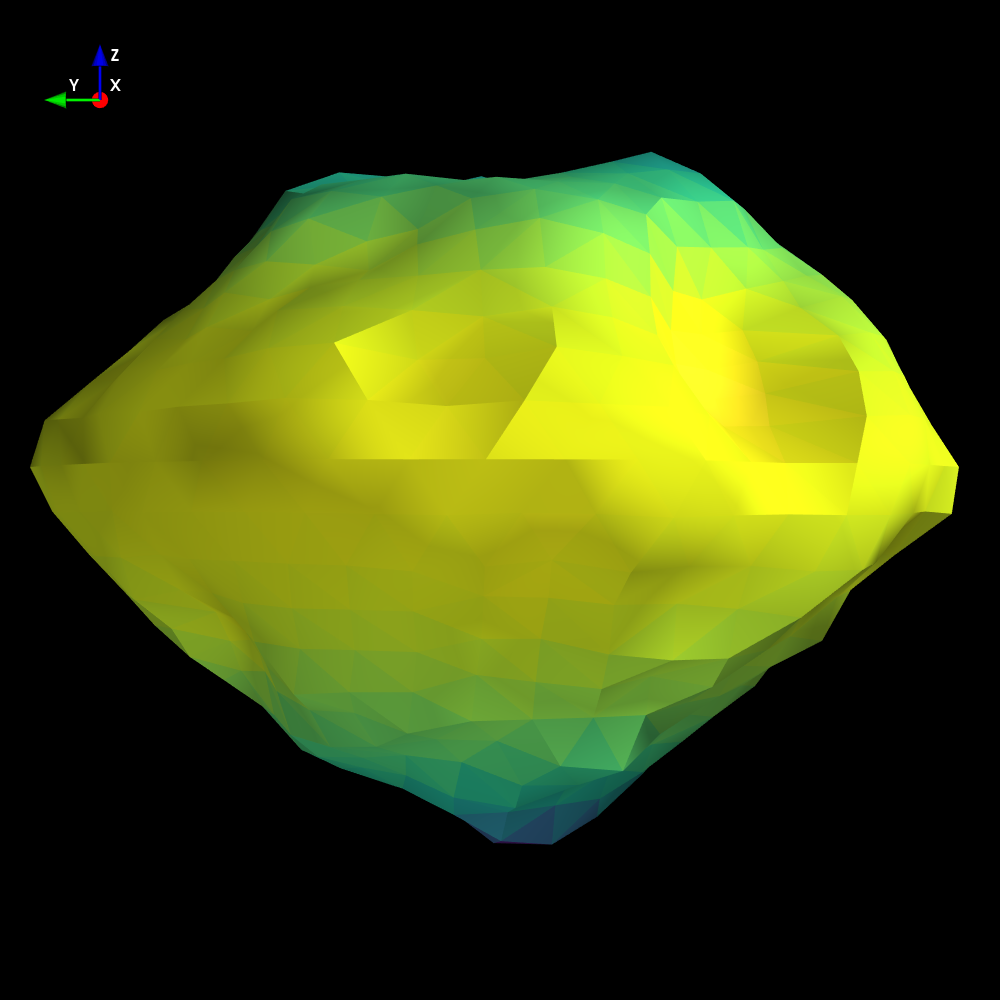} & 
   \includegraphics[width=0.26\textwidth ]{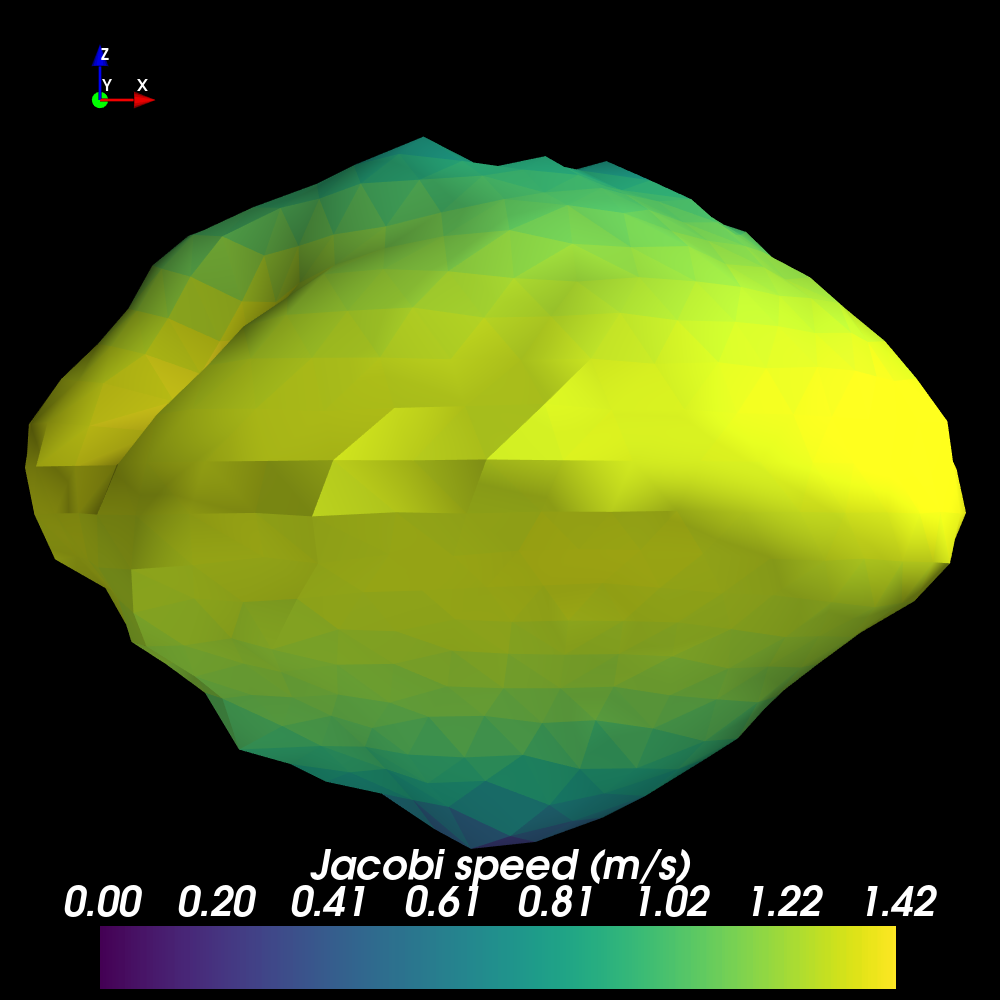} &
   \includegraphics[width=0.26\textwidth ]{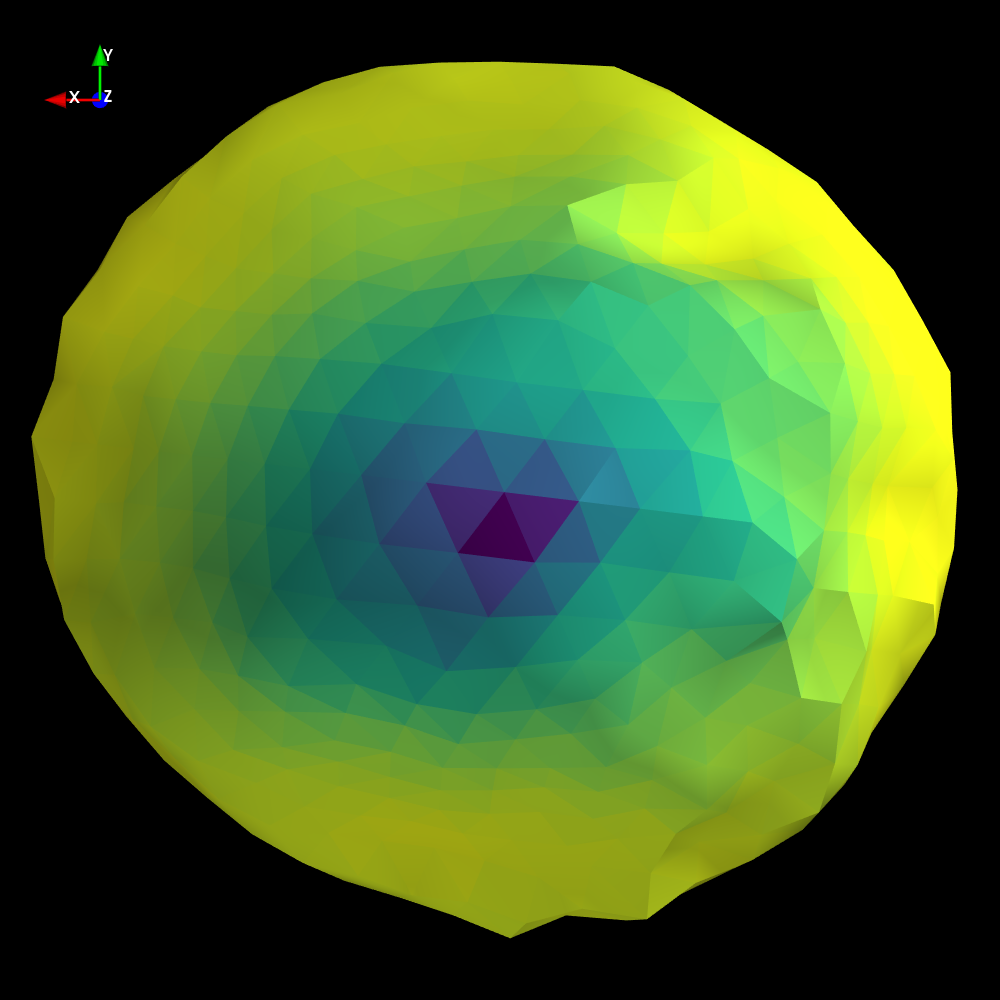}\\
  \end{tabular}
 \caption{Map of the Jacobi speeds on Phaethon's surface. The value corresponds to the expected velocity of an object when migrated from the highest geopotential point without energy loss.
 \label{fig:jacobi}}
\end{figure*}

Here, we present the figures that show the relative differences of geopotential between the mascon and polyhedron models (Figs. \ref{fig:pot_diff_simple} and \ref{fig:pot_diff_core}) and the map of the surface accelerations (Figs. \ref{fig:surf_acc} and \ref{fig:net_acc}), slopes (Fig. \ref{fig:slope}) and Jacobi speeds (Fig. \ref{fig:jacobi}). 

\FloatBarrier

\FloatBarrier

\section{Model parameter influence on stability} \label{app:model_slopes}

Table \ref{tab:model_slopes} lists the minimum, mean and maximum slopes of Phaethon after varying the bulk density, size and smoothness of the model. We note that a slope value of more than $90\degr$ indicates that cohesionless material on the region will be detached from the asteroid \citep{2019MNRAS.482.4243Y}.

\begin{table*}[h!]
 \caption{Slopes for different model parameters}              
 \label{tab:model_slopes}      
 \centering                                      
 \begin{tabular}{cccc|ccc}          
 \hline\hline                        
 \multicolumn{4}{c}{Model parameters} & \multicolumn{3}{c}{Slope (\degr)} \\
 Bulk density (g/cm\textsuperscript{3}) & $\Delta x, \Delta y$ (\%) & $\Delta z$ (\%) & Smoothness & Min. & Mean & Max.\\    
 \hline                                   
    \multirow{15}{*}{1.13} &  0  &  0  & 100  & 3.09  & 41.69 & 172.19 \\
         &     &     & 300  & 1.52  & 40.57 & 171.95 \\
         &     &     & 1000 & 0.96  & 39.68 & 167.33 \\
         & -3  &  -6 & 100  & 3.03  & 42.81 & 173.03 \\
         &     &     & 300  & 1.39  & 42.17 & 169.33 \\
         &     &     & 1000 & 1.34  & 41.37 & 172.16 \\
         &     &  +6 & 100  & 3.34  & 37.31 & 147.86 \\
         &     &     & 300  & 1.43  & 36.51 & 144.95 \\
         &     &     & 1000 & 1.81  & 35.59 & 141.06 \\
         & +3  &  -6 & 100  & 2.64  & 45.33 & 172.51 \\
         &     &     & 300  & 0.70  & 44.80 & 172.43 \\
         &     &     & 1000 & 1.22  & 44.20 & 173.98 \\
         &     &  +6 & 100  & 2.86  & 39.79 & 158.91 \\
         &     &     & 300  & 1.93  & 39.19 & 167.74 \\
         &     &     & 1000 & 0.85  & 38.35 & 170.65 \\
    \hline
    \multirow{15}{*}{1.58} &  0  &  0  & 100  & 0.31  & 18.10 & 52.00 \\
         &     &     & 300  & 0.37  & 16.84 & 51.38 \\
         &     &     & 1000 & 0.05  & 15.90 & 52.88 \\
         & -3  &  -6 & 100  & 0.32  & 17.59 & 53.93 \\
         &     &     & 300  & 0.44  & 16.68 & 53.88 \\
         &     &     & 1000 & 0.48  & 15.79 & 55.60 \\
         &     &  +6 & 100  & 0.92  & 18.23 & 51.73 \\
         &     &     & 300  & 1.09  & 17.29 & 51.43 \\
         &     &     & 1000 & 0.37  & 16.39 & 48.38 \\
         & +3  &  -6 & 100  & 0.41  & 17.37 & 51.39 \\
         &     &     & 300  & 0.55  & 16.53 & 51.02 \\
         &     &     & 1000 & 0.64  & 15.69 & 51.49 \\
         &     &  +6 & 100  & 0.31  & 17.85 & 49.92 \\
         &     &     & 300  & 0.40  & 16.93 & 49.76 \\
         &     &     & 1000 & 0.78  & 16.02 & 51.48 \\
    \hline
    \multirow{15}{*}{2.03} &  0  &  0  & 100  & 0.35  & 15.08 & 46.72 \\
         &     &     & 300  & 0.29  & 13.82 & 47.10 \\
         &     &     & 1000 & 0.68  & 12.63 & 47.44 \\
         & -3  &  -6 & 100  & 0.26  & 14.63 & 48.46 \\
         &     &     & 300  & 0.58  & 13.60 & 48.50 \\
         &     &     & 1000 & 0.16  & 12.42 & 49.85 \\
         &     &  +6 & 100  & 0.22  & 15.51 & 46.72 \\
         &     &     & 300  & 0.27  & 14.48 & 46.39 \\
         &     &     & 1000 & 0.40  & 13.32 & 48.38 \\
         & +3  &  -6 & 100  & 0.19  & 14.19 & 46.71 \\
         &     &     & 300  & 0.17  & 13.14 & 46.68 \\
         &     &     & 1000 & 0.49  & 11.99 & 45.89 \\
         &     &  +6 & 100  & 0.40  & 15.08 & 45.31 \\
         &     &     & 300  & 0.52  & 14.0  & 45.44 \\
         &     &     & 1000 & 0.26  & 12.86 & 46.30 \\    
 \hline
 \hline                                             
 \end{tabular}
 \tablefoot{Higher smoothness values correspond to smoother models, with a smoothness value of 100 representing the rough (nominal) model, 1000 indicating the smooth model, and 300 providing an intermediate level of smoothness.}
\end{table*}

\end{appendix}
\end{document}